\documentclass[10pt]{amsart}

\usepackage[margin=1.5in]{geometry}
\usepackage{colortbl}

\newtheorem{theorem}{Theorem}[section]

\newtheorem{proposition}{Proposition}[section]

\theoremstyle{definition}

\theoremstyle{remark}
\newtheorem{remark}[theorem]{Remark}

\numberwithin{equation}{section}

\usepackage{graphicx}
\usepackage{subcaption}
\usepackage{multirow}
\usepackage{bbm}
\usepackage{amssymb}

\begin{document}

\title{BGK models for inert mixtures: comparison and applications}

\author[S. Boscarino]{Sebastiano Boscarino}
\address{Sebastiano Boscarino\\
Department of Mathematics and Computer Science\\
University of Catania\\
95125 Catania, Italy} \email{boscarino@dmi.unict.it}

\author[S. Y. Cho]{Seung Yeon Cho}
\address{Seung Yeon Cho\\
Department of Mathematics and Computer Science\\
University of Catania\\
95125 Catania, Italy}
\email{chosy89@skku.edu}

\author[M. Groppi]{Maria Groppi}
\address{Maria Groppi\\
Department of Mathematical, Physical and Computer Sciences, University of Parma\\ Parco Area delle Scienze 53/A,
I–43124 Parma, Italy} \email{maria.groppi@unipr.it}

\author[G. Russo]{Giovanni Russo}
\address{Giovanni Russo\\
	Department of Mathematics and Computer Science\\
	University of Catania\\
	95125 Catania, Italy} \email{russo@dmi.unict.it}

\begin{abstract}
Consistent BGK models for inert mixtures are compared, first in their kinetic behavior and then versus the hydrodynamic limits that can be derived in different collision-dominated regimes.
The comparison is carried out both analytically and numerically, for the latter using an asymptotic preserving semi-Lagrangian scheme for the BGK models. Application to 
the plane shock wave in a binary mixture of noble gases is also presented.
\end{abstract}
\maketitle
\section{Introduction}
Since the seminal  paper of Bhatnagar, Gross and Krook \cite{BGK}, BGK models of the Boltzmann equation \cite{Cer,Kogan} play the role of simpler and effective modeling  tools to describe the dynamics of rarefied gases. Their importance is much more evident when gas mixtures are taken into account. However, the extension to an arbitrary mixture of monoatomic gases is not trivial, since interactions and exchanges between different components, as well as conservation of total momentum and energy, must be properly considered.\\
The first mathematically rigorous contribution appeared in \cite{AAP}, where the authors proposed a BGK model characterized by a single global operator for each species, able to reproduce the correct  exchange rates for momemtum and kinetic energy among species of the Boltzmann equations for mixtures, assuming intermolecular potentials of Maxwell molecules type.
Later, several BGK models have been introduced, characterized by different structures of the BGK operators (see for instance \cite{HAACK,KM,KPP} and the reference therein). In \cite{GRS2} the authors consider a BGK model for inert mixtures whose single global collision operator for each species allows to correctly reproduce the conservation of total momentum and kinetic energy; moreover, in the same paper a comparison among this BGK model and the one  proposed in \cite{AAP} is presented and discussed. For both BGK models in \cite{AAP} and \cite{GRS2} the possibility to include simple chemical reactions has been considered, and their extension to bimolecular chemical reactions in mixtures of monoatomic gases can be found in  \cite{PoFGS} and \cite{BGS,GRjS}, respectively.\\
Recently, in \cite{BBGSP1} a further different consistent BGK-type model is built up, that mimics the structure of the Boltzmann equations for mixtures, namely in which the collision operator for each species is a sum of bi-species BGK operators. This last model is  consistent and well--posed as the one in \cite{AAP}, but in addition allows to deal with general intermolecular potentials. The exchange rates for momentum and energy of each BGK operator coincide by construction with the corresponding exchange rates of each Boltzmann binary integral operator, getting thus exact conservations.
The structure of the collision operator in this case allows to consistently derive evolution equations for the main macroscopic fields in different hydrodynamic regimes, according to the dominant collisional phenomenon \cite{BBGSP2,BGM2}. \\
This paper is aiming at comparing the behaviors of the above mentioned BGK models for inert mixtures of monoatomic gases in various regimes, from kinetic to hydrodynamic, and to highlight their analogies and discrepancies. In addition, two different Navier-Stokes hydrodynamic limits, characterized by global velocity and temperature or multi-velocity and multi-temperature, respectively, and obtained from the consistent BGK model in \cite{BBGSP1}, are numerically tested with respect to their capability to reproduce the correct dynamics of realistic mixtures of noble gases, with particular attention to the case of mixtures of heavy and light particles (the so called $\varepsilon$-mixtures \cite{Galkin}). For the numerical solutions to BGK models, semi-Lagrangian methods proposed in \cite{CBGR,GRS2} will be adopted together with a conservative reconstruction technique \cite{BCRY2,BCRY3}, which enables us to capture the correct behaviors of hydrodynamic limit models. As regards details and properties of semi-Lagrangian schemes (for single gas BGK model), we refer to papers regarding the construction of high order semi-Lagrangian schemes \cite{BCRY,GRS,RS}, treatment of boundary problems \cite{GRS1,RF} and convergence analysis \cite{BCRY1,RSY,RY}.
	\\
	The paper is organized as follows. In section 2, we briefly recall the classical Boltzmann equation for inert gas mixtures. Section 3 is devoted to the description of the three BGK models considered in this paper. Next, in section 4, we study analytically the discrepancy between BGK models at the level of collision operators. Then, in section 5, we present hydrodynamic limits at Navier-Stokes level that can be obtained from the kinetic BGK models. In section 6, we perform numerical experiments in which we investigate the discrepancy of the three BGK models, compare them with their two corresponding Navier-Stokes limits, and study the Riemann problem and the steady shock structure in binary mixtures of noble gases, with different mass ratios.

\section{Kinetic Boltzmann-Type Equations}
Let us consider a mixture of $L$ monoatomic inert gases. Under the assumption that $s$-th gas has a mass $m_s>0$, its dynamics can be described through the distribution functions $f_s({\bf x},{\bf v},t)$, $s=1,\cdots,L$ defined on phase space $({\bf x},{\bf v})\in \mathbb{R}^3 \times \mathbb{R}^3$ at time $t>0$, whose evolution is governed by the Boltzmann-type equations:
\begin{align}\label{Boltzmann eqn}
\frac{\partial {f_s}}{\partial t} + {\bf v}\cdot \nabla_{\textbf{x}} f_s = Q_{s},
\end{align}
where $Q_s$ is the collision term of the $s$-th species, which collects bi-species collision operators between $s$-th and other $k$-th gases: 
\[
Q_s=\sum_{k=1}^N Q_{sk}(f_s,f_k).
\]
The binary collision operator $Q_{sk}$ can be cast as
\begin{align}\label{Qsk}
Q_{sk}(f_s,f_k)= \int_{\mathbb{R}^3 \times \mathbb{S}^2} d{\bf w}\,d{\boldsymbol{\omega}}\, g_{sk}
\left(|\bf{y}|,\hat{\bf{y}}\cdot {\boldsymbol{\omega}}\right)
 \bigg[ f_s({\bf v}')f_k({\bf w}')-f_s({\bf v})f_k({\bf w})\bigg]
\end{align}
with a non-negative scattering kernel $g_{sk}$ depending on intermolecular potentials. Here we use the integration variable ${\bf w} \in \mathbb{R}^3$, a unit vector on a sphere $\boldsymbol{\omega} \in \mathbb{S}^2$, the relative velocity ${\bf y}:={\bf v}-{\bf w}$ and its unit vector $\hat{{\bf y}}:={\bf y}/|{\bf y}|$. The other two variables ${\bf v}'$ and ${\bf w}'$ are post collisional velocities of $s$-th and $k$-th gases whose masses are $m_s$, $m_k$ and pre-collisional velocities are ${\bf v}$, ${\bf w}$, respectively:
\[
{\bf v}'= \frac{m_s {\bf v} + m_k{\bf w}}{m_s + m_k} + \frac{m_k}{m_s + m_k}|{\bf y}|\boldsymbol{\omega},
\quad 
{\bf w}'= \frac{m_s {\bf v} + m_k{\bf w}}{m_s + m_k} - \frac{m_s}{m_s + m_k}|{\bf y}|\boldsymbol{\omega}.
\]
(for more details, see for instance \cite{Cer} and references therein).

The distribution function $f_s$, $s=1,\dots,L$, can be used to reproduce $s$-species macroscopic quantities such as number density $n_s$, average velocity $u_s$, absolute temperature $T_s$:
\begin{align*}
n_s=\langle f_s,1 \rangle, \quad n_s u_s=\langle f_s,{\bf v} \rangle, \quad 3n_sK_BT_s=m_s\langle f_s,|{\bf v}-u_s|^2 \rangle 
\end{align*}
where
\begin{align*}
\langle f, h\rangle := \int_{\mathbb{R}^3} dv f({\bf v})h({\bf v}).
\end{align*}
Similarly, global macroscopic variables such as  number density $n$, mass density $\rho$,  velocity $u$ and temperature $T$ of the mixture can be obtained as follows:
\begin{align*}
n=\sum_{s=1}^L n_s,\quad \rho=\sum_{s=1}^L \rho_s,\quad \rho_s = m_s n_s,\quad s=1,\cdots,L\cr
u= \frac{1}{\rho} \sum_{s=1}^L \rho_s u_s, \quad 3nK_BT=3\sum_{s=1}^L n_sK_BT_s + \sum_{s=1}^L \rho_s|u_s-u|^2	
\end{align*}
The equilibrium solution to \eqref{Boltzmann eqn} is given by the Maxwellian which shares a common velocity $u$ and temperature $T$:
\[
f_s^{eq} = n_s M\left({\bf v};u,\frac{K_B T}{m_s}\right)
\]
where
\begin{align}\label{M ref}
		M\left({\bf v};a,b\right) \equiv \left(\frac{1}{2\pi b}\right)^{3/2} \exp \left(- \frac{1}{2b}\left|{\bf v}-a\right|^2\right), \quad a \in \mathbb{R}^3,\, b >0.
\end{align}

The structure of the collision operators $Q_s$ allows to guarantee fundamental properties of the Boltzmann equations for inert gas mixtures: conservation laws,  
uniqueness of equilibrium solutions, H- theorem.

\section{BGK-type models}
In this section we briefly recall the three different BGK models for eqns. (2.1)-(2.2) which  have been compared in this paper.

\subsection{The BGK Model of Andries, Aoki and Perthame (AAP model)}
In \cite{AAP}, a BGK-type model was proposed, that allows to reproduce the same exchange rate in momentum and energy of the Boltzmann equation \eqref{Boltzmann eqn}. In this model the collision term $Q_s$ \eqref{Qsk} of the Boltzmann equation is replaced by a relaxation operator which drives the evolution towards an attracting auxiliary Maxwellian $M^s$, depending on fictitious parameters. In spite of such structural change, this BGK model still satisfies the main properties of the Boltzmann equation such as conservation law, $H$-theorem, indifferentiability principle. The scaled model equations are described by 
\begin{align}\label{bgk AAP}
\frac{\partial{f_s}}{\partial{t}} + {\bf v} \cdot \nabla_{\textbf{x}}{f_s} =  \frac{\nu_s}{\varepsilon}\left(n_{s}M^{s}-f_s\right),\quad s=1, \cdots, L,
\end{align}
where $\varepsilon$ is the Knudsen number, $\nu_s$ is the collision frequency for $s$-species gas and $M^{s}$ is the attracting Maxwellian:
\begin{align*} 
		M^{s}=M\left({\bf v};u^{s},\frac{K_BT^{s}}{m_s}\right),
\end{align*}
where $M$ is defined in \eqref{M ref}. Notice that $M^{s}$ is defined in terms of fictitious parameters $u^s, T^s$ (different from the actual fields $u_s, T_s$), which, under the assumption of Maxwell molecules interaction potential, are explicit functions of the actual moments of the distribution function as follows:
\begin{align}
\begin{split}\label{usTs}
u^s&=u_s +  \frac{1}{m_sn_s\nu_s}\sum_{k=1}^{L}\xi^{sk}u_k,\cr
T^{s}&= T_s - \frac{m_s}{3K_B} \left(|u^s|^2-|u_s|^2\right)  + \frac{2}{3n_s K_B\nu_s}\sum_{k=1}^{L}\gamma^{sk}T_k\cr
&\quad + \frac{2}{3n_s K_B\nu_s}\sum_{k=1}^{L}\nu_1^{sk}\frac{m_sm_kn_sn_k}{(m_s + m_k)^2}\left(m_su_s + m_ku_k\right)\left(u_k-u_s\right),
\end{split}
\end{align}
where
\begin{align}\label{xi gamma}
\begin{split}
\xi_{sk}&= \nu_1^{sk}\frac{m_sm_kn_sn_k}{m_s + m_k} -\delta_{sk} \sum_{r=1}^L \nu_1^{sr}\frac{m_s m_r n_s n_r}{m_s + m_r}\cr
\gamma^{sk}&= 3K_B\nu_1^{sk}\frac{m_sm_kn_sn_k}{(m_s + m_k)^2} -\delta_{sk} \sum_{r= 1}^L 3K_B\nu_1^{sr}\frac{m_s m_r n_s n_r}{(m_s + m_r)^2}.
\end{split}
\end{align}
The collision frequency $\nu_\ell^{sk}$ is defined by
\begin{align*}
\nu_\ell^{sk}= 2\pi |g| \int_0^\pi  g(\omega) (1-\cos \omega)^\ell \sin \omega d\omega,
\end{align*}
and satisfies $\nu_\ell^{sk} \leq 2 \nu_0^{sk}$ for $\ell=1,2$. We remark that this model is well defined with the choice
\begin{align*}
\nu_s = \sum_{k=1}^L \nu_0^{sk} n_k,
\end{align*}
which guarantees the positivity of temperature. Moreover, for consistency,
we hereafter assume the following case:
	\begin{align}\label{collision fre rel}
	\nu_0^{sk}= \nu_1^{sk} =: \lambda_{sk}.
	\end{align}

\subsection{The BGK model preserving global conservations (GS model)}

Another BGK-type model with one attracting Maxwellian for each species has been proposed in \cite{BGS,GRS2}. The fictitious parameters are adjusted to impose the same conservation laws of the Boltzmann equation \eqref{Boltzmann eqn}, namely species number densities, global momentum, total kinetic energy. In \cite{BGS,GRS2}, the model has been originally designed to describe a bimolecular reversible chemical reaction in a four species mixture; the model has been then adapted to a general $L$ species inert gas mixture in \cite{GRS2}. The scaled model reads 
\begin{align}\label{bgk GS}
	\frac{\partial{f_s}}{\partial{t}} + {\bf v} \cdot \nabla_{\textbf{x}}{f_s} =  \frac{\nu_s}{\varepsilon}\left(n_{s}M^{s}_{GS}-f_s\right).
\end{align}
where $\nu_s$ is the collision frequency for $s$-species gas and $M^{s}_{GS}$ is the attracting Maxwellian:
\begin{align}\label{MGS}
		M_{GS}^{s}=M \left({\bf v};\bar{u},\frac{K_B\bar{T}}{m_s}\right),
\end{align}
where $M$ is defined in \eqref{M ref}.
Note that it depends only on the auxiliary parameters $\bar{u}$ and $\bar{T}$, which are determined by imposing the conservation of total momentum and energy:
\begin{align*}
	\sum_{s=1}^L\int_{\mathbb{R}^3} m_s \begin{pmatrix}
		{\bf v}\\|{\bf v}|^2/2
	\end{pmatrix} \left(n_{s}M^{s}_{GS}-f_s\right) \,d{\bf v}&= 0,	
\end{align*}
Consequently, we obtain the following representation of $\bar{u}$ and $\bar{T}$ in terms of the actual macroscopic fields $u_s$ and $T_s$:
\begin{align*}
	\begin{split} 
		\bar{u}&=\frac{\sum_{s=1}^L \nu_s m_s n_s u_s}{\sum_{s=1}^L \nu_s m_s n_s},
	\end{split}
\end{align*}
\begin{align*}
	\begin{split} 
		\bar{T}&= \frac{\sum_{s=1}^L \nu_s n_s \left(m_s \left(|u_s|^2 - |\bar{u}|^2 \right) + 3K_BT_s\right) }{3K_B\sum_{s=1}^L\nu_s n_s}.
	\end{split}
\end{align*}
In \cite{GRS2}, it is proved that the positivity of auxiliary temperature $\bar{T}$ is guaranteed and the H-theorem holds for the space homogeneous case.

\subsection{A general consistent BGK model for inert gas mixtures (BBGSP model)}
In \cite{BBGSP1}, authors introduce a different BGK-type model whose BGK operators $Q_s$ mimic the structure of the Boltzmann ones, namely are sums of bi-species operators $Q_{sk}$, each of them prescribing the same exchange rates of the corresponding term of the Boltzmann equations.   This model also satisfies the main qualitative properties of Boltzmann equation such as conservation laws, $H$-theorem, indifferentiability principle. The scaled equations read
\begin{align}\label{bgk bbgsp}
\frac{\partial{f_s}}{\partial{t}} + {\bf v} \cdot \nabla_{\textbf{x}}{f_s} =  \frac{1}{\varepsilon}\sum_{k = 1}^{L} \nu_{sk}\left(n_{s}M_{sk}-f_s\right),\quad s=1, \cdots, L,
\end{align}
with
\[
M_{sk}=M\left({\bf v};u_{sk},\frac{K_BT_{sk}}{m_s}\right),
\]
where $M$ is defined in \eqref{M ref}.
The auxiliary parameters $u_{sk}$ and $T_{sk}$ are defined by
\begin{align}
\begin{split}\label{uskTsk}
u_{sk}&= (1-a_{sk})u_s + a_{sk}u_k\cr
T_{sk}&= (1-b_{sk})T_s + b_{sk}T_k +  \frac{\gamma_{sk}}{K_B}\left|u_{s}-u_{k}\right|^2
\end{split}
\end{align}
with
\begin{align}\label{abgamma original}
a_{sk}= \frac{\lambda_{sk} n_k m_k}{\nu_{sk}(m_s + m_k)}, \quad 
b_{sk}= \frac{2a_{sk} m_s}{m_s + m_k}, \quad \gamma_{sk}= \frac{m_s a_{sk}}{3}\left( \frac{2m_k}{m_s + m_k} -a_{sk}\right).
\end{align}
In \cite{BBGSP1}, authors proved that the positivity of $T_{sk}$ is guaranteed by
\begin{align*} 
	T_s>0, \quad T_k>0, \quad \nu_{sk} \geq \frac{1}{2}\lambda_{sk} n_k.
\end{align*}
Considering this, throughout this paper, we set  $\nu_{sk}=\nu_{0}^{sk}n_k = \lambda_{sk} n_k$. This implies
\begin{align*}
a_{sk}= \frac{m_k}{m_s+m_k} , \quad 
b_{sk}= \frac{2a_{sk} m_s}{m_s + m_k}, \quad \gamma_{sk}= \frac{m_s a_{sk}}{3}\left( \frac{2m_k}{m_s + m_k} -a_{sk}\right).
\end{align*}

\section{Discrepancy between BGK-type models}\label{sec discrepancy}
In this section, our goal is to check the discrepancy between AAP model \eqref{bgk AAP} and BBGSP model \eqref{bgk bbgsp}. For this, we start from multiplying the two models \eqref{bgk AAP} and \eqref{bgk bbgsp} by $\displaystyle \frac{\varepsilon}{n_s}$, and subtract the resulting equations. After then, we expand two Maxwellians $M^s$, $M_{sk}$ around $u_s$ and $T_s$ to obtain
\begin{align*}
	&\nu_s M^s - \sum_{k=1}^L \nu_{sk}M_{sk}\cr
		&=\sum_{k=1}^L \nu_{sk} \Bigg[ \left( M_s+ \frac{\partial M}{\partial u}\bigg|_{(u,T)=(u_s,T_s)}(u^s-u_s) + \frac{\partial M}{\partial T}\bigg|_{(u,T)=(u_s,T_s)}(T^s-T_s) +\text{h.o.t.} \right) \cr
		&\qquad\quad\quad- \left(M_s+ \frac{\partial M}{\partial u}\bigg|_{(u,T)=(u_s,T_s)}(u_{sk}-u_s) + \frac{\partial M}{\partial T}\bigg|_{(u,T)=(u_s,T_s)}(T_{sk}-T_s) +\text{h.o.t.} \right)\Bigg],
\end{align*}
where
\begin{align*}
M_s&:=M\left({\bf v};u_{s},\frac{K_BT_{s}}{m_s}\right),\cr
\frac{\partial M}{\partial u}\bigg|_{(u,T)=(u_s,T_s)}&:=\frac{m_s({\bf v}-u_s)}{K_BT_s}M_s,\cr
\frac{\partial M}{\partial T}\bigg|_{(u,T)=(u_s,T_s)}&:=\left(-\frac{3}{2T_s}+\frac{m_s|{\bf v}-u_s|^2}{2K_BT_s^2}\right)M_s.
\end{align*}
Then, we have
\begin{align*}
	\nu_s M^s - \sum_{k=1}^L \nu_{sk}M_{sk} 
	&= \left(\frac{\partial M}{\partial u}\bigg|_{(u,T)=(u_s,T_s)}\right) \mathcal{E}_u +  \left(\frac{\partial M}{\partial T}\bigg|_{(u,T)=(u_s,T_s)}\right)\mathcal{E}_T + \text{h.o.t.}
\end{align*}
where
\begin{align}\label{leading order terms}
	\begin{split}
		\mathcal{E}_u&:=\sum_{k=1}^L \nu_{sk} \left( u^s-u_{sk}\right),\quad 
		\mathcal{E}_T:=\sum_{k=1}^L \nu_{sk}\left(T^s-T_{sk}\right).
	\end{split}
\end{align}
Here $\mathcal{E}_u$ and $\mathcal{E}_T$ are the contributions of leading order errors with respect to the derivative of $u$ and $T$, and all the remainders are denoted by h.o.t..

In the following Proposition, we provide the explicit forms of $\mathcal{E}_u$ and $\mathcal{E}_T$ (the proof can be found in Appendix \ref{Appendix A1}.)
\begin{proposition}\label{prop error}
	Suppose that $M_s$ in \eqref{bgk AAP} and $M_{sk}$ in \eqref{bgk bbgsp} are sufficiently smooth with respect to macroscopic variables, velocity and temperature. Assuming that $\nu_s=\sum\nu_{sk}$, the leading error terms $\mathcal{E}_U$ and $\mathcal{E}_T$ in \eqref{leading order terms} are given by
	\begin{align*}
		(1)~ \mathcal{E}_u&=0\cr
		(2)~ \mathcal{E}_T&= \frac{m_s}{3K_B}\sum_{k=1}^L \nu_{sk}  (a_{sk})^2 |u_s-u_k|^2 - \frac{m_s}{3K_B} \left(  \frac{1}{\nu_s}\sum_{r=1}^{L}\nu_{sr}a_{sr}(u_r-u_s) \right)\cdot \left( \sum_{r=1}^{L}\nu_{sr}a_{sr}(u_r-u_s) \right).
	\end{align*}
\end{proposition}


\begin{remark}\label{remark 4.1}
	In a similar manner, we can compare the AAP model \eqref{bgk AAP} and the GS model \eqref{bgk GS} as follows:
	\begin{align*}
			\nu_s M^s - \nu_{s}M_{GS}^s 
			&= \left(\frac{\partial M}{\partial u}\bigg|_{(u,T)=(u_s,T_s)}\right) \bar{\mathcal{E}}_u +  \left(\frac{\partial M}{\partial T}\bigg|_{(u,T)=(u_s,T_s)}\right)\bar{\mathcal{E}}_T + \text{h.o.t.,}
	\end{align*}
	where
	\begin{align}\label{leading order terms GS}
		\begin{split}
			\bar{\mathcal{E}}_u&:=\nu_{s} \left( u^s-\bar{u}\right),\quad 
			\bar{\mathcal{E}}_T:= \nu_{s}\left(T^s-\bar{T}\right).
		\end{split}
	\end{align}
	Note that $\mathcal{E}_u$ in this case does not vanish:
	\begin{align*}
		\bar{\mathcal{E}}_u&=\sum_{r\neq s}\left(\nu_1^{sr}\frac{m_rn_r}{m_s + m_r}-\frac{\nu_s\nu_r m_r n_r}{\sum_{r=1}^L \nu_r m_r n_r}\right)(u_r-u_s).
	\end{align*}
	Due to the complexity of $\bar{\mathcal{E}}_T$, we provide its form in Appendix \ref{appendix AAPBG}. This analysis shows a more pronounced discrepancy between AAP and GS models; it will be confirmed and quantified in the next section at Navier-Stokes level, and discussed later in the numerical tests in section \ref{sec test 1}.
\end{remark}





\section{Hydrodynamic limits at Navier–Stokes (NS) level}
Here we describe the Navier-Stokes asymptotics that can be derived from the different BGK models to $\mathcal{O}(\varepsilon)$. 
\subsection{NS equations with global velocity and temperature}\label{sec NS equations with single velocity and temperature}
In \cite{BBGSP2}, the hydrodynamic limit of the BBGSP model \eqref{bgk bbgsp}-\eqref{abgamma original} at the Navier-Stokes level is derived using the Chapman-Enskog expansion in a collision dominated regime. The equations for macroscopic variables $n$, $u$ and $T$, obtained as $\varepsilon$-order closure of the macroscopic equations (moments of the BGK ones), are given by
\begin{align}\label{NSE}
\begin{split}
	&\frac{\partial n_s}{\partial t} + \nabla \cdot(n_s u) + \varepsilon \nabla \cdot(n_s u_s^{(1)})=0, \quad s=1,\cdots,L\cr
&\frac{\partial}{\partial t}(\rho u) + \nabla \cdot(\rho u \otimes u) + \nabla(n K_BT) + \varepsilon \nabla \cdot(P^{(1)})=0,\cr
&\frac{\partial}{\partial t}\left(\frac{1}{2} \rho|u|^2 + \frac{3}{2}nK_BT\right) + \nabla \cdot\left[\left(\frac{1}{2}\rho |u|^2 + \frac{5}{2}nK_BT\right)u\right] + \varepsilon \nabla \cdot(P^{(1)}\cdot u) + \varepsilon \nabla \cdot q^{(1)}=0, 
\end{split}
\end{align}
where $u_s^{(1)}$, $P^{(1)}$, $q^{(1)}$ are first order corrections with respect to $\varepsilon$. The diffusion velocity $u_s^{(1)}$ takes the following form:
\begin{align}\label{us1}
	u_s^{(1)}&= \sum_{k=1}^L \frac{L_{sk}}{\rho_s\rho_k}\nabla(n_kK_BT)
\end{align}
where the symmetric matrix $L$ is computed as 
\begin{align*}
	L&=\tilde{\textsf{M}}^{-1}\Omega,\quad \Omega_{sk}=\rho_s \delta_{sk} - \frac{\rho_s\rho_k}{\rho},\cr 
	\tilde{\textsf{M}}_{sk}&=\textsf{M}_{sk}-\frac{1}{2}\kappa, \quad \text{where} \, \kappa =  \min_{s\ne k} \textsf{M}_{sk}
\end{align*}
\begin{align}\label{MAAP}
	\textsf{M}_{sk}&=\frac{\lambda_{sk}\rho_s}{m_s+m_k} -\delta_{sk} \sum_{r=1}^{L}\frac{\lambda_{sr}\rho_{r}}{m_s+m_{r}}.
\end{align}
The first order corrections for pressure tensor $P^{(1)}$ is of the following form:
\begin{align}\label{P1}
	P_{\alpha \beta}^{(1)} &= - \mu \left( \frac{\partial u_\alpha}{\partial x_\beta} + \frac{\partial u_\beta}{\partial x_\alpha} - \frac{2}{3}\nabla\cdot u \delta_{\alpha\beta}\right),\quad 1 \leq \alpha,\beta \leq 3,
\end{align}
where the viscosity coefficient $\mu$ is given by 
\begin{align}\label{viscmu}
	\mu:= \sum_{s=1}^L \frac{n_s K_B T}{\sum_{s=1}^L \nu_{sk}^{(0)} }.
\end{align}
Here we denote by 
$\nu_{sk}^{(0)}$ the leading order in the expansion of the collision frequency $\nu_{sk}$.
The heat flux $q^{(1)}$ is given by 
\begin{align}\label{q1}
	q^{(1)} &= \frac{5}{2}K_BT \sum_{s=1}^L n_su_s^{(1)} - \lambda \nabla T,
\end{align}
where $\lambda$ is the thermal conductivity coefficient:
\begin{align}\label{thcond}
	\lambda=\frac{5}{2} K_B^2 T \sum_{s=1}^L \frac{n_s}{m_s\sum_{k=1}^L\nu_{sk}^{(0)}}.
\end{align}
For detailed description of this model, we refer to \cite{BBGSP2}. It is remarkable that such results are in 
complete agreement with those obtained from the AAP model \eqref{bgk AAP}-\eqref{usTs} \cite{AAP}. This is not surprising since  
these results are indeed exact for the Boltzmann equations with Maxwell molecules.

The same structure of the Navier-Stokes equations (\ref{NSE}), with first order corrections (\ref{us1}),(\ref{P1}) and (\ref{q1}), is reproduced also by the $\varepsilon$-order asymptotics of the GS model  (\ref{bgk GS})-(\ref{MGS}). However, the matrix $\textsf{M}$ involved in the Fick's law (\ref{us1}) for diffusion velocities is different from (\ref{MAAP}); indeed, in case of BGK model  (\ref{bgk GS})-(\ref{MGS}), such matrix accounts for all mechanical interactions via the inverse relaxation times $\nu_s$, whereas for the other BGK models considered above only the bi-species collision frequencies $\nu^1_{sk}=\lambda_{sk}$ are involved. Its expression for GS model is given by \cite{Bisi Spiga Proc}
$$\textsf{M}^{GS}_{sk}=\frac{\nu_s \nu_k}{\sum_{r=1}^{L}\rho_k \nu_k} \rho_s-\nu_s\delta_{sk}$$
and consequently the diffusion velocities $u_s^{(1)}$ in the Navier-Stokes equations are quantitatively different from the previous ones.
As regards the transport coefficients, namely viscosity $\eta$ and thermal conductivity $\lambda$, they are exactly given by (\ref{viscmu}) and (\ref{thcond}), respectively, also for the BGK model (\ref{bgk GS})-(\ref{MGS}).

\subsubsection{Representation of NS equations for $n_s$, $u$ and $T$}
The system \eqref{NSE} can be rewritten in the following form, which is more convenient for its numerical treatment:
\begin{align}\label{NSE rewrite}
\begin{split}
\frac{\partial n_s}{\partial t} &=- \nabla \cdot(n_s u) - \varepsilon \nabla \cdot(n_s u_s^{(1)}), \quad s=1,\cdots,L\cr
\frac{\partial u}{\partial t}&=\frac{u}{\rho} \left(  \nabla \cdot\left(\sum_{i=1}^L\varepsilon \rho_s u_s^{(1)}\right)\right) - u\nabla \cdot u -  \frac{\nabla(n K_B T)}{\rho} - \frac{\varepsilon \nabla \cdot(P^{(1)})}{\rho}\cr
\frac{\partial T}{\partial t}&=
-\frac{|u|^2}{3nK_B} \left(  \nabla \cdot\left(\sum_{i=1}^L\varepsilon \rho_s u_s^{(1)}\right)\right) + \frac{2u}{3nK_B} \cdot \left( \varepsilon \nabla \cdot(P^{(1)})\right) +\frac{T}{n}\left(  \nabla \cdot\left(\sum_{s=1}^L\varepsilon n_s u_s^{(1)}\right) \right)\cr 
&\quad-\left( \nabla T\right) \cdot u - \frac{2}{3}T \nabla \cdot u - \frac{2}{3nK_B}\varepsilon \nabla \cdot(P^{(1)}\cdot u) - \frac{2}{3nK_B}\varepsilon \nabla \cdot q^{(1)}.
\end{split}
\end{align}

\subsection{NS equations with multi-velocity and temperature}
An interesting point of \eqref{bgk bbgsp} is the possibility of allowing different hydrodynamic limits, thanks to the structure of the BBGSP collision operators as a sum of bispecies relaxation terms. In \cite{BGM2}, the authors consider the case in which intra-species collisions are the dominant process in the evolution of the mixture. This occurs for instance in the so called $\varepsilon$-mixtures of heavy and light gases  \cite{Galkin}, where molecules with very disparate masses exchange energy more slowly than molecules of the same species, and also in  some applications to plasmas and astrophysics \cite{VK}. In this case it is possible to define a proper Knudsen number and obtain the adimensional scaled equations:	
	\begin{align}\label{bgk bbgsp multi}
	\frac{\partial{f_s}}{\partial{t}} + {\bf v} \cdot \nabla_{\textbf{x}}{f_s} =  \frac{1}{\varepsilon} \nu_{ss}\left(n_{s}M_{ss}-f_s\right) + \frac{1}{\kappa}\sum_{k\neq s}^{L} \nu_{sk}\left(n_{s}M_{sk}-f_s\right).
	\end{align}
	This scaling leads to a Navier-Stokes model of multi-velocity and multi-temperature for $s$-species gases. In this case, the equations for macroscopic variables $n_s$, $u_s$ and $T_s$ are given by
	\begin{align}\label{NSE multi}
	\begin{split}
	&\frac{\partial n_s}{\partial t} + \nabla \cdot(n_s u_s) =  0,\cr
	&\frac{\partial}{\partial t}(\rho_s u_s) + \nabla \cdot(\rho_s u_s \otimes u_s) + \nabla(n_s K_B T_s) + \varepsilon \nabla \cdot(P_s^{(1)})= \frac{1}{\kappa}\sum_{k\neq s }^{L}\mathcal{R}_{sk},\cr
	&\frac{\partial}{\partial t}\left(\frac{1}{2} \rho_s|u_s|^2 + \frac{3}{2}n_sK_BT_s\right) + \nabla \cdot\left[\left(\frac{1}{2}\rho_s |u_s|^2 + \frac{5}{2}n_sK_BT_s\right)u_s\right] \cr
	&\qquad\qquad\qquad\qquad\qquad\qquad + \varepsilon \nabla \cdot(P_s^{(1)}\cdot u_s) + \varepsilon \nabla \cdot q_s^{(1)}=\frac{1}{\kappa}\sum_{k\neq s }^{L}\mathcal{S}_{sk},  
	\end{split}
	\end{align}
	for $s=1,\cdots,L$, with
	\begin{align*}
	\begin{split}
	\mathcal{R}_{sk}&= \lambda_{sk}m_{sk}n_sn_k(u_k-u_s),\cr 
	\mathcal{S}_{sk}&= \lambda_{sk}\frac{m_{sk}}{m_s + m_k}n_sn_k\left[(m_su_s+m_ku_k)\cdot (u_k-u_s) + 3K_B(T_k - T_s)\right]\cr
	m_{sk} &= \frac{m_sm_k}{m_s + m_k}.
	\end{split}
	\end{align*}
Here the pressure tensor is given by
	\begin{align*}
	P_s^{(1)} &= -\frac{n_sK_BT_s}{\nu_{ss}^{(0)}} \mathcal{H}_s 
	+ \frac{1}{\nu_{ss}^{(0)}} \sum_{k \neq s}^L \nu_{sk}^{(0)}(a_{sk}^{(0)})^2 m_sn_s \left[ (u_s-u_k) \otimes (u_s-u_k) - \frac{1}{3}|u_s-u_k|^2I\right]\cr
	\mathcal{H}_{s,\alpha \beta}&= \frac{\partial u_{s,\alpha}}{\partial x_\beta} + \frac{\partial u_{s,\beta}}{\partial x_\alpha} -\frac{2}{3}\cdot \nabla u_s \delta_{\alpha \beta},
	\end{align*}
	and the heat flux vector takes the following form: 
	\begin{align*} 
	q_s^{(1)}&= -\frac{5}{2}\frac{n_sK_B^2T_s}{n_s \nu_{ss}^{(0)}}\nabla T_s + 5\frac{m_sn_s}{\nu_{ss}^{(0)}}\sum_{k \neq s}^L \frac{\nu_{sk}^{(0)} (a_{sk}^{(0)})^2}{m_s + m_k} K_B(T_k-T_s)(u_k-u_s)\cr
	&\quad +\frac{1}{3}\frac{m_sn_s}{\nu_{ss}^{(0)}}\sum_{k \neq s}^L \nu_{sk}^{(0)} (a_{sk}^{(0)})^2 \left(\frac{5m_k}{m_s + m_k}-a_{sk}^{(0)}\right) \left|u_k-u_s\right|^2(u_k-u_s).
	\end{align*}

Note that multi-temperature Euler equations can be obtained by putting $\varepsilon=0$ in \eqref{NSE multi}. We refer to \cite{Simic1,Simic2} where multi-temperature Euler equations are described in the framework of Extended Thermodynamics.

It is worth noticing that, according to \cite{Simic3}, shock structure in Helium-Argon mixtures can be better reproduced by the multi-temperature description. In numerical tests, we will also numerically deal with the Helium-Argon mixtures.

\subsubsection{Representation of NS equations for $n_s$, $u$ and $T$}
We rewrite \eqref{NSE multi} as follows (for detail see Appendix \ref{App 4}.)
	\begin{align}\label{NSE multi rewrite}
	\begin{split}
	\frac{\partial n_s}{\partial t} &=- \nabla \cdot(n_s u), \quad s=1,\cdots,L\cr
	\frac{\partial u_s}{\partial t}&=- u_s\nabla \cdot u_s -  \frac{\nabla(n_s K_BT_s)}{\rho_s} - \frac{\varepsilon \nabla \cdot(P_s^{(1)})}{\rho_s} + \frac{1}{\rho_s}\sum_{k\neq s }^{L}\mathcal{R}_{sk}\cr
	\frac{\partial T_s}{\partial t}&=
	\frac{2u_s}{3n_sK_B} \cdot \left( \varepsilon \nabla \cdot(P_s^{(1)})-\sum_{k\neq s }^{L}\mathcal{R}_{sk}\right) -\left( \nabla T_s\right) \cdot u_s - \frac{2}{3}T_s \nabla \cdot u_s \cr
	&\quad - \frac{2}{3n_sK_B}\varepsilon \nabla \cdot(P_s^{(1)}\cdot u) - \frac{2}{3n_sK_B}\varepsilon \nabla \cdot q_s^{(1)} + \frac{2}{3 n_sK_B} \sum_{k\neq s }^{L}\mathcal{S}_{sk}.
	\end{split}
	\end{align}



\section{Numerical tests}
In this section, we present several numerical examples. First, we numerically check the discrepancies between the three BGK-type models for inert gas mixtures given by \eqref{bgk AAP}, \eqref{bgk GS} and \eqref{bgk bbgsp}. Second, with reference to the scaled model \eqref{bgk bbgsp multi} we approximate the two corresponding systems of NS equations \eqref{NSE} and \eqref{NSE multi}. Third, we consider a binary mixture of noble gases with large mass ratio, in which the multi-velocity and temperature description \eqref{bgk bbgsp multi} and \eqref{NSE multi} may explain better the behavior of gases. Finally, we study the structure of a stationary shock wave for a binary mixture of noble gases.

To compute numerical solutions to \eqref{bgk AAP} and \eqref{bgk GS}, we consider a semi-Lagrangian method introduced in \cite{GRS2}. Since the method has been introduced with a non-conservative reconstruction, to make the method conservative we adopt a technique introduced in \cite{BCRY2,BCRY3}. In particular, we make use of Q-CWENO23 and Q-CWENO35 reconstructions, which are based on CWENO23 \cite{LPR2} and CWENO35 \cite{CPSV}, respectively. For details, we refer to \cite{BCRY2,BCRY3}. For \eqref{bgk bbgsp}, we use a conservative semi-Lagrangian method introduced in \cite{CBGR}. For the time discretization, we consider an implicit Runge-Kutta method (DIRK) and a backward difference formula (BDF). In particular, we here consider a second order DIRK method and a third order BDF3 method in \cite{CBGR}. 

Note that we perform numerical simulations based on Chu reduction \cite{chu} as in \cite{CBGR,GRS2}, that allows to reduce the problem from 3D to 1D in velocity and space, under suitable symmetry assumptions. Below we list the name of schemes, which will be used in this section:
\begin{enumerate}
	\item RK2-QCWENO23: DIRK2 with Q-CWENO23.
	\item BDF3-QCWENO35: BDF3 with Q-CWENO35.
\end{enumerate}

For discretization of the space and velocity (1D) domain, we use $N_x$ and $N_v+1$ grid points with uniform mesh sizes $\Delta x$ and $\Delta v$, respectively. Based on this, we will use grid points
$x_i$ and $v_j$ over computation domain $[x_{min},x_{max}] \times [v_{min},v_{max}]$.
To fix a time step $\Delta t$, we use a CFL number defined by
\[
\text{CFL}= \max\left\{|v_{min}|,|v_{max}|\right\}\frac{\Delta t}{\Delta x}.
\]

To compute the solutions of NS equations \eqref{NSE} and \eqref{NSE multi}, we instead solve \eqref{NSE rewrite} and \eqref{NSE multi rewrite} with spectral methods, which make the treatment of the many derivatives appearing there easier. Let us consider a general representation of the two NS equations:
\begin{align}\label{general pde}
	U_t=F(U,U_x,U_{xx}),
\end{align}
where $U\equiv U(x,t)$ is assumed to be smooth and periodic on the spatial domain. Given values of $\{U_i^n\}$, we first compute Fourier coefficients $\hat{U}_k$, $k=-N_x/2, \cdots, N_x/2-1$ using the fast Fourier transform (FFT). Then, we compute the $n$th spatial derivatives of functions involved in equations of interest using the inverse fast Fourier transform of $\{(jk)^n\hat{U}_k\}$ where $j$ denotes the imaginary number. For the time integration in \eqref{general pde}, we adopt an explicit RK4 scheme. 

\subsection{Comparison among the BGK models}\label{sec test 1}
Here we numerically investigate the discrepancy between the three BGK-type models for gas mixture. For this, as in \cite{ADGG,GRS2}, we consider a mixture of four monoatomic gases whose molecular masses are given by
\begin{align}\label{mass values}
	m_1= 58.5,\quad  m_2 = 18,\quad  m_3 = 40,\quad  m_4 = 36.5.
\end{align}
We use the following mechanical collision frequencies:
\begin{align}\label{lambda values}
	\nu_0^{11}= 5,\quad \nu_0^{12}= 6,\quad \nu_0^{13}=2,\quad \nu_0^{14}=7& \cr 
	\nu_0^{22}= 4,\quad \nu_0^{23}=5,\quad \nu_0^{24}=8& \cr 
	\nu_0^{33}=4,\quad \nu_0^{34}=3& \cr 
	\nu_0^{44}=6& \cr 			
\end{align}
with $\nu_0^{sk}= \nu_0^{ks}$ for $s, k = 1,\dots, 4$. Since we are comparing the three BGK models, the only common NS limit is the one with global velocity and temperature.

 
\subsubsection{Smooth initial data with large variance in velocity}\label{sec large variance}
In section \ref{sec discrepancy}, Proposition \ref{prop error} shows that the discrepancy between the AAP model \eqref{bgk AAP} and the BBGSP model \eqref{bgk bbgsp} becomes apparent when there is a large variance in the macroscopic velocities of gases. To show these aspects, we set as initial data Maxwellians whose macroscopic fields are given by
\begin{align}\label{initial acc2}
\begin{split}
n_0^s(x)=\frac{1}{m_s},\quad 
T_0^s(x)=\frac{4}{\sum_{s=1}^4 n_0^s},\qquad\qquad\qquad\qquad\cr
u_0^s(x)= \frac{\eta_s}{\sigma_s} \left[   \exp\left(-\left(\sigma_s x- 1 + \frac{s}{3} \right)^2\right) + \exp\left(-\left(\sigma_s x+ 3 - \frac{s}{10} \right)^2\right)\right]
\end{split}
\end{align}
where $\sigma_s = (10, 13, 16, 9)$ and $\eta_s=(-30,-10,10,30)$, for $s = 1, \cdots, 4$.
We impose periodic boundary conditions on the space domain $[-1,1]$ and truncate velocity domain by $[-15,15]$. 
We compute numerical solutions using RK2-QCWENO23 for $N_x=200$ and $N_v=60$. We first check the discrepancy of the three models at final time $t_f=0.04$. We use CFL $= 0.2$ up to $t=0.004$ and CFL$=2$ for $t \in [0.004,0.04]$ in order to be able to resolve the relaxation towards local equilibrium.

\begin{figure}[htbp]
	\centering
	\begin{subfigure}[b]{0.48\linewidth}
		\includegraphics[width=1\linewidth]{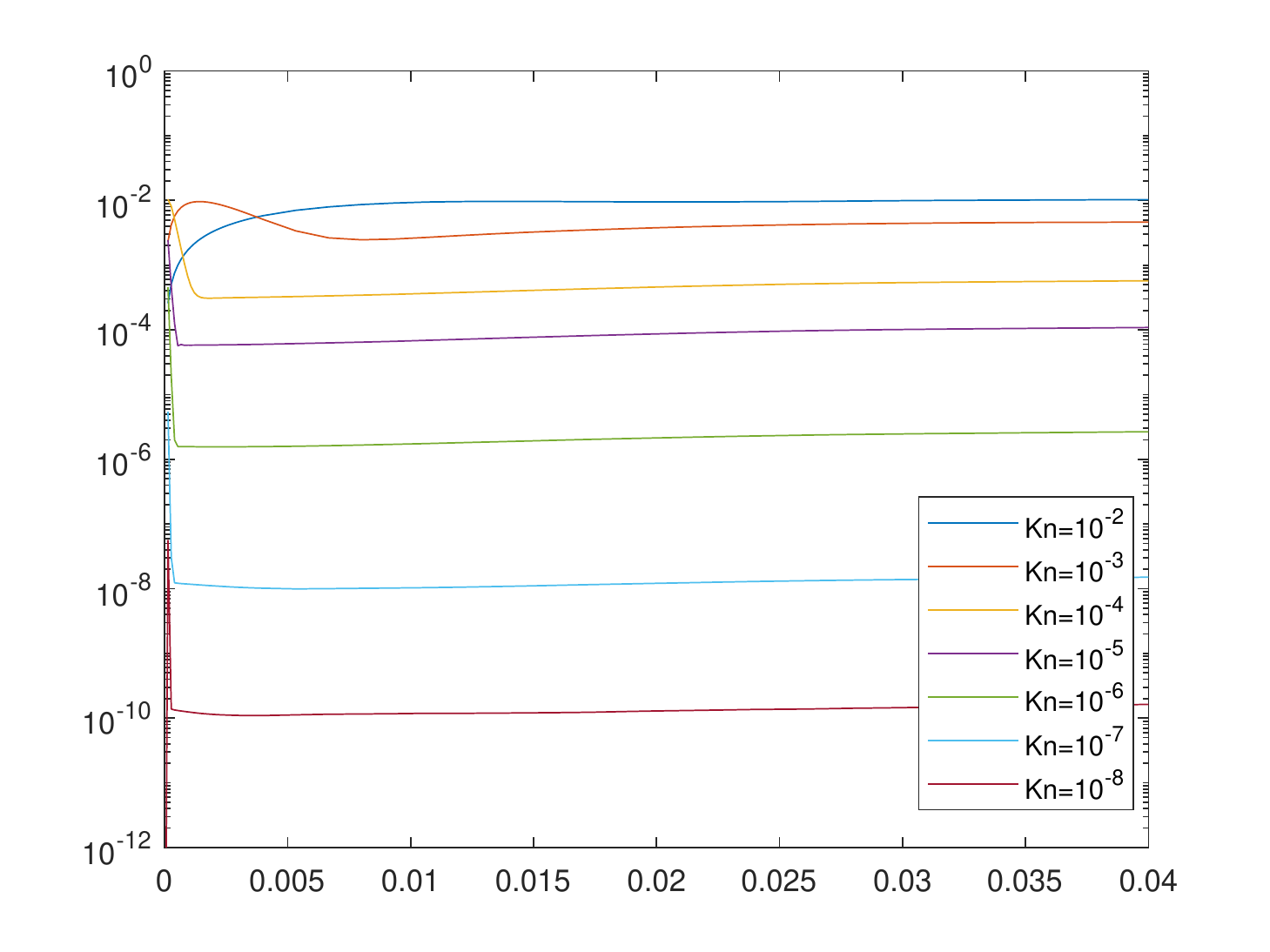}
		\subcaption{AAP and BBGSP models}\label{discrepancy error AAP BBGSP}
	\end{subfigure}
	\begin{subfigure}[b]{0.48\linewidth}
	\includegraphics[width=1\linewidth]{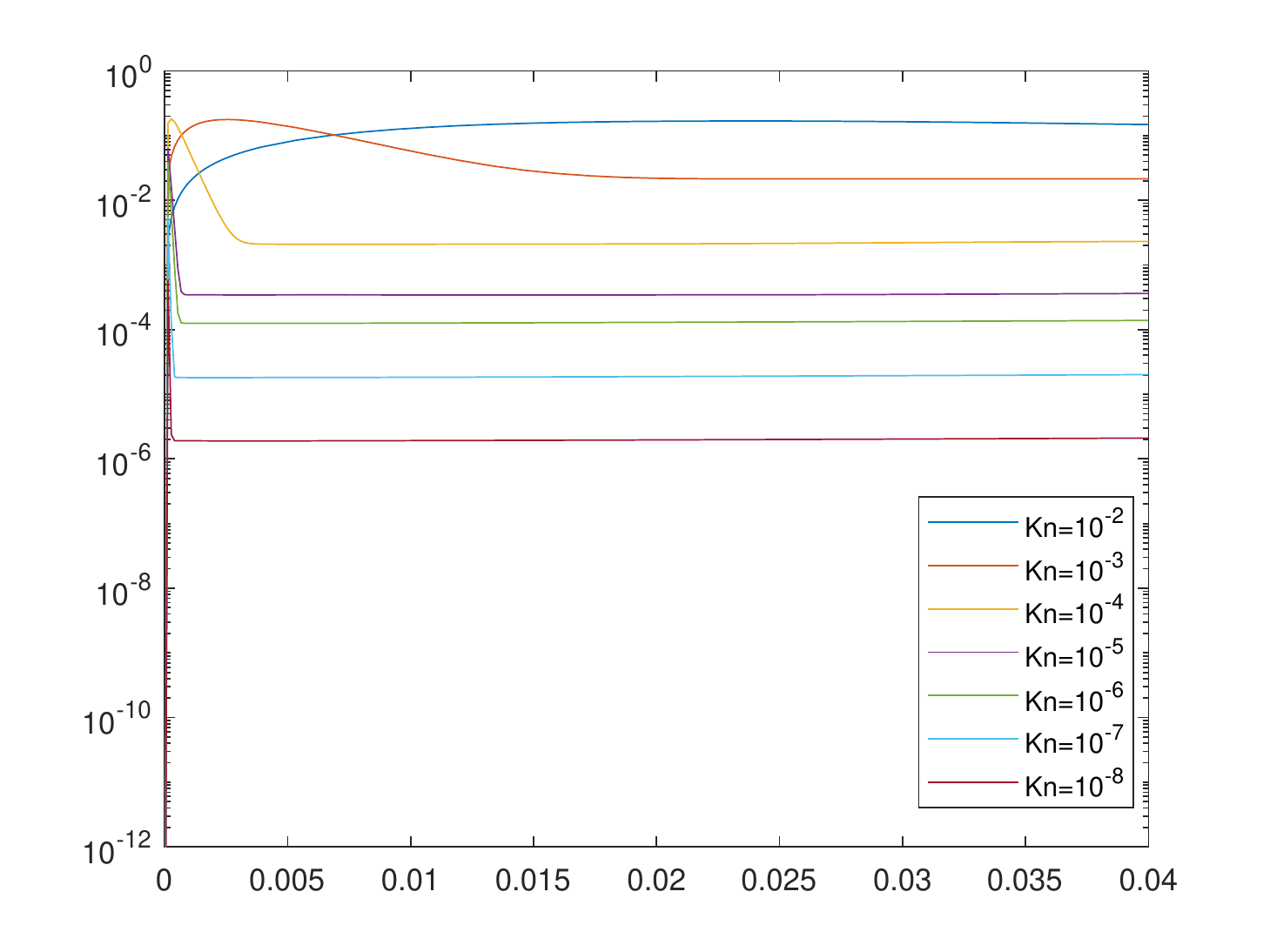}
	\subcaption{AAP and GS models}\label{discrepancy error AAP GS}
\end{subfigure}
	\caption{Time evolution of relative $L^1$-norm of the differences in the distribution functions $g_1$ between BGK models for various values of $\varepsilon$.
}\label{discrepancy error}
\end{figure}

We first measure the discrepancy of three BGK-type models for different values of $\varepsilon=10^{-q}$, $2\leq q \leq 8$, and plot the differences of two solutions to the AAP model \eqref{bgk AAP} and the BBGSP model \eqref{bgk bbgsp} using relative $L^1$-norm in Figure \ref{discrepancy error AAP BBGSP}. Here we compare the quantity $g_1(x,v,t)\equiv \int_{\mathbb{R}^2}f\left(x,{\bf v},t\right)\,dv_2\,dv_3$, which is based on the Chu reduction (for details, we refer to \cite{chu}). Similarly, we report numerical results relevant to the comparison of the two models AAP \eqref{bgk AAP} and GS \eqref{bgk GS} in Figure \ref{discrepancy error AAP GS}. 
Figure \ref{discrepancy error AAP BBGSP} shows that the differences between the AAP and BBGSP models are of order $\mathcal{O}\left(\varepsilon\right)$ for relatively large values of $\varepsilon \in [10^{-1},10^{-3}]$, while they become of order $\mathcal{O}\left(\varepsilon^2\right)$ for small values of $\varepsilon \in [10^{-4},10^{-6}]$. On the contrary, we can only observe differences of order $\mathcal{O}\left(\varepsilon\right)$ between AAP and GS models in Figure \ref{discrepancy error AAP GS}. These numerical evidences support the analytical result that the three models share the same Euler limit, and in addition AAP \eqref{bgk AAP} and BBGSP \eqref{bgk bbgsp} models have the same hydrodynamic limits at the NS level. On the contrary, the GS model \eqref{bgk GS} has a quantitatively different hydrodynamic limit, as shown in section \ref{sec NS equations with single velocity and temperature}.
Although we obtain similar results for global velocity $u$ and temperature $T$, we omit them here for brevity.

\begin{figure}[!htbp]
	\centering
	\begin{subfigure}[b]{0.43\linewidth}
		\includegraphics[width=1\linewidth]{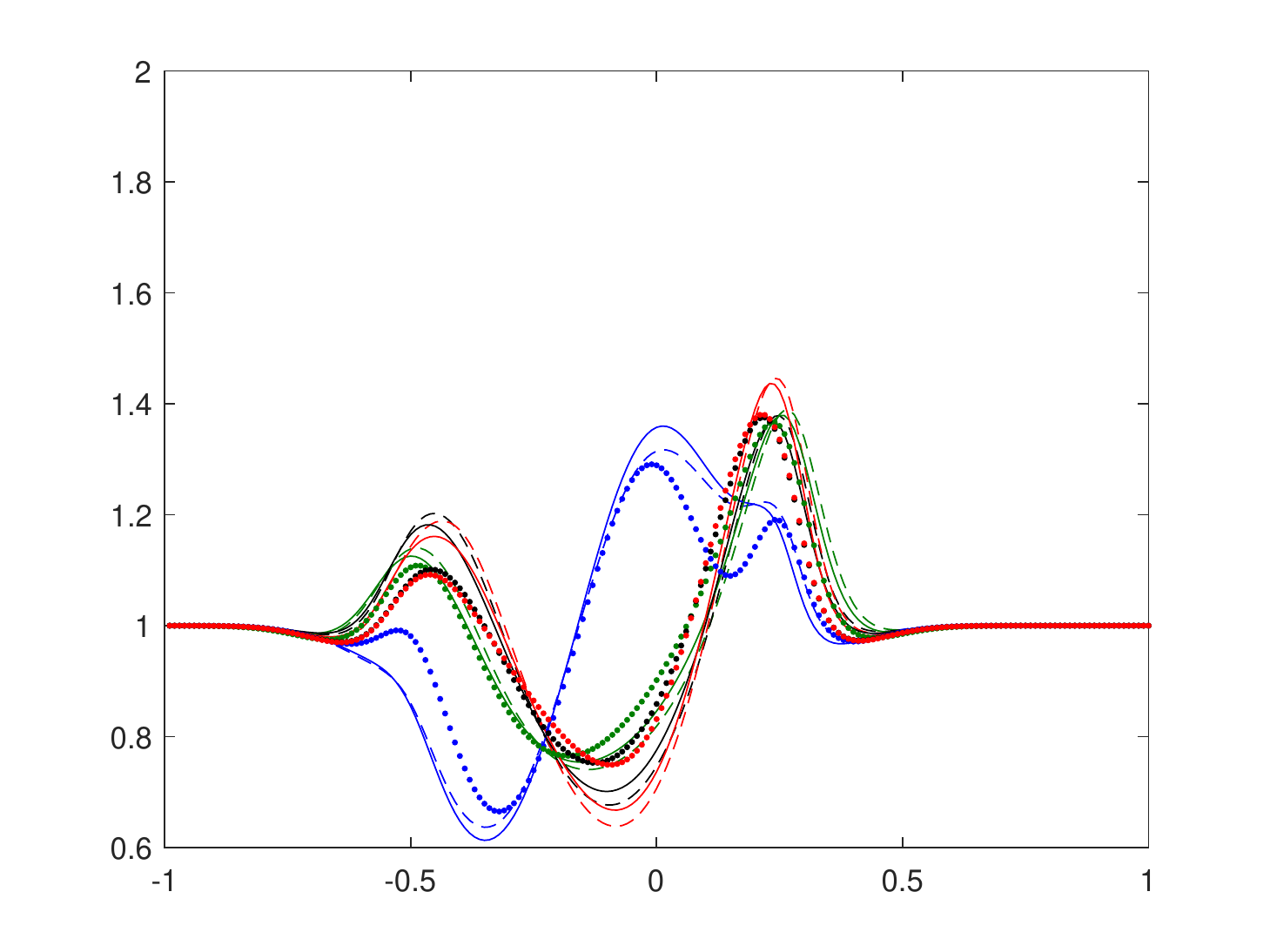}
		\subcaption{Density $\rho_s$, $s=1,2,3,4$}
	\end{subfigure}
	\begin{subfigure}[b]{0.43\linewidth}
		\includegraphics[width=1\linewidth]{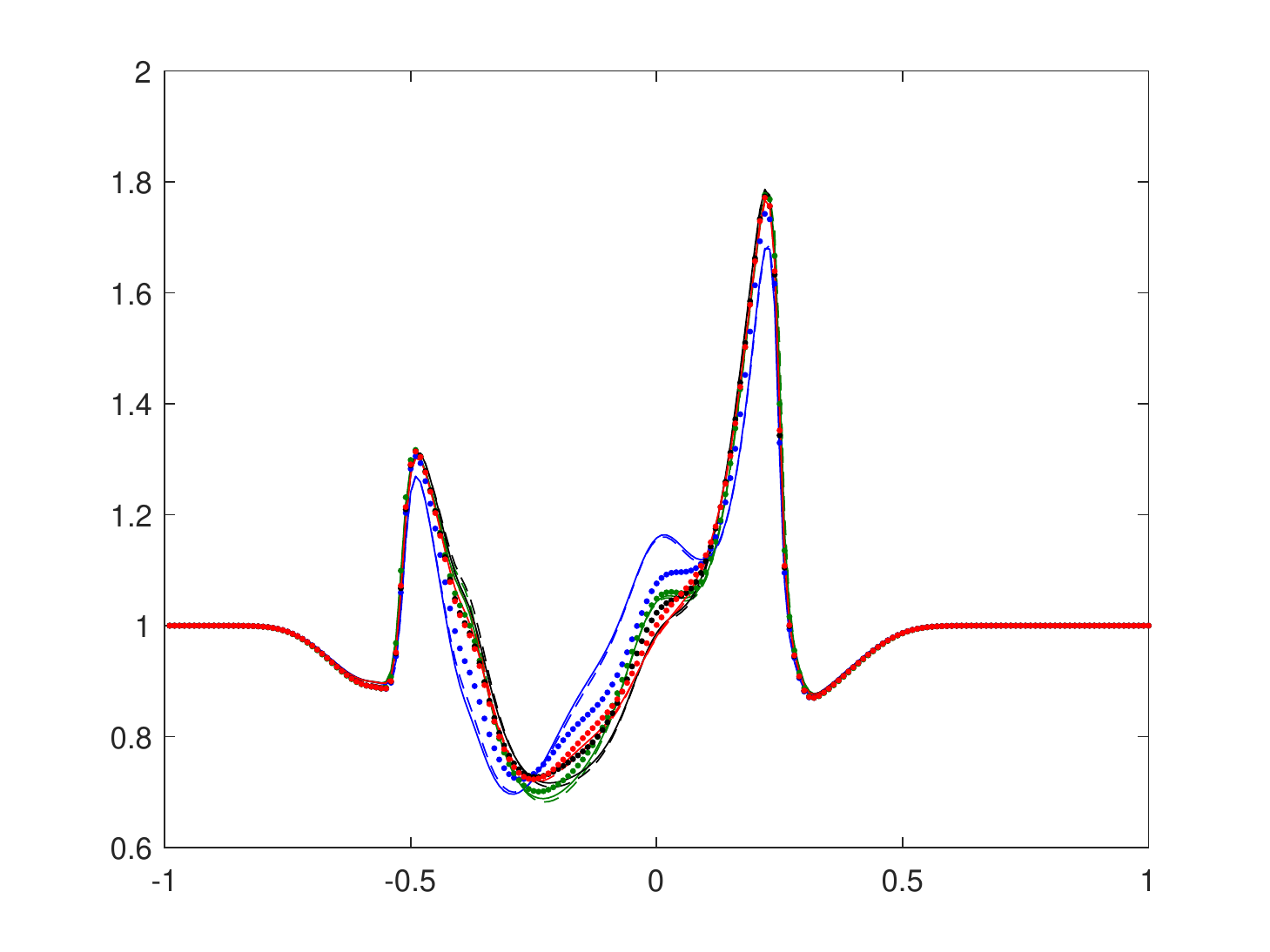}
		\subcaption{Density $\rho_s$, $s=1,2,3,4$}
	\end{subfigure}
	\begin{subfigure}[b]{0.43\linewidth}
		\includegraphics[width=1\linewidth]{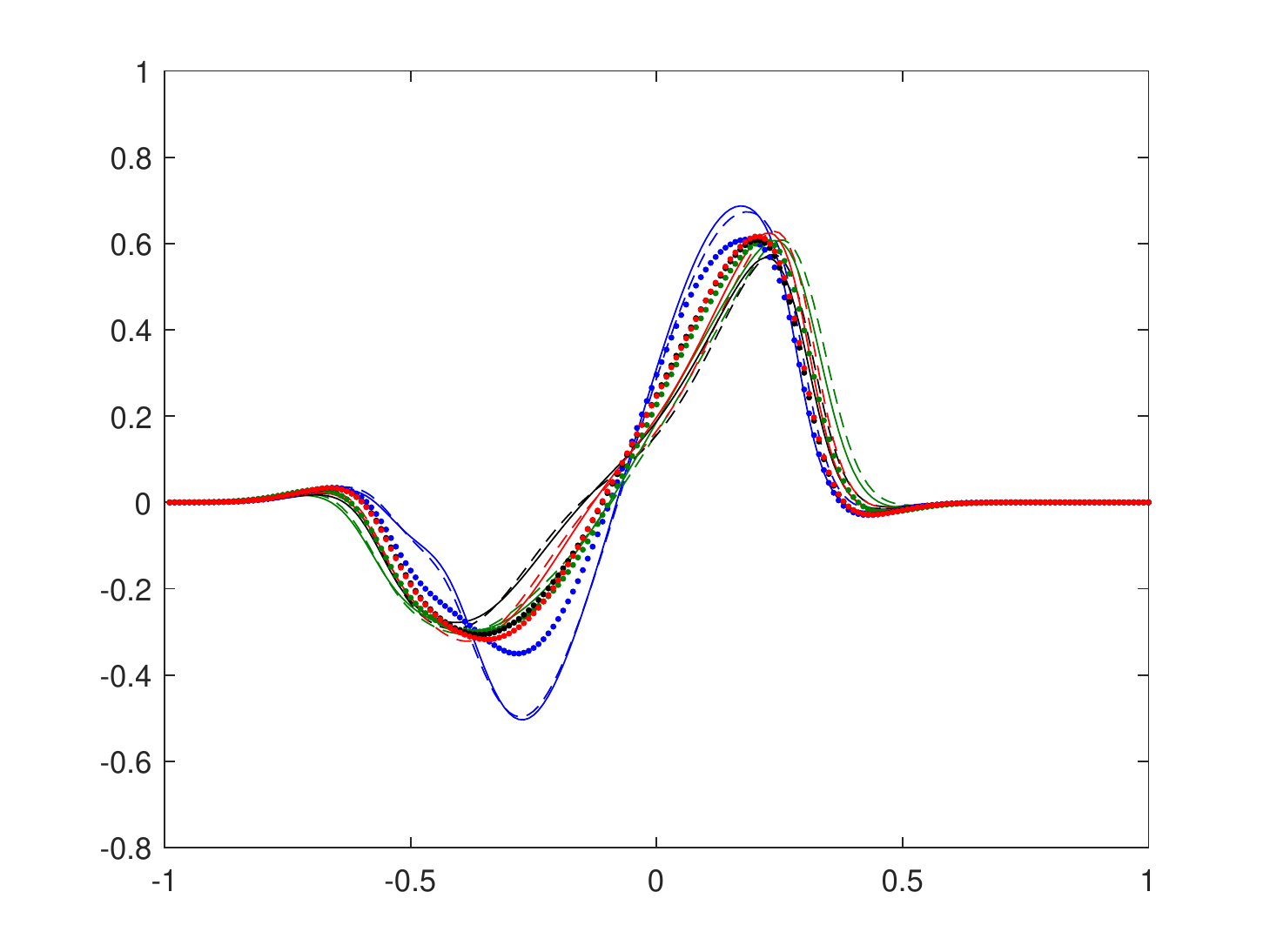}
		\subcaption{Velocity $u_s$, $s=1,2,3,4$}
	\end{subfigure}
	\begin{subfigure}[b]{0.43\linewidth}
		\includegraphics[width=1\linewidth]{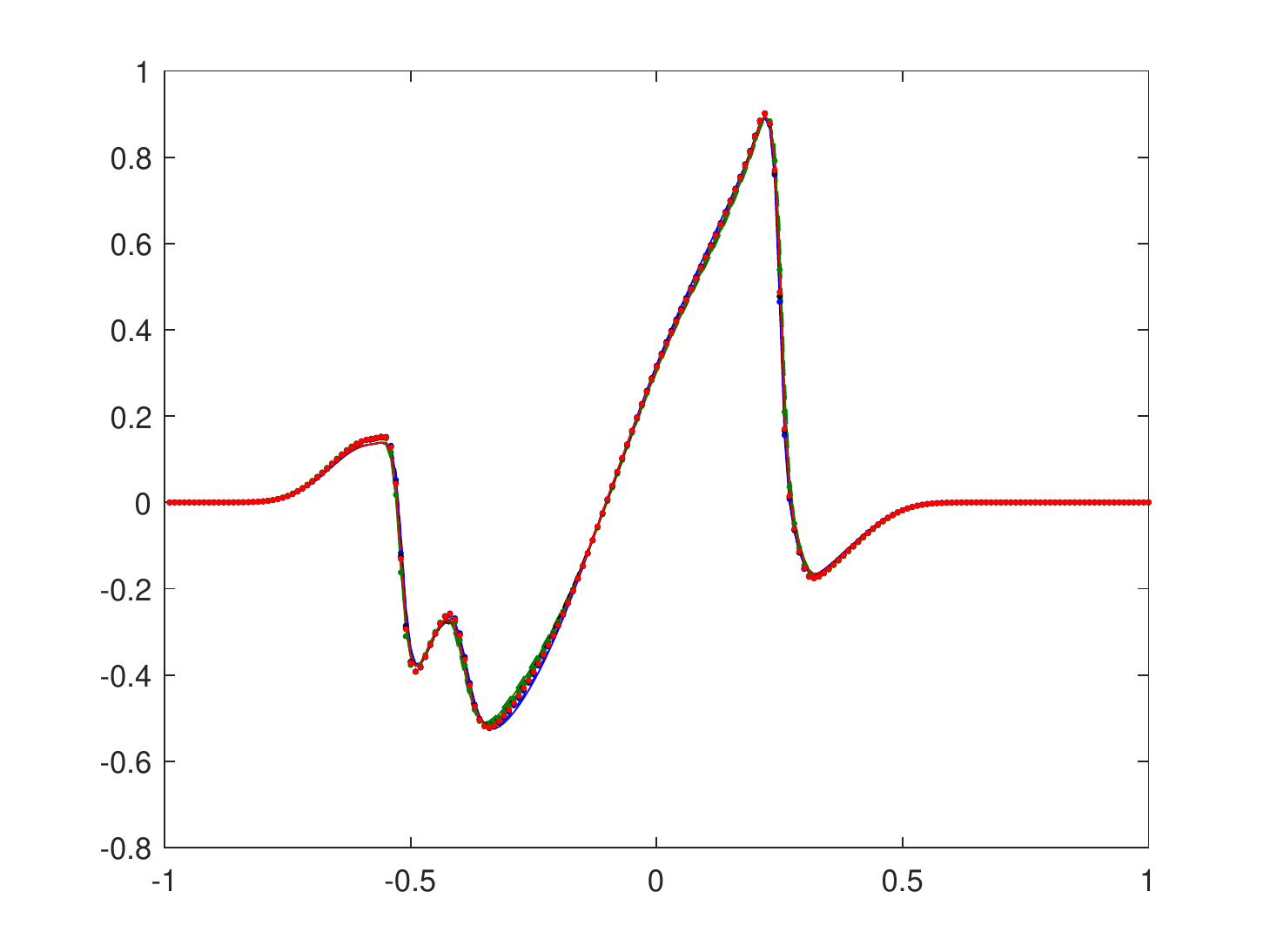}
		\subcaption{Velocity $u_s$, $s=1,2,3,4$}
	\end{subfigure}
	\begin{subfigure}[b]{0.43\linewidth}
		\includegraphics[width=1\linewidth]{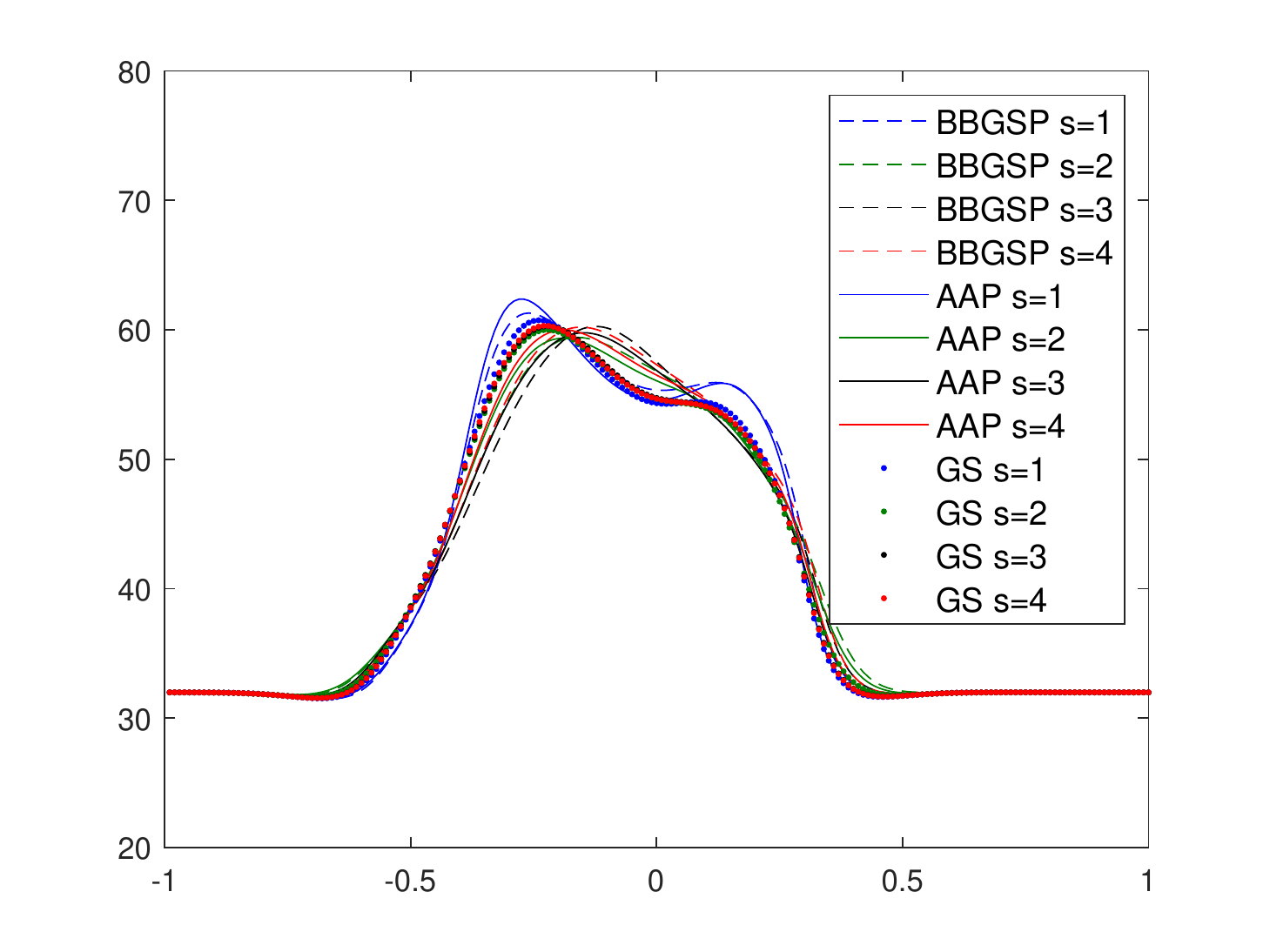}
		\subcaption{Temperature $T_s$, $s=1,2,3,4$}
	\end{subfigure}			
	\begin{subfigure}[b]{0.43\linewidth}
		\includegraphics[width=1\linewidth]{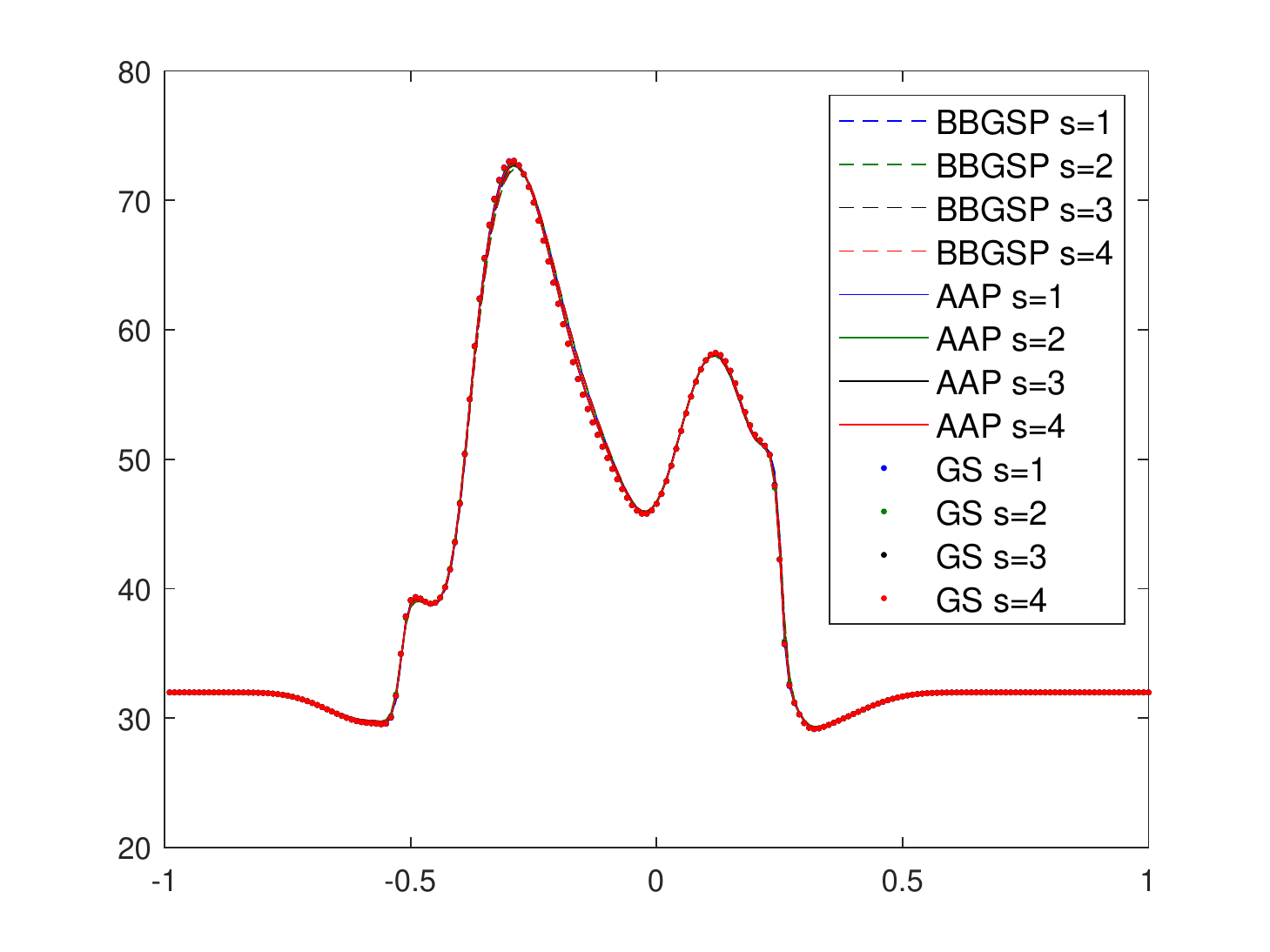}
		\subcaption{Temperature $T_s$, $s=1,2,3,4$}\label{test2 fig F}
	\end{subfigure}			
	\caption{Comparison of the three BGK models for $\varepsilon=10^{-2}$ (Left) and $\varepsilon=10^{-3}$ (Right) with initial data in \eqref{initial acc2}.}\label{fig test2 01}
\end{figure}
\begin{figure}[htbp]
	\centering
	\begin{subfigure}[b]{0.43\linewidth}
		\includegraphics[width=1\linewidth]{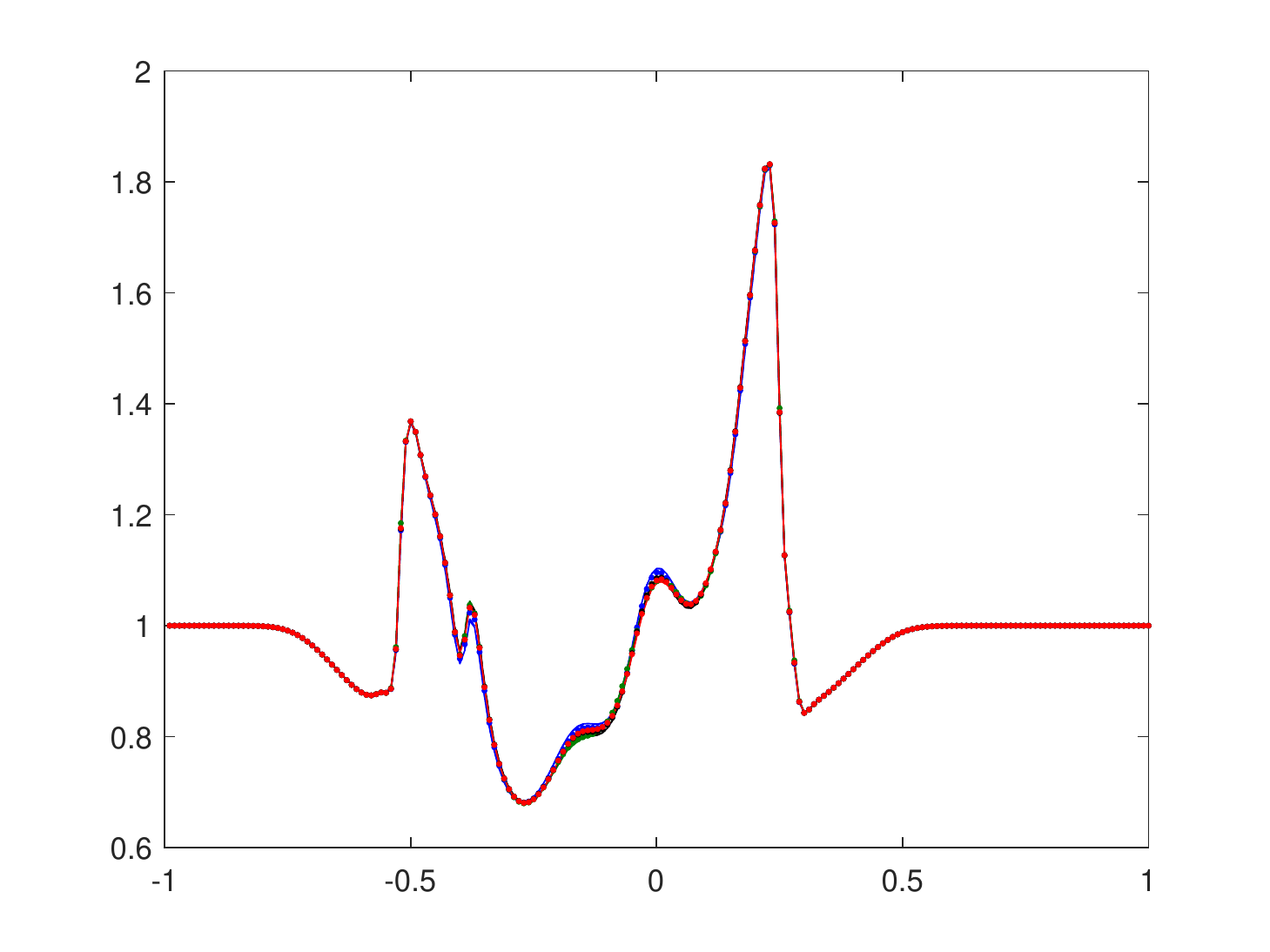}
		\subcaption{Density $\rho_s$, $s=1,2,3,4$}\label{fig test2 23 a}
	\end{subfigure}
	\begin{subfigure}[b]{0.43\linewidth}
		\includegraphics[width=1\linewidth]{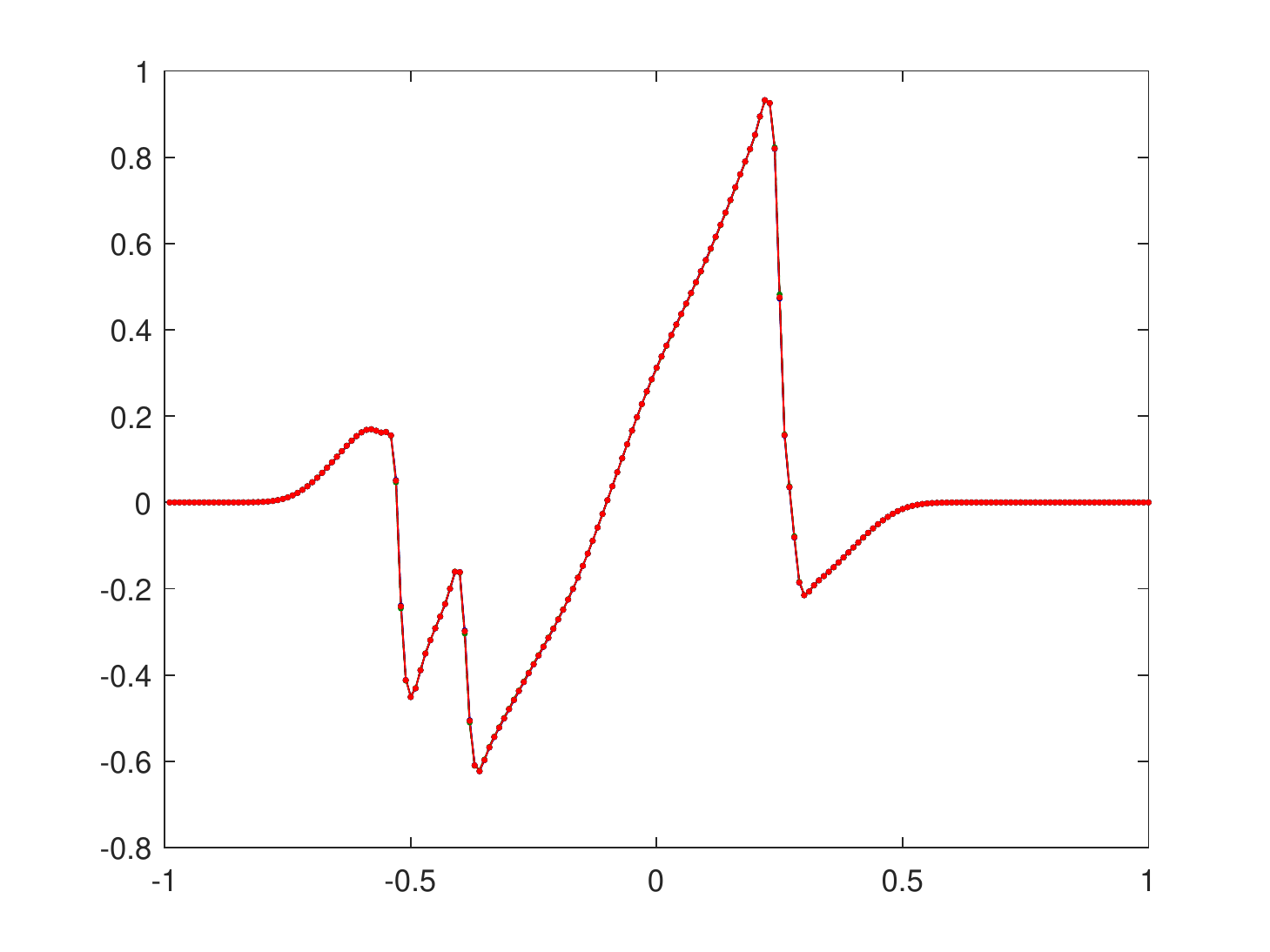}
		\subcaption{Velocity $u_s$, $s=1,2,3,4$}
	\end{subfigure}
	\caption{Comparison of the three BGK models for $\varepsilon=10^{-4}$ with initial data in \eqref{initial acc2}.}\label{fig test2 23}
\end{figure}

In Figs \ref{fig test2 01}-\ref{fig test2 23}, we plotted the numerical solutions of the four gases obtained with the three BGK methods at $t_f=0.2$, using CFL$=0.2$ up to $t=0.02$ and CFL$=2$ for $t\in[0.02,0.2]$. The results show that some difference appears for $\varepsilon=10^{-2}$, especially between GS models and the other ones, which instead remain closer to each other,
but all species velocities and temperatures are close to global velocity and temperature already from $\varepsilon=10^{-3}$, while densities equalize slower. Although we took smooth initial data, it is noticeable that shocks appear as time flows around $x=0.27$. 
Figure \ref{fig test2 23 a} shows even that  densities are almost overlapped for $\varepsilon=10^{-4}$. Here we omit the profiles of temperature since they show similar trends of Figure~\ref{test2 fig F}.




\subsection{Comparison with NS equations}
Here we focus on the  different hydrodynamic limits that can be obtained from the BGK model \eqref{bgk bbgsp} in order to highlight their peculiar behaviors. 

\subsubsection{Case 1: global velocity and temperature}\label{test single NSE comparison}
In this test, we check the capability to capture hydrodynamic limit \eqref{NSE} by solving model \eqref{bgk bbgsp}. For this, we consider the same test proposed in \cite{ADGG,GRS2}.
Here we consider the mixture of four monoatomic gases with molecular masses in \eqref{mass values} and collision frequencies in \eqref{lambda values}. 

We set initial data as Maxwellians whose macroscopic fields are 
\begin{align}\label{initial acc}
\begin{split}
n_0^s(x)=\frac{1}{m_s},\quad 
T_0^s(x)=\frac{4}{\sum_{s=1}^4 n_0^s},\qquad\qquad\qquad\qquad\cr
u_0^s(x)= \frac{s}{\sigma_s} \left[   \exp\left(-\left(\sigma_s x- 1 + \frac{s}{3} \right)^2\right) + \exp\left(-\left(\sigma_s x+ 3 - \frac{s}{10} \right)^2\right)\right]
\end{split}
\end{align}
for $s = 1, \cdots, 4$, where $\sigma_s = (10, 13, 16, 19)$. 

To confirm the agreement between the limiting solutions to \eqref{bgk bbgsp} and the solutions to NS equations \eqref{NSE}, we consider various choices of Knudsen numbers $\varepsilon=10^{-k}$, $k=2,3,4$. Here we compute numerical solutions to \eqref{bgk bbgsp} using $N_x=500$ and $N_v=60$. We take CFL$=0.2$ up to $t=0.02$, and CFL$=2$ in $t\in[0.02,0.2]$. 
\begin{figure}[htbp]
	\centering
	\begin{subfigure}[b]{0.43\linewidth}
		\includegraphics[width=1\linewidth]{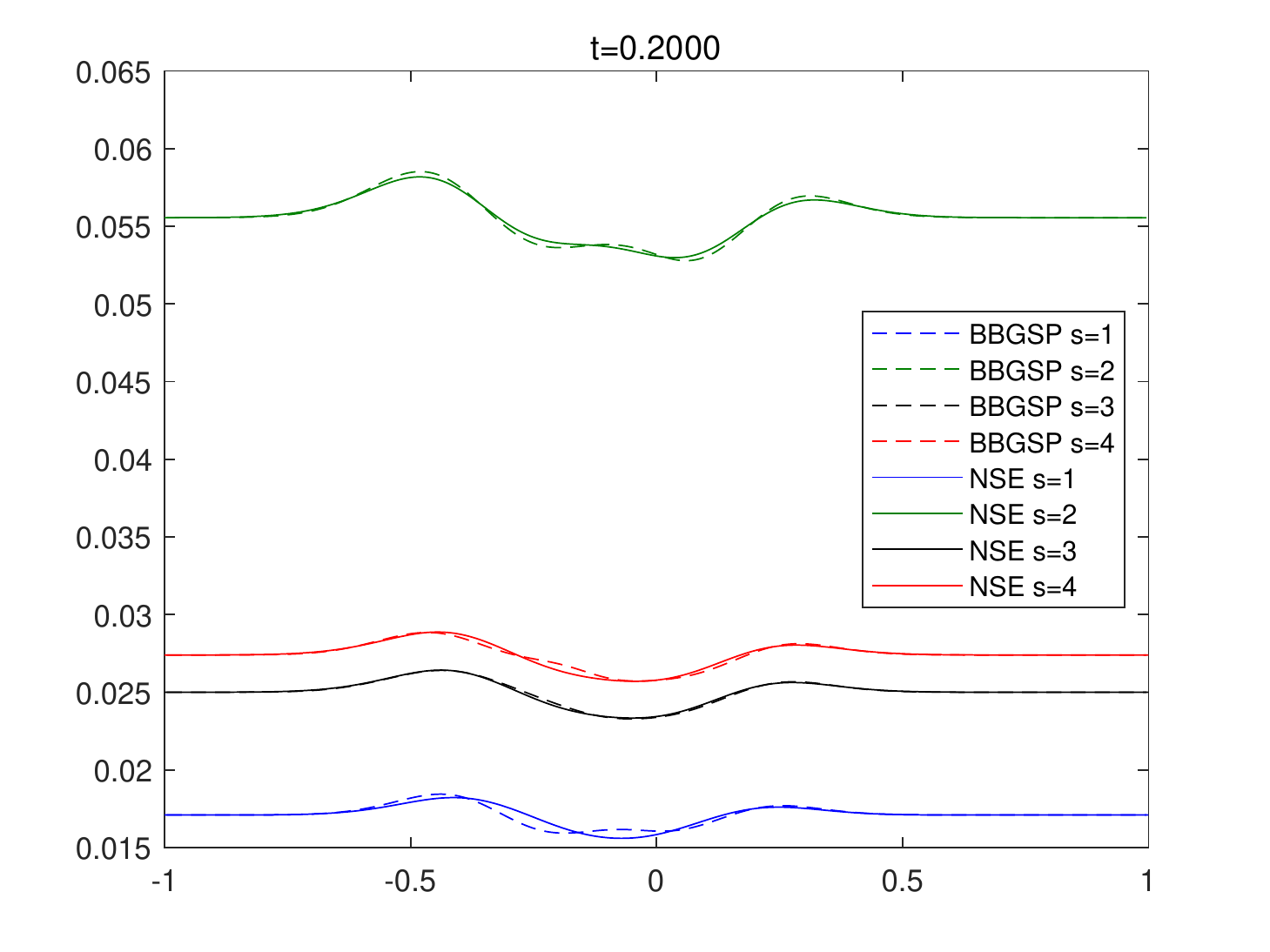}
		\subcaption{Number density $n_s$, $s=1,2,3,4$}
	\end{subfigure}
	\begin{subfigure}[b]{0.43\linewidth}
		\includegraphics[width=1\linewidth]{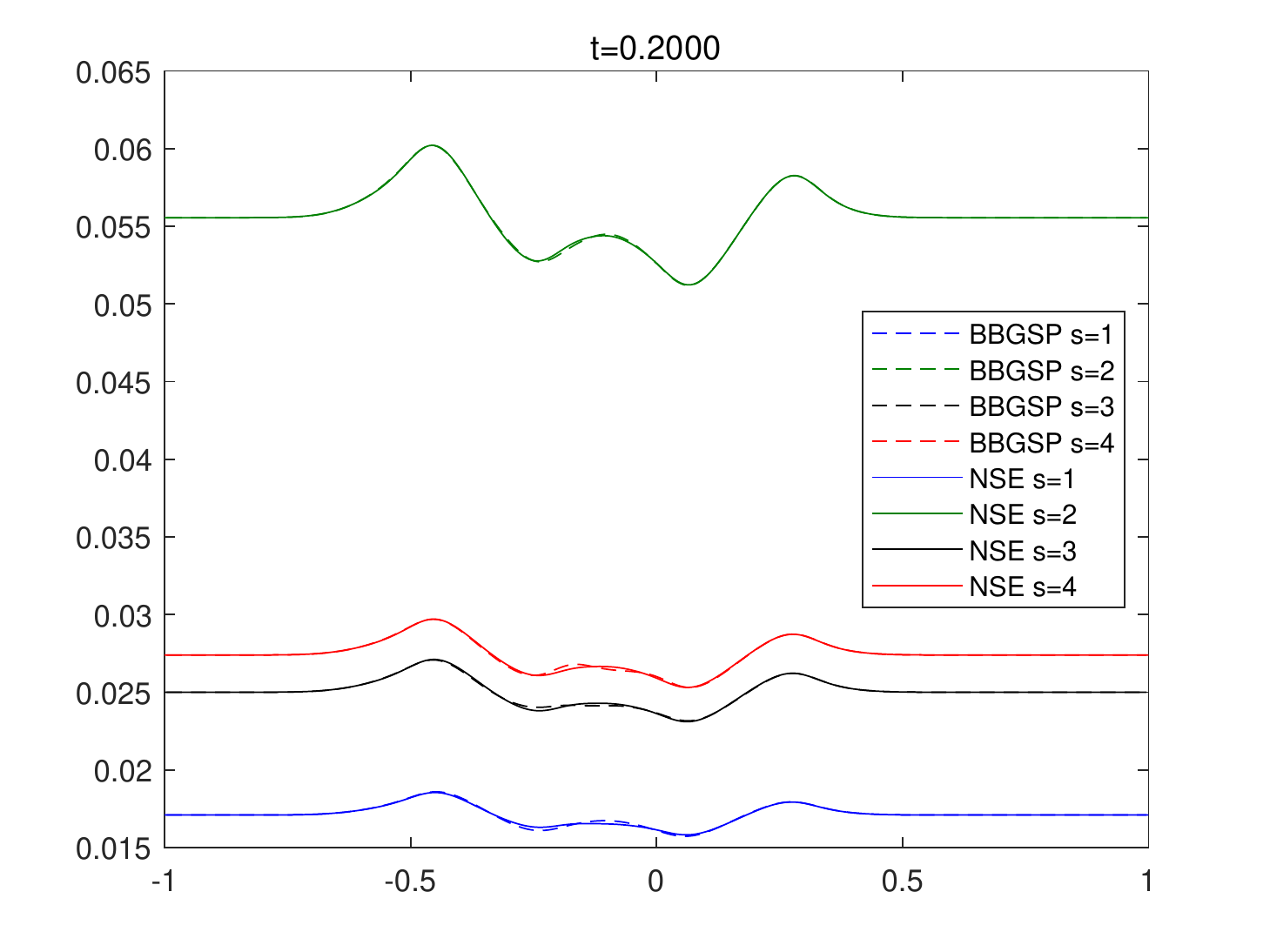}
		\subcaption{Number density $n_s$, $s=1,2,3,4$}
	\end{subfigure}
	\begin{subfigure}[b]{0.43\linewidth}
	\includegraphics[width=1\linewidth]{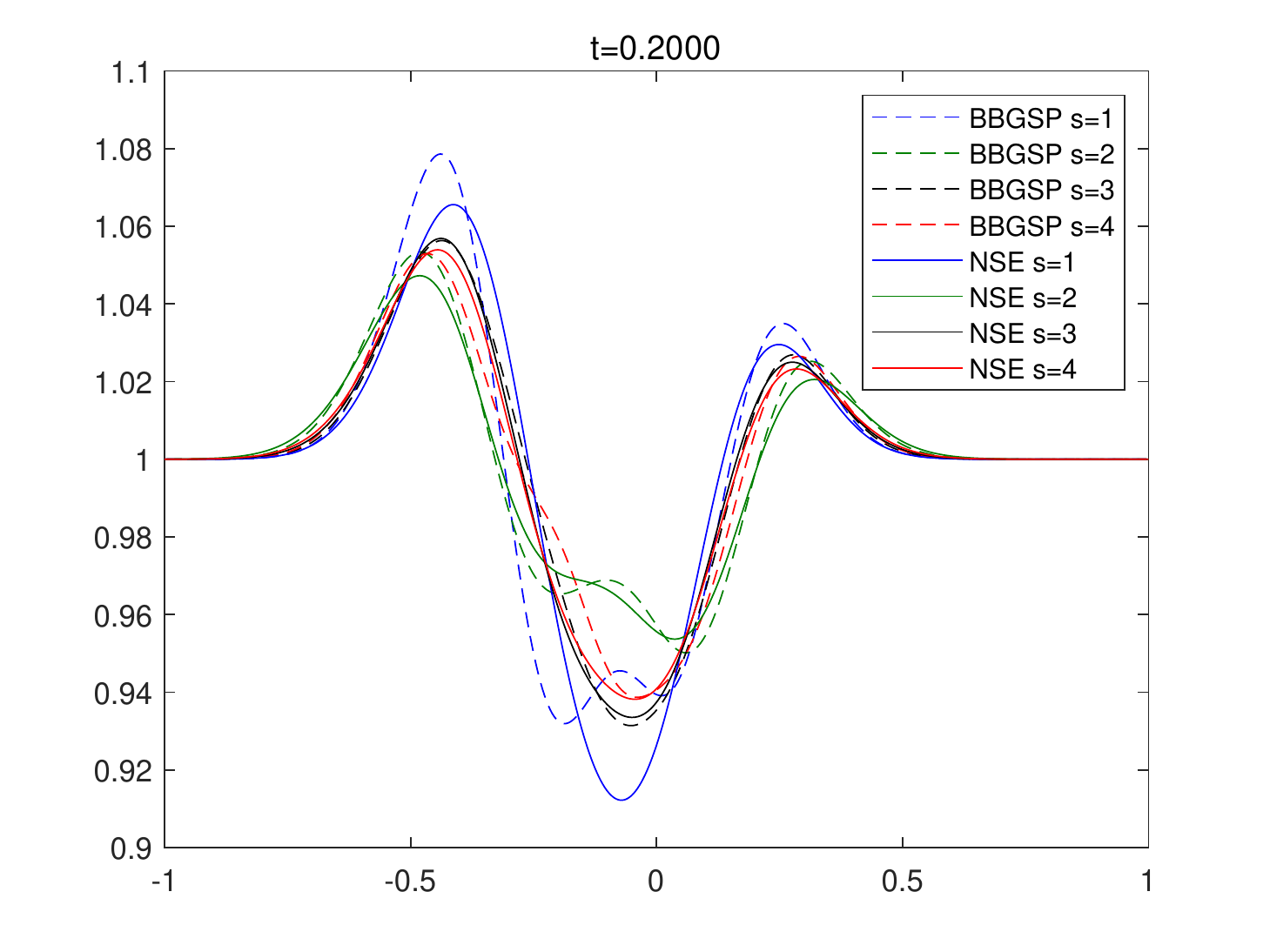}
	\subcaption{Density $\rho_s$, $s=1,2,3,4$}
\end{subfigure}
\begin{subfigure}[b]{0.43\linewidth}
	\includegraphics[width=1\linewidth]{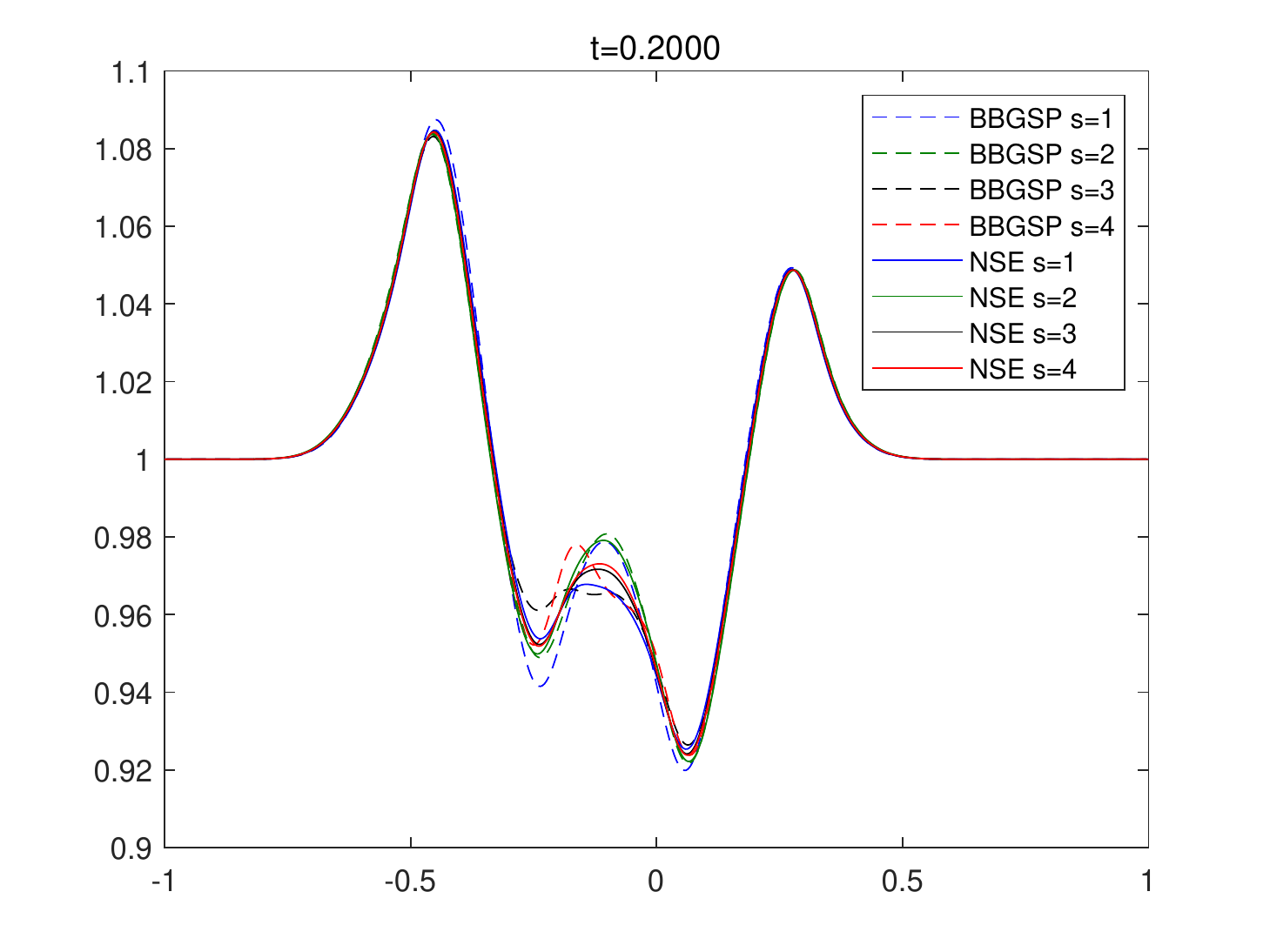}
	\subcaption{Density $\rho_s$, $s=1,2,3,4$}
\end{subfigure}
	\begin{subfigure}[b]{0.43\linewidth}
		\includegraphics[width=1\linewidth]{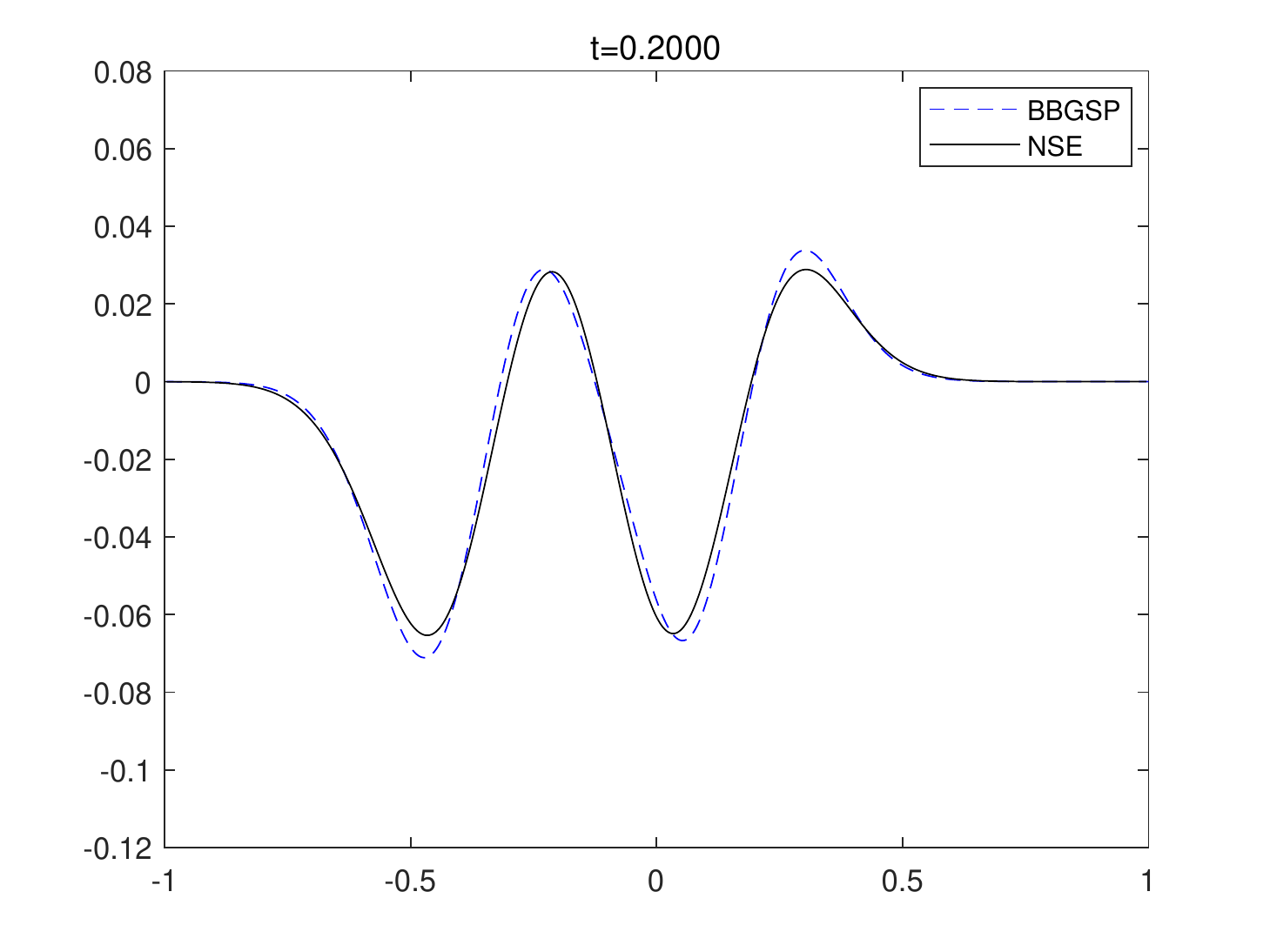}
		\subcaption{Velocity $u$}
	\end{subfigure}
	\begin{subfigure}[b]{0.43\linewidth}
		\includegraphics[width=1\linewidth]{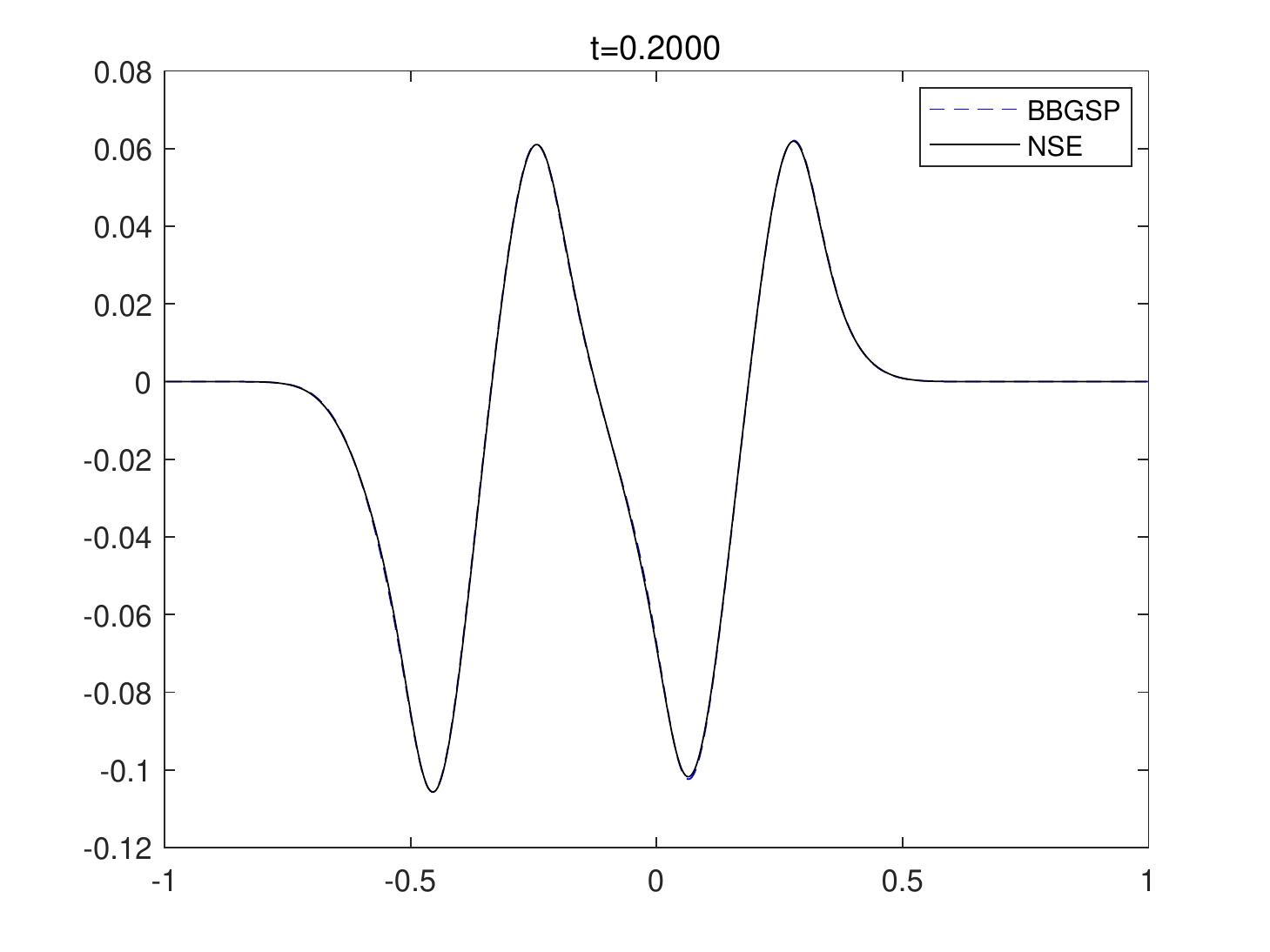}
		\subcaption{Velocity $u$}
	\end{subfigure}
	\begin{subfigure}[b]{0.43\linewidth}
		\includegraphics[width=1\linewidth]{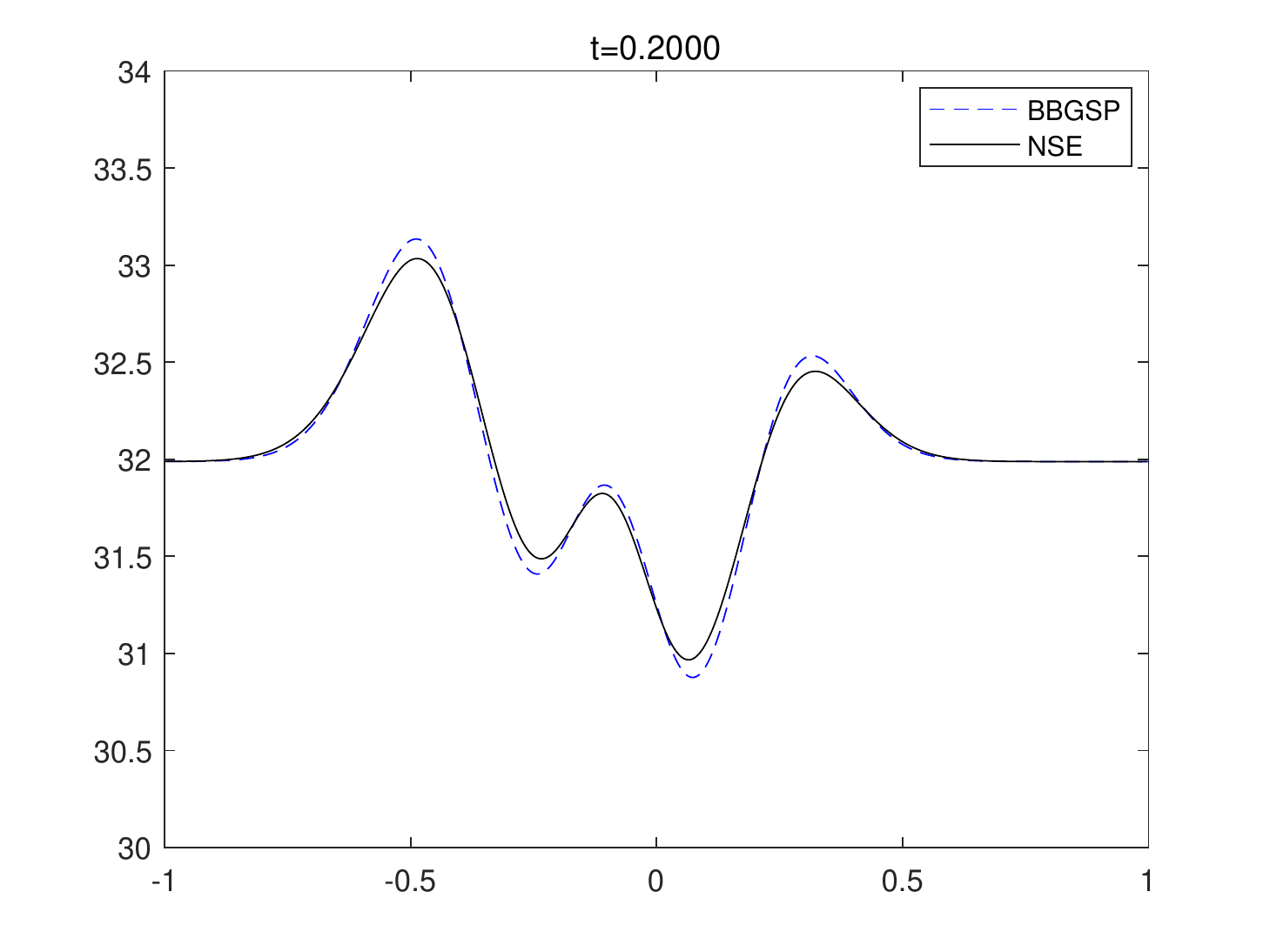}
		\subcaption{Temperature $T$}
	\end{subfigure}			
	\begin{subfigure}[b]{0.43\linewidth}
		\includegraphics[width=1\linewidth]{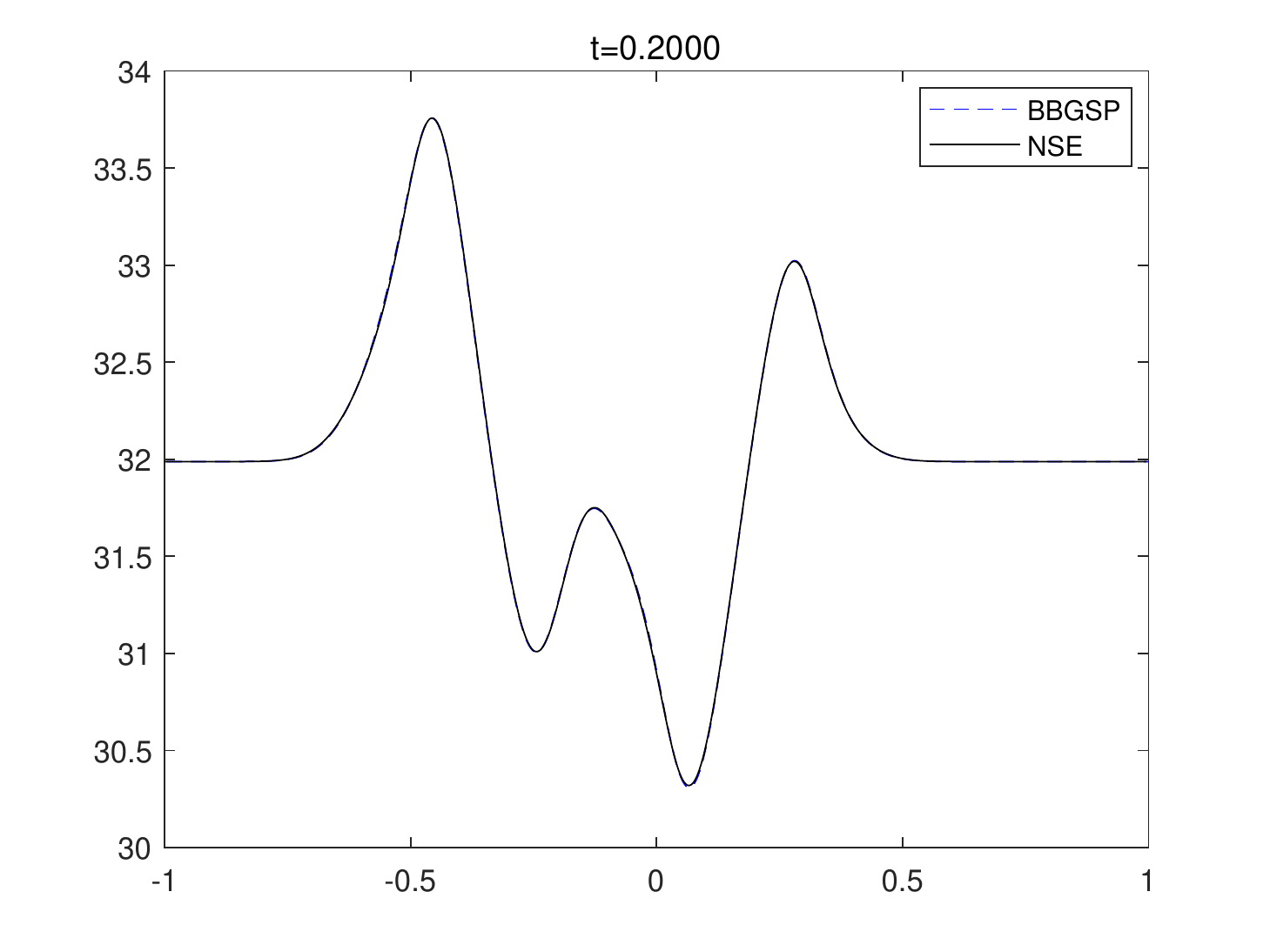}
		\subcaption{Temperature $T$}
	\end{subfigure}			
	\caption{Comparison of BGK model \eqref{bgk bbgsp} and NS equations \eqref{NSE} for $\varepsilon=10^{-2}$ (Left) and $\varepsilon=10^{-3}$ (Right) with initial data in \eqref{initial acc}.	
	}
	\label{fig NSE 01}
\end{figure}

\begin{figure}[htbp]
	\centering
	\begin{subfigure}[b]{0.43\linewidth}
		\includegraphics[width=1\linewidth]{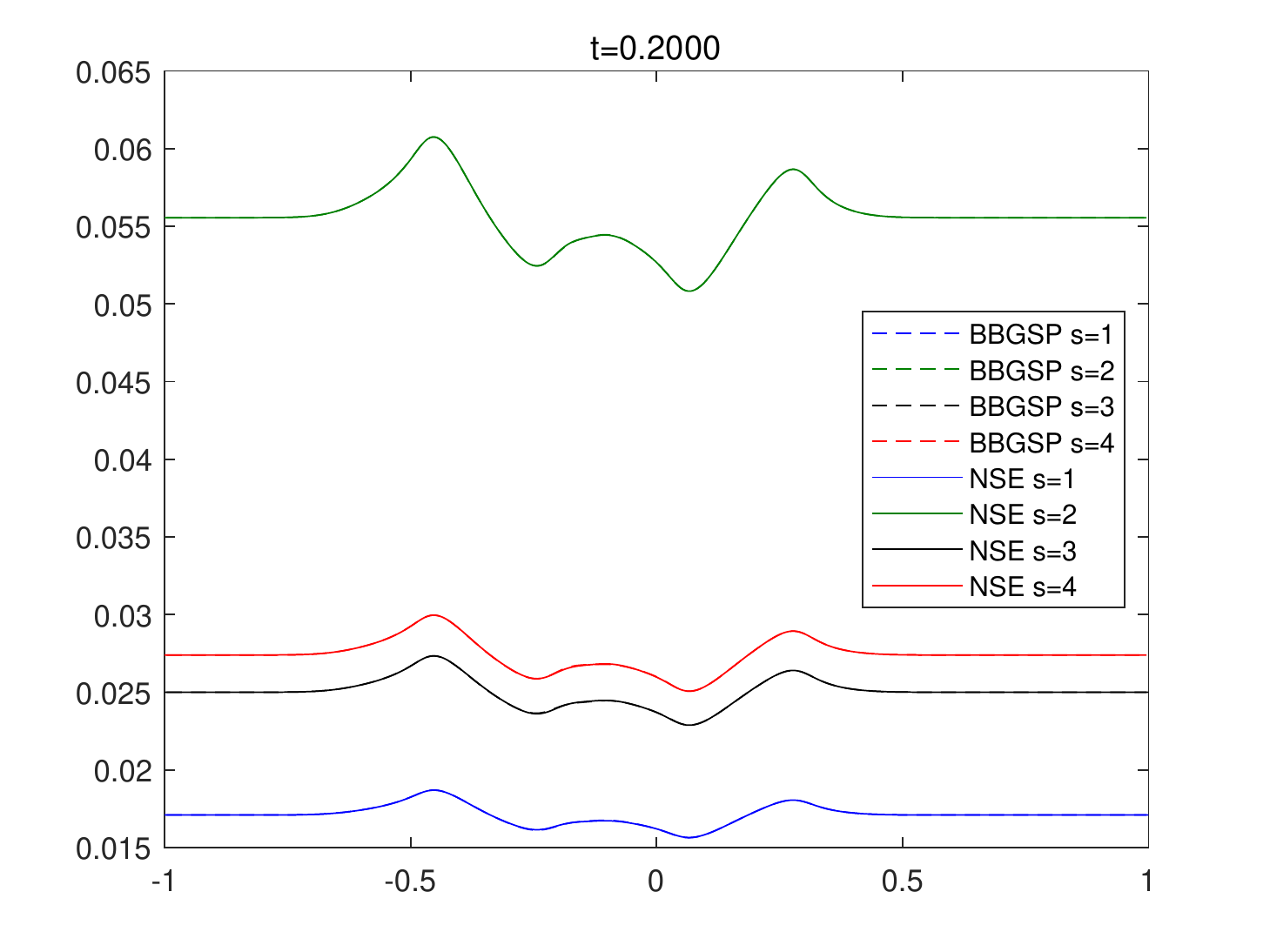}
		\subcaption{Number density $n_s$, $s=1,2,3,4$}
	\end{subfigure}
	\begin{subfigure}[b]{0.43\linewidth}
		\includegraphics[width=1\linewidth]{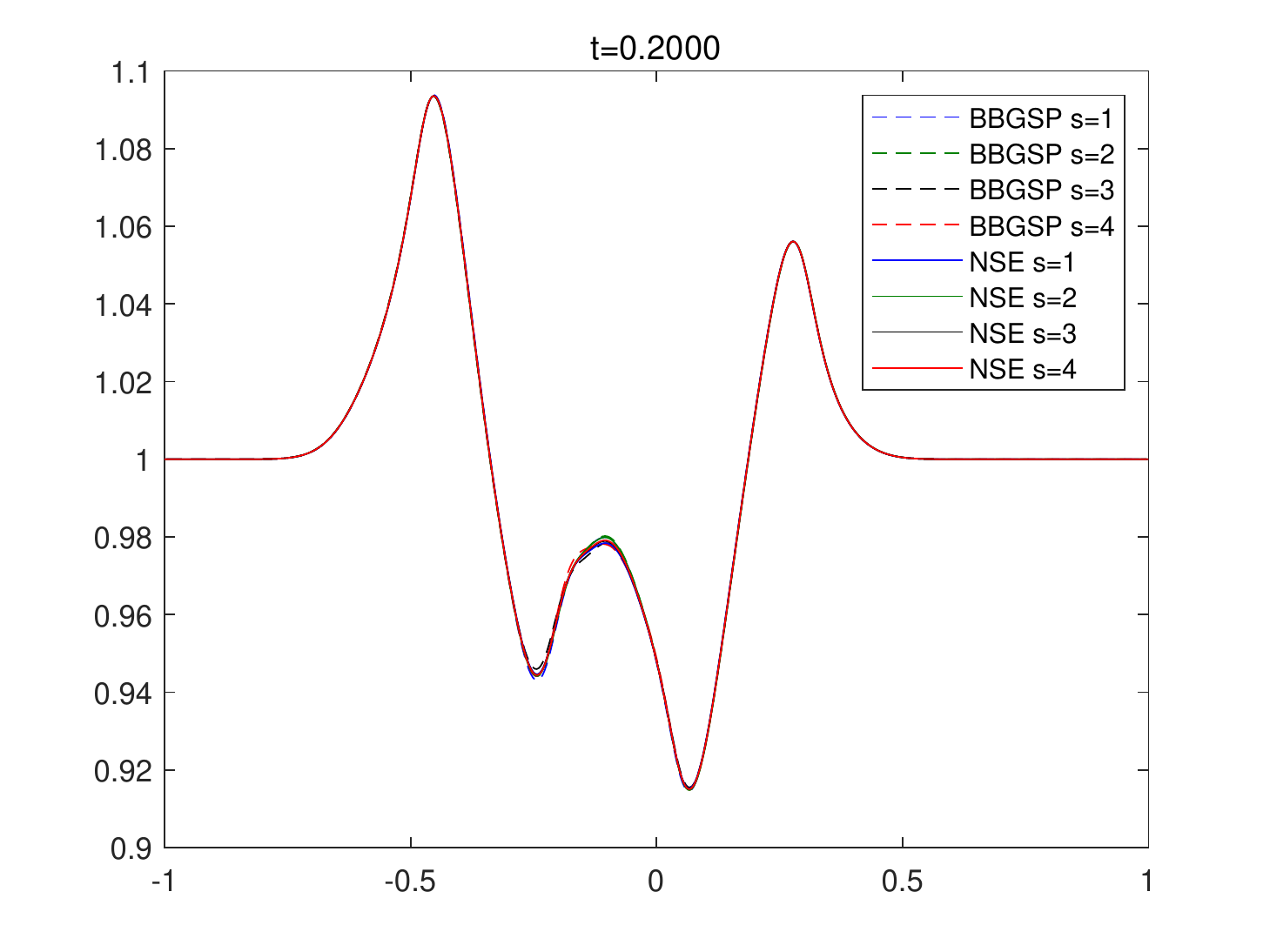}
		\subcaption{Density $\rho_s$, $s=1,2,3,4$}
	\end{subfigure}
	\begin{subfigure}[b]{0.43\linewidth}
		\includegraphics[width=1\linewidth]{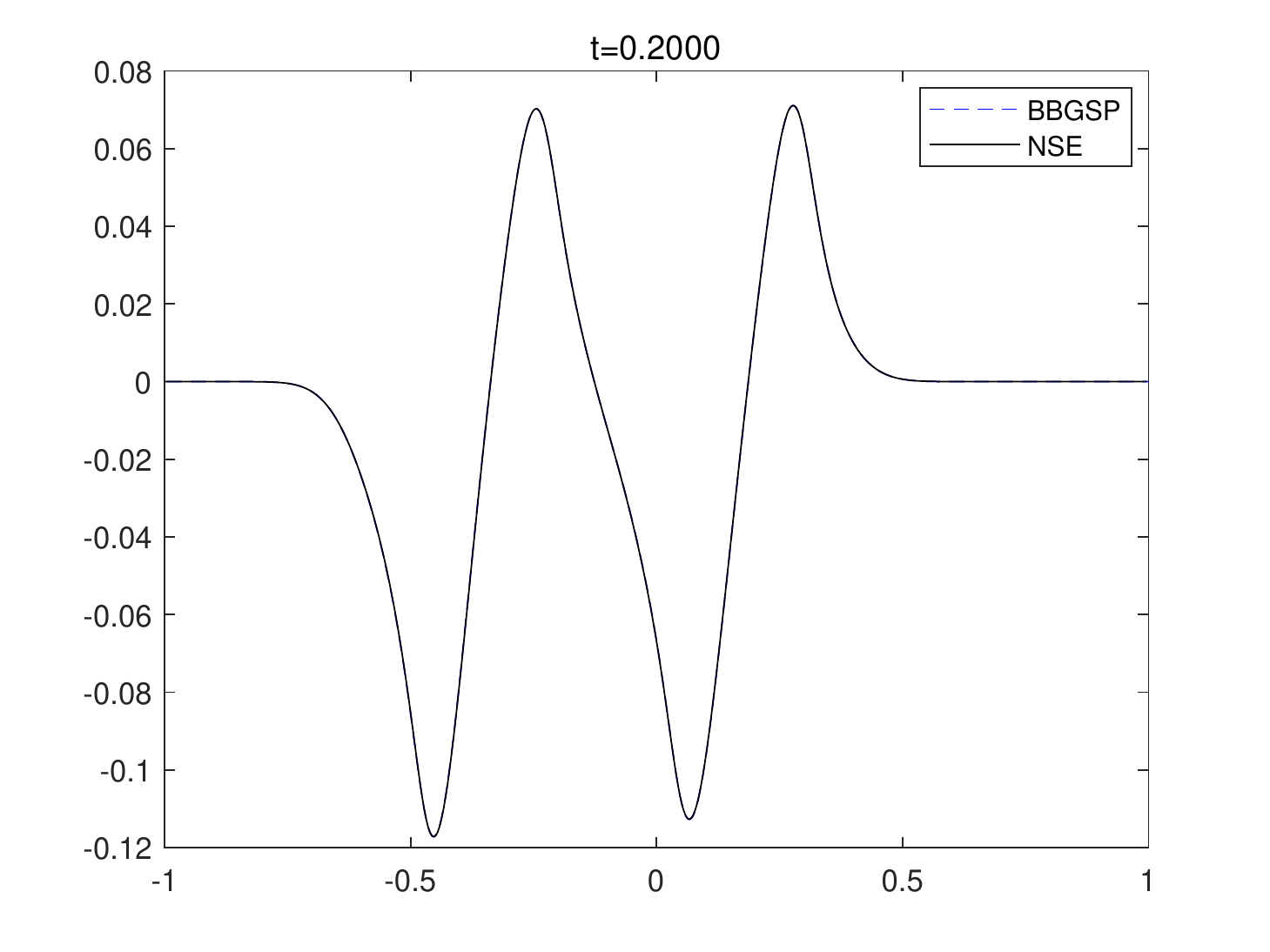}
		\subcaption{Velocity $u$}
	\end{subfigure}
	\begin{subfigure}[b]{0.43\linewidth}
		\includegraphics[width=1\linewidth]{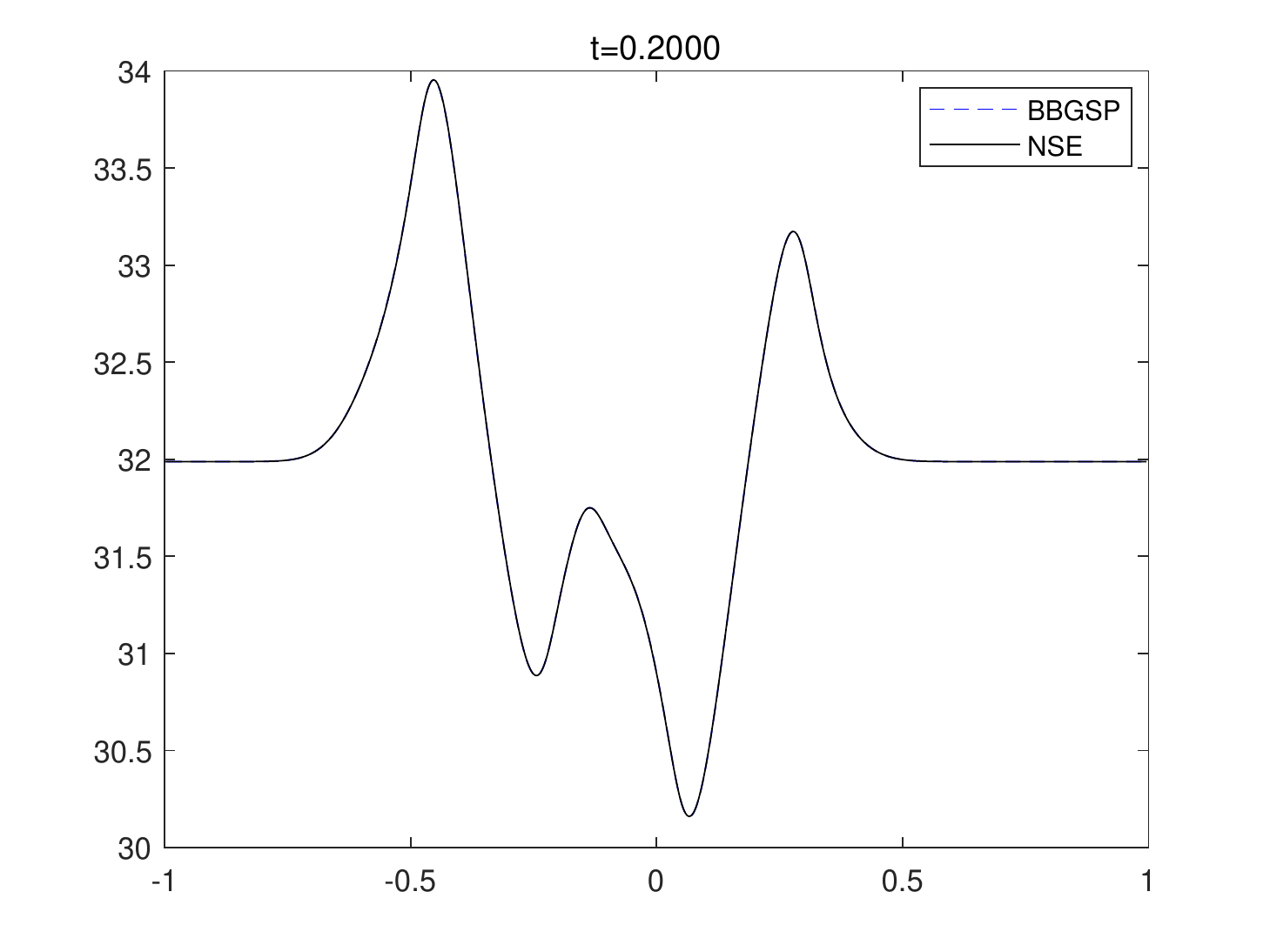}
		\subcaption{Temperature $T$}
	\end{subfigure}			
	\caption{
		Comparison of BGK model \eqref{bgk bbgsp} and NS equations \eqref{NSE} for $\varepsilon=10^{-4}$ with initial data in \eqref{initial acc}.
	}\label{fig NSE 23}
\end{figure}

In Figures \ref{fig NSE 01}, we observe that for these parameters and initial data BGK solutions \eqref{bgk bbgsp} and NS solutions \eqref{NSE} are quite different for $\varepsilon=10^{-2}$. However, both solutions give similar values of macroscopic quantities as we take smaller Knudsen numbers. For $\varepsilon=10^{-3}$, global velocity $u$ and temperature $T$ are almost overlapped, contrary to species densities. In Figure \ref{fig NSE 23}, we note that even species densities become identical for a sufficiently small Knudsen number $\varepsilon=10^{-4}$. 

\subsubsection{Case 2: multi velocity and temperature}
Here we consider the case in which intra-species collisions are dominant, and hence we impose $\kappa\neq \varepsilon$ in \eqref{bgk bbgsp multi} so that the scaled version of the BBGSP model \eqref{bgk bbgsp} leads to the multi-velocity and multi-temperature NS equations \eqref{NSE multi} for small values of $\varepsilon$. Here, we take Knudsen number $\varepsilon=10^{-k}$, $k=2,3,4$ and set $\kappa=1$. We compute numerical solutions with the same numerical setting of sect.~\ref{test single NSE comparison}.


\begin{figure}[htbp]
	\centering
	\begin{subfigure}[b]{0.43\linewidth}
		\includegraphics[width=1\linewidth]{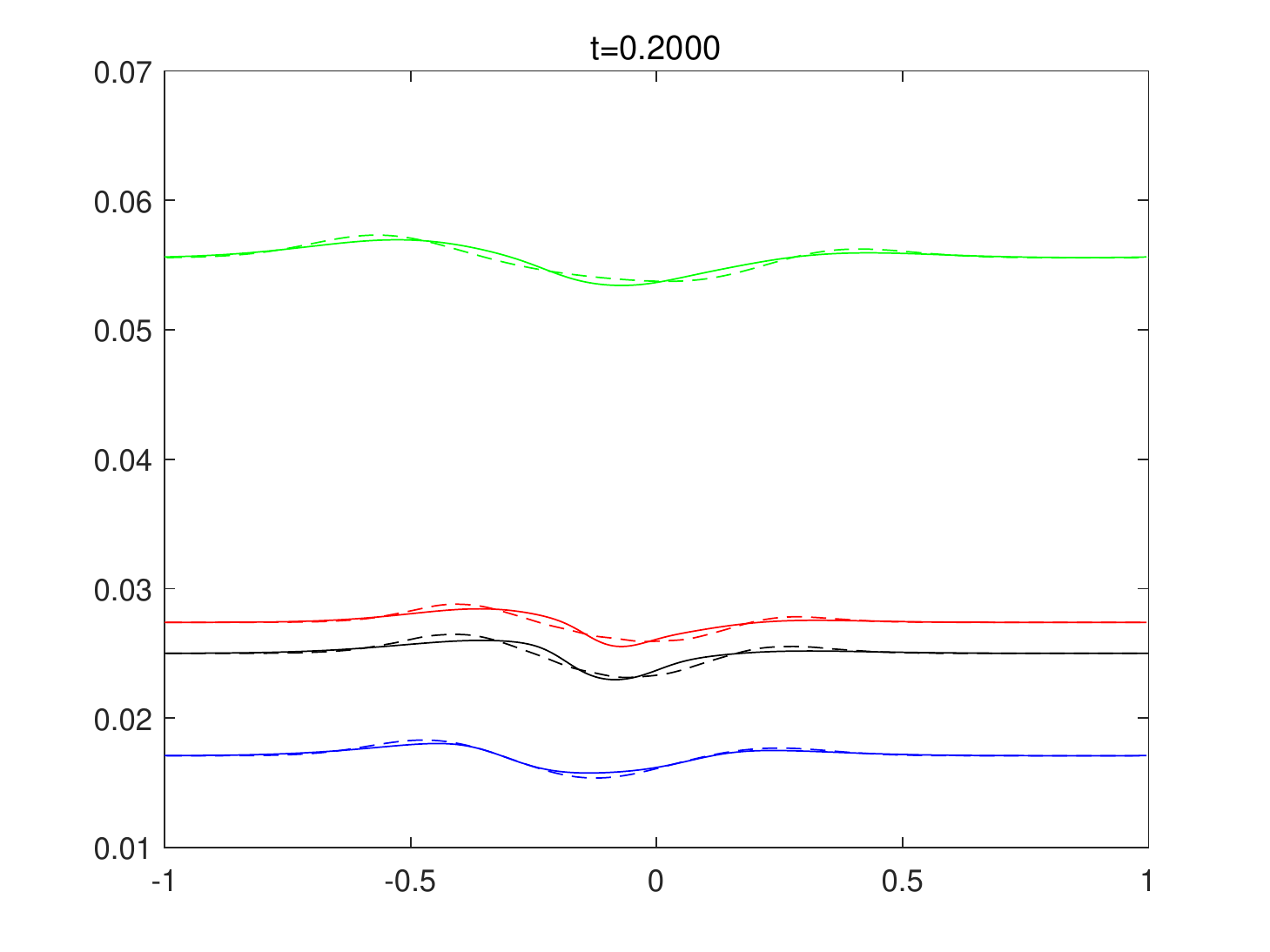}
		\subcaption{Number density $n_s$, $s=1,2,3,4$}
	\end{subfigure}
	\begin{subfigure}[b]{0.43\linewidth}
		\includegraphics[width=1\linewidth]{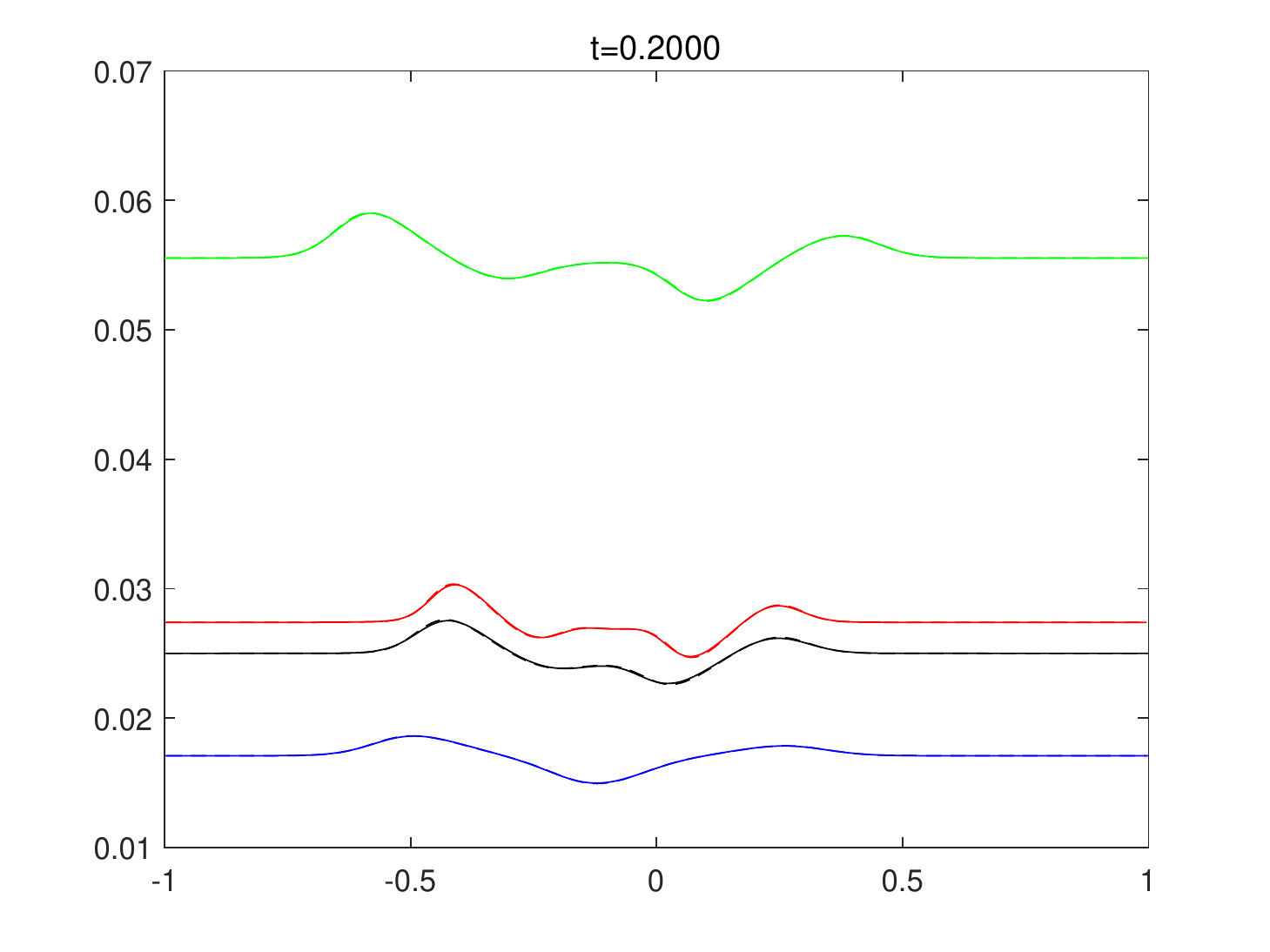}
		\subcaption{Number density $n_s$, $s=1,2,3,4$}
	\end{subfigure}
	\begin{subfigure}[b]{0.43\linewidth}
		\includegraphics[width=1\linewidth]{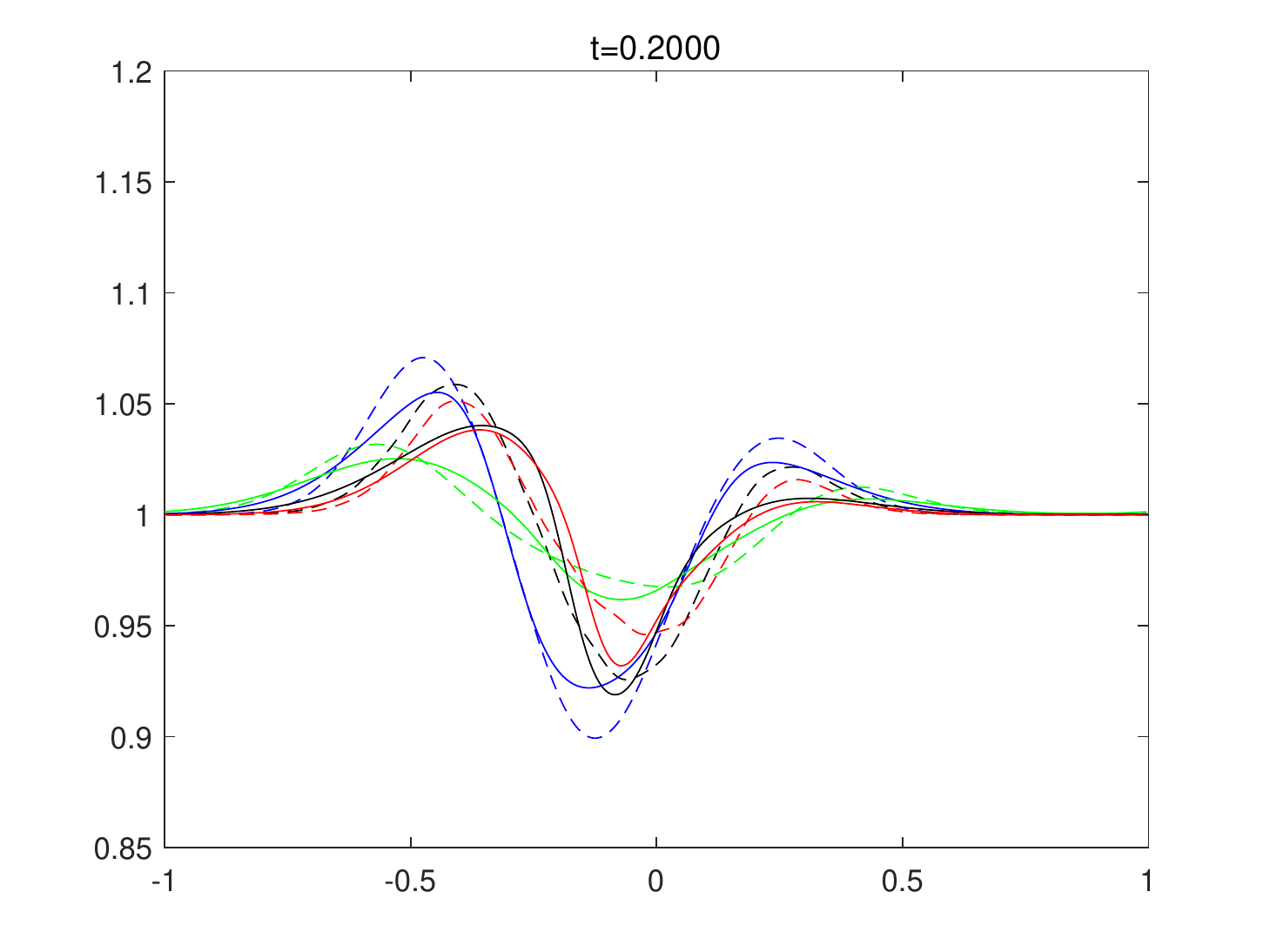}
		\subcaption{Density $\rho_s$, $s=1,2,3,4$}
	\end{subfigure}
	\begin{subfigure}[b]{0.43\linewidth}
		\includegraphics[width=1\linewidth]{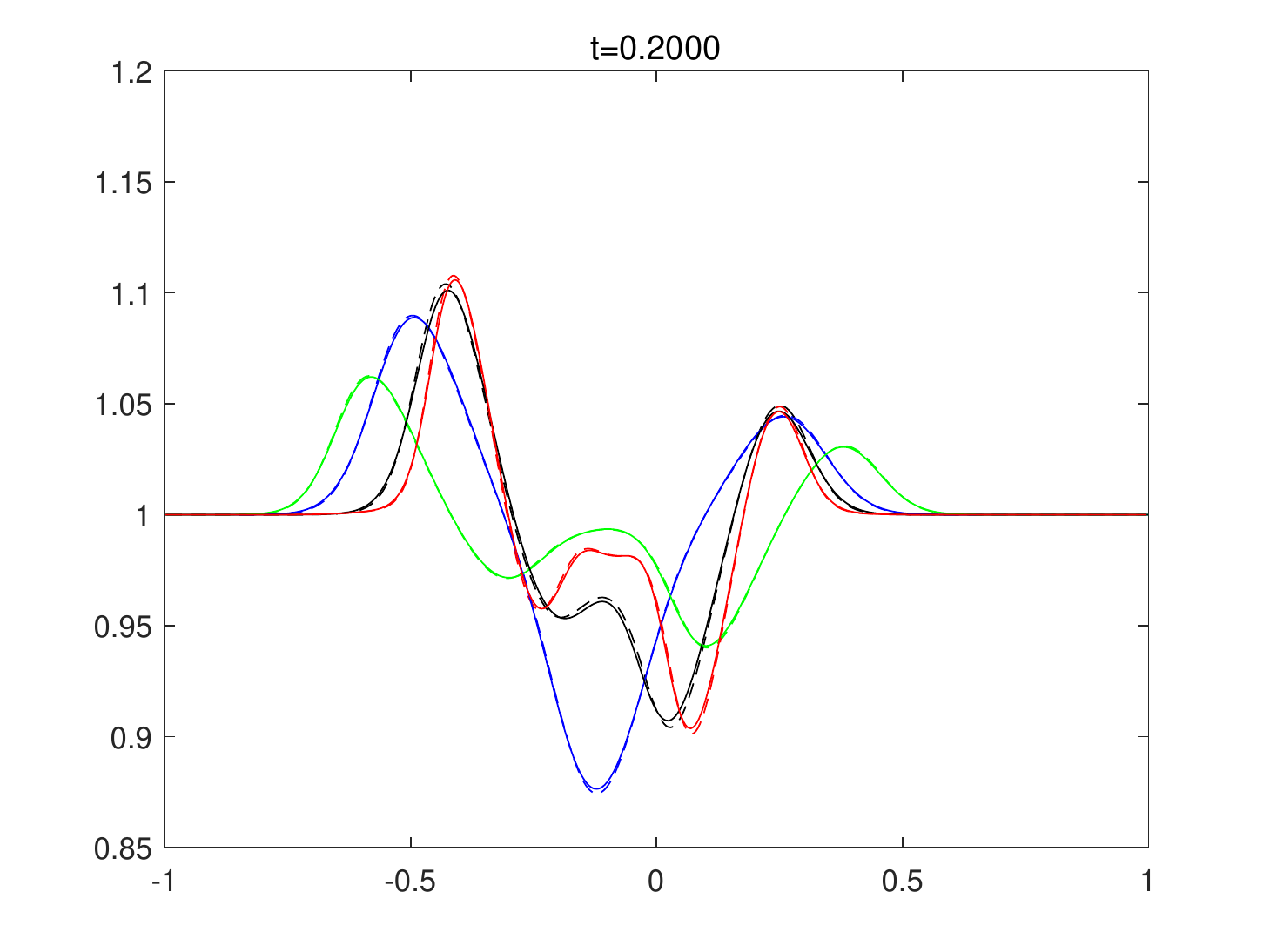}
		\subcaption{Density $\rho_s$, $s=1,2,3,4$}
	\end{subfigure}
	
	\begin{subfigure}[b]{0.43\linewidth}
		\includegraphics[width=1\linewidth]{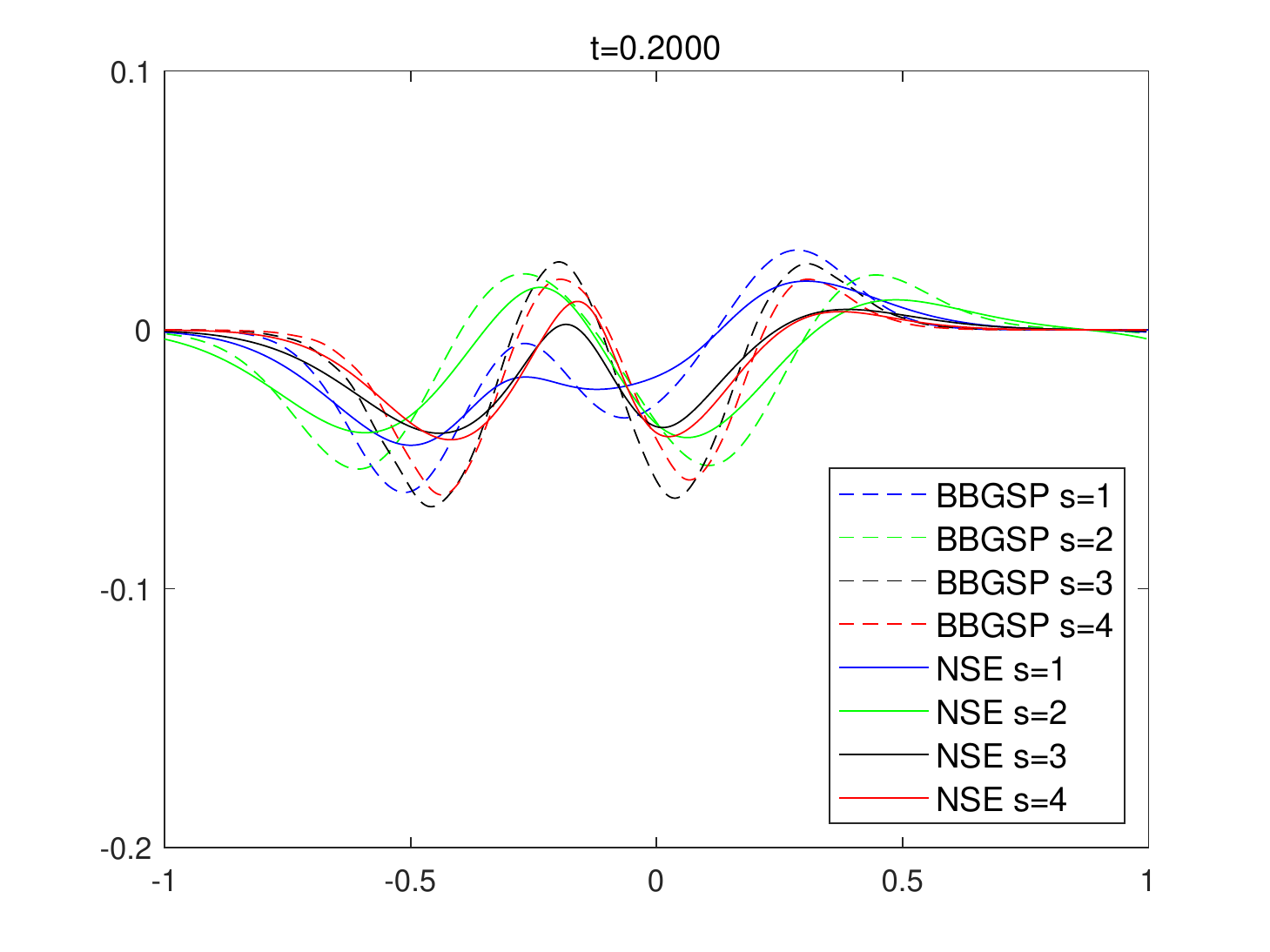}
		\subcaption{Velocity $u_s$, $s=1,2,3,4$}
	\end{subfigure}
	\begin{subfigure}[b]{0.43\linewidth}
		\includegraphics[width=1\linewidth]{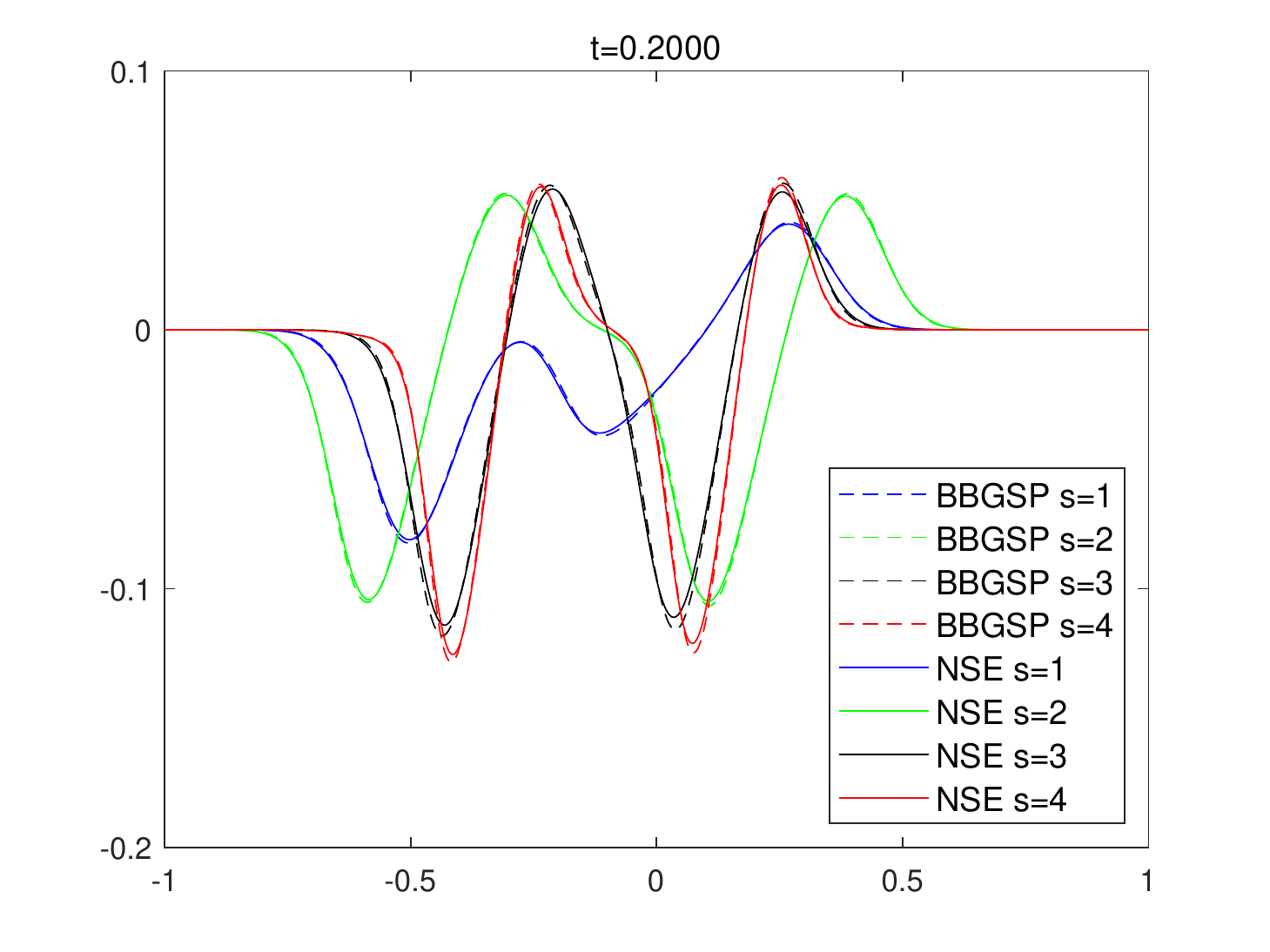}
		\subcaption{Velocity $u_s$, $s=1,2,3,4$}
	\end{subfigure}
	\begin{subfigure}[b]{0.43\linewidth}
		\includegraphics[width=1\linewidth]{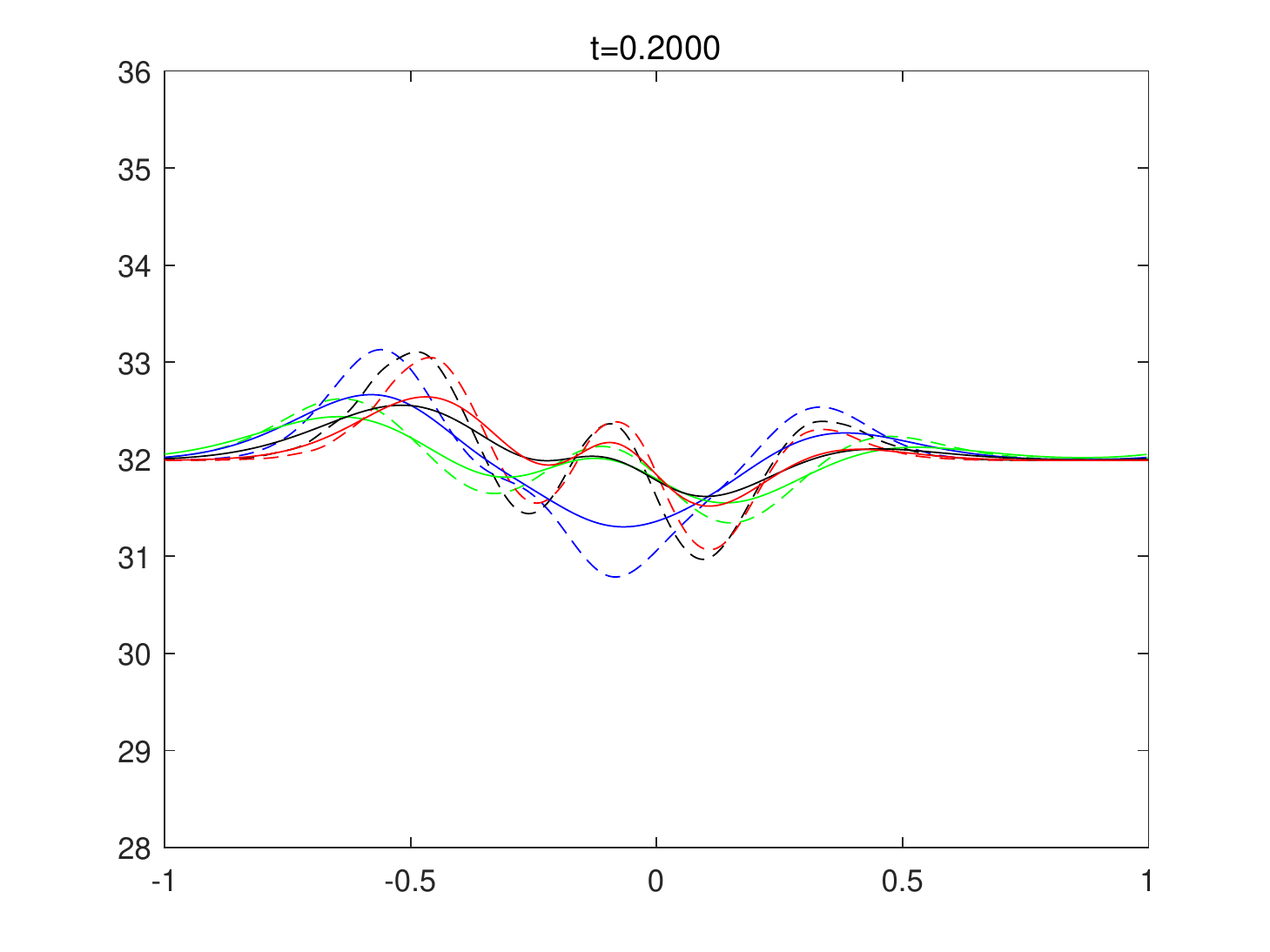}
		\subcaption{Temperature $T_s$, $s=1,2,3,4$}
	\end{subfigure}			
	\begin{subfigure}[b]{0.43\linewidth}
		\includegraphics[width=1\linewidth]{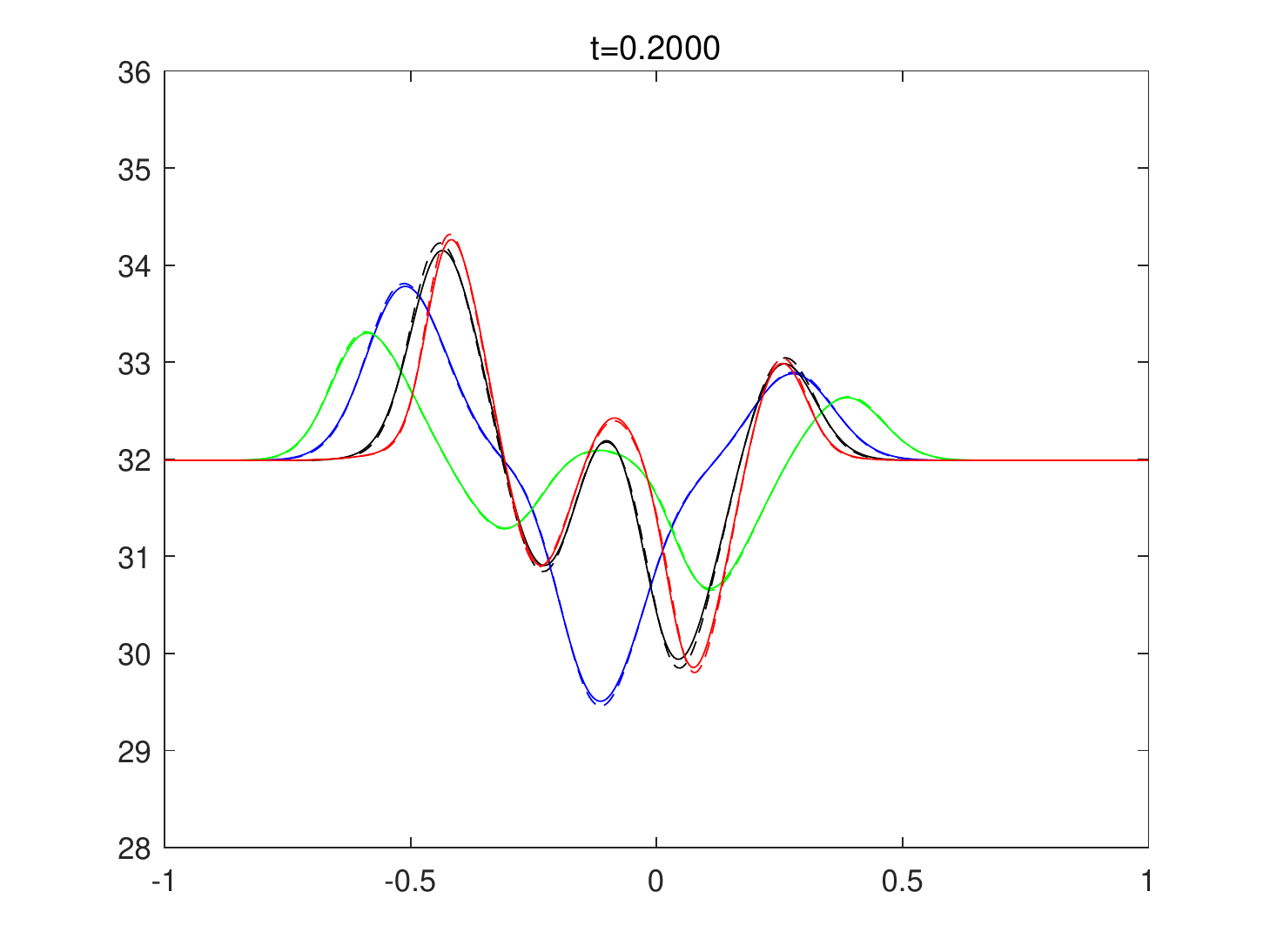}
		\subcaption{Temperature $T_s$, $s=1,2,3,4$}
	\end{subfigure}			
	\caption{
		Comparison of the scaled BBGSP model \eqref{bgk bbgsp multi} and Navier-Stokes equations \eqref{NSE multi} with $\kappa=1$ for $\varepsilon=10^{-2}$ (Left) and $\varepsilon=10^{-3}$ (Right) with initial data in \eqref{initial acc}.
	}\label{fig NSE multi 01}
\end{figure}

\begin{figure}[htbp]
	\centering
	\begin{subfigure}[b]{0.43\linewidth}
		\includegraphics[width=1\linewidth]{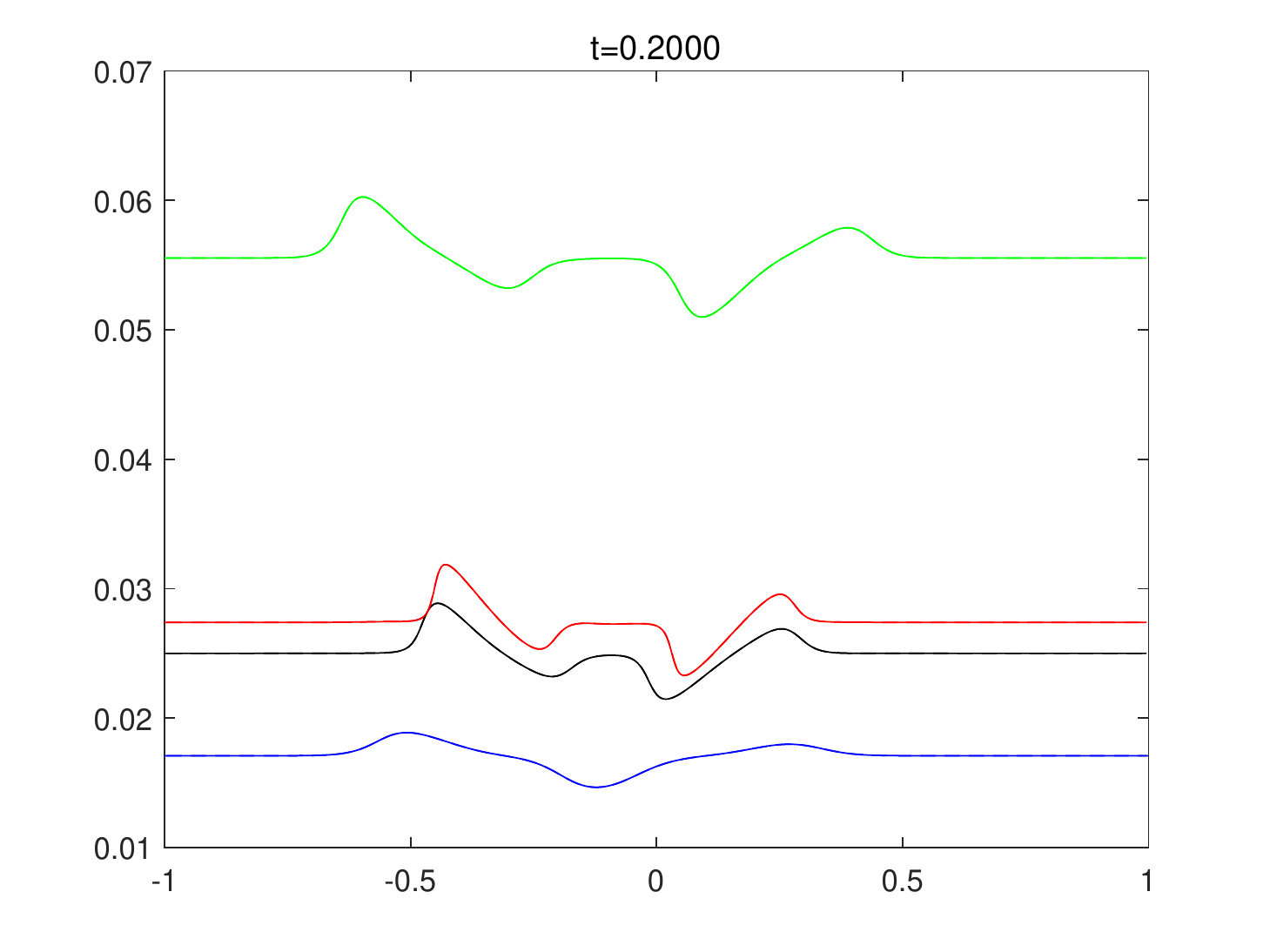}
		\subcaption{Number density $n_s$, $s=1,2,3,4$}
	\end{subfigure}
	\begin{subfigure}[b]{0.43\linewidth}
		\includegraphics[width=1\linewidth]{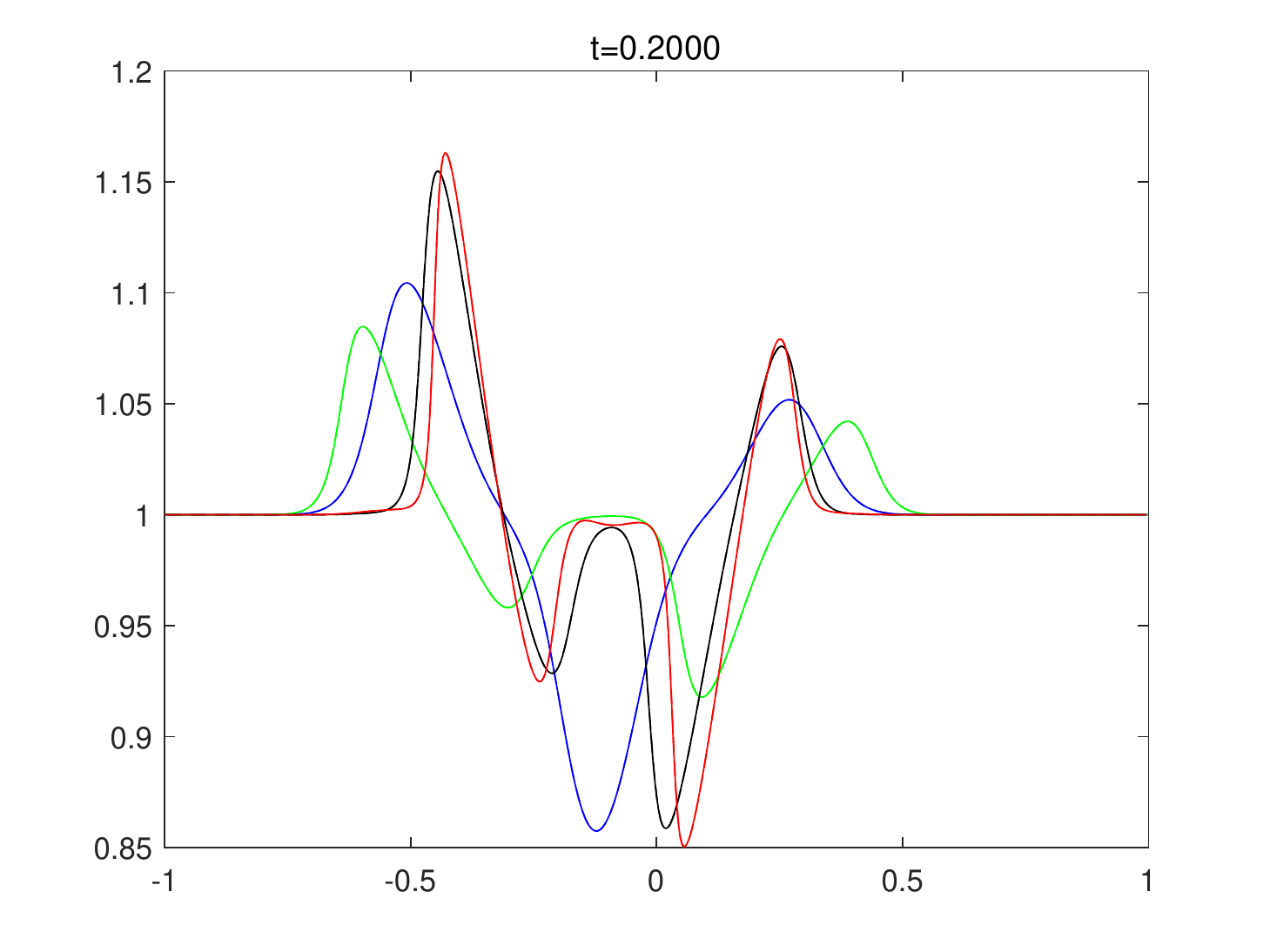}
		\subcaption{Density $\rho_s$, $s=1,2,3,4$}
	\end{subfigure}
	
	\begin{subfigure}[b]{0.43\linewidth}
		\includegraphics[width=1\linewidth]{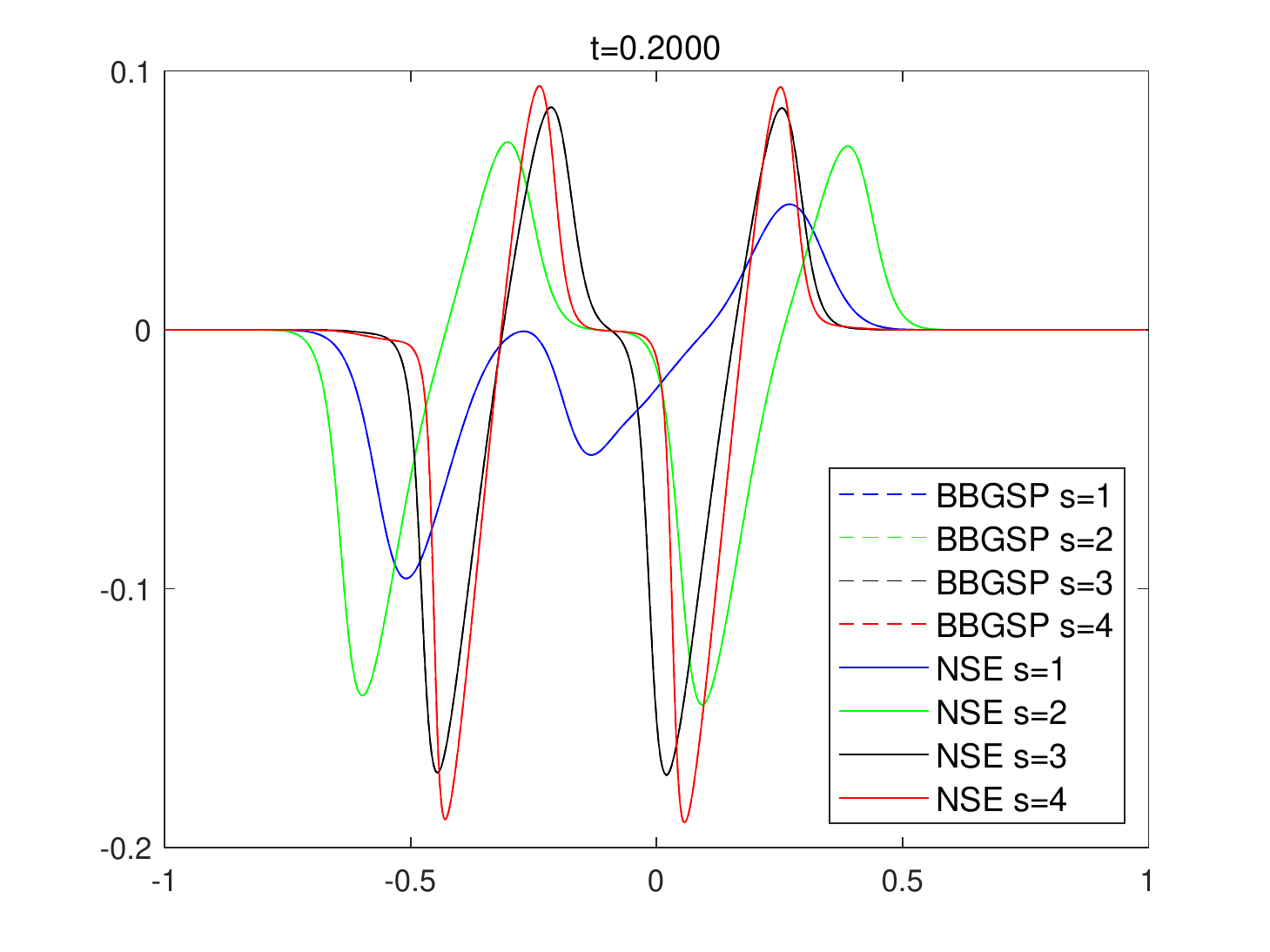}
		\subcaption{Velocity $u_s$, $s=1,2,3,4$}
	\end{subfigure}
	\begin{subfigure}[b]{0.43\linewidth}
		\includegraphics[width=1\linewidth]{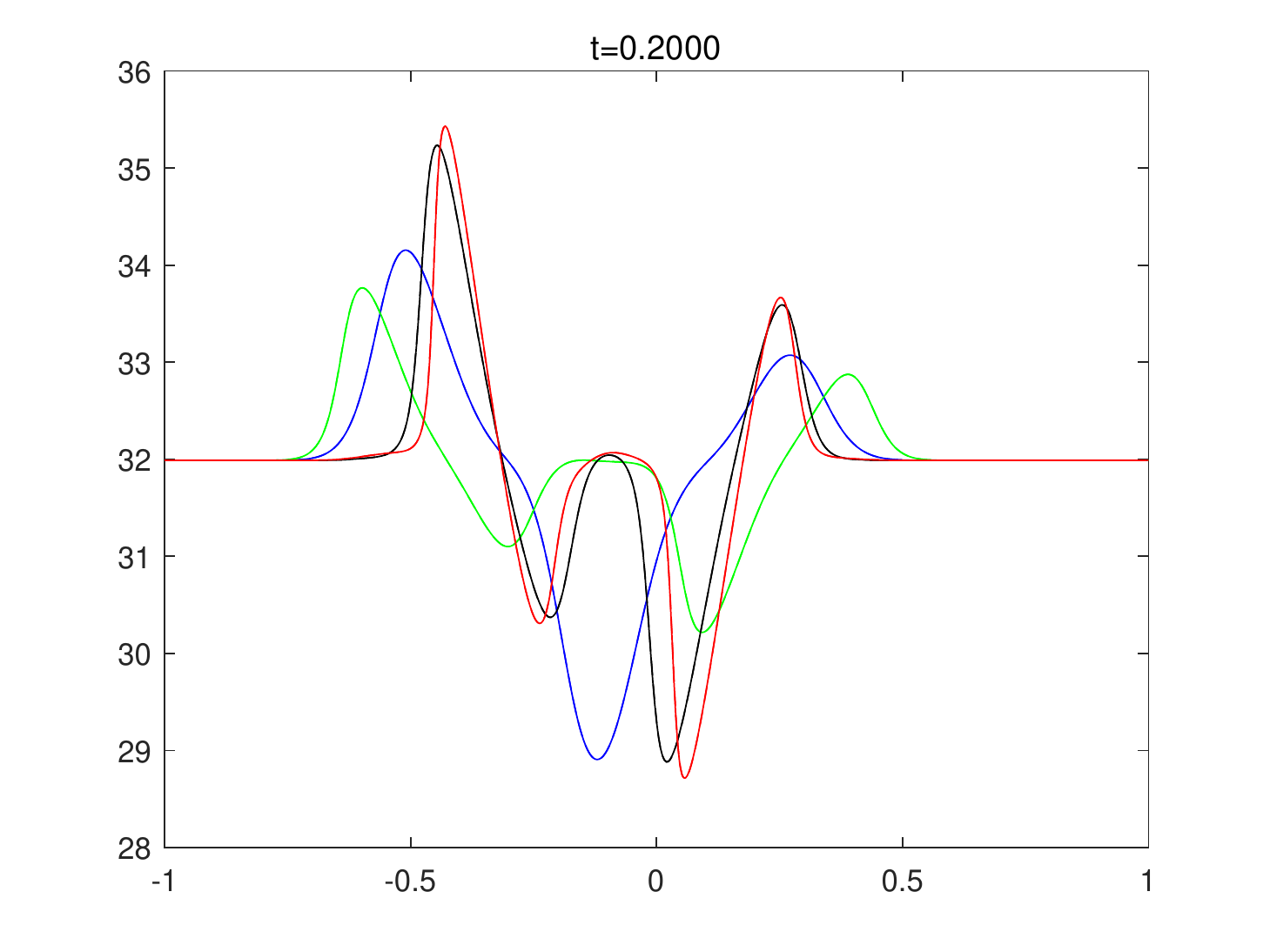}
		\subcaption{Temperature $T_s$, $s=1,2,3,4$}
	\end{subfigure}			
	\caption{
		Comparison of the scaled BBGSP model \eqref{bgk bbgsp multi} and Navier-Stokes equations \eqref{NSE multi} with $\kappa=1$ for $\varepsilon=10^{-4}$ with initial data in \eqref{initial acc}.		
	}\label{fig NSE multi 23}
\end{figure}

In Figures \ref{fig NSE multi 01}-\ref{fig NSE multi 23}, we can observe, contrary to the previous case (that can be reproduced by taking $\kappa=\varepsilon$), a clear separation between macroscopic quantities of the different species; moreover, the behavior of the BGK solutions of \eqref{bgk bbgsp multi} are similar to those of  multi-velocity and multi-temperature NS description \eqref{NSE multi}. We note that in Figure \ref{fig NSE multi 23} for $\varepsilon=10^{-4}$ the kinetic and macroscopic solutions show very good agreement in each species velocity and temperature.

\subsection{Shock problem for binary gas mixture}\label{sec real parameter}
In this test, we consider a shock problem for binary mixture of two noble gases: Helium (He) and Argon (Ar), for which mass ratio becomes relatively large. We aim at checking in this real case which Navier-Stokes description (for global or species macroscopic fields) is better capable to capture the behavior of the mixture. 

To perform the realistic simulation, we consider molecular mass of He and Ar as follows:
\begin{align*}
	m_1^m\approx 4\times 10^{-3} kg/mol\, (He),\quad  m_2^m\approx 40 \times 10^{-3} kg/mol\, (Ar).
\end{align*}
In view of this, we rewrite the BBGSP model  \eqref{bgk bbgsp} in terms of mole. We divide $f_s$ by the Avogadro's number $N_a\approx 6.02 \times 10^{23}$. Now, we denote the molecular density by 
\begin{align*}
	f_{s}^m(\textbf{x},{\bf v},t):=\frac{f(\textbf{x},{\bf v},t)}{N_a}
\end{align*}
for $s=1,2$.
Then, we can rewrite \eqref{bgk bbgsp} as
\begin{align}
	\frac{\partial{f_s^{m}}}{\partial{t}} + {\bf v} \cdot \nabla_{\textbf{x}}{f_s^m} =  \frac{1}{\varepsilon}\sum_{k = 1}^{L} \nu_{sk}\left(n_{s}M_{sk}-f_s^m\right),\quad M_{sk}=M({\bf v};u_{sk}^m,\frac{RT_{sk}^m}{m_s^m}),
\end{align}
with macroscopic quantities at the level of mole:
\begin{align*}
	n_s^m=\langle f_s^m,1 \rangle, \quad n_s^m u_s^m=\langle f_s^m,{\bf v} \rangle, \quad 3n_s^mRT_s^m=m_s^m\langle f_s^m,|{\bf v}-u_s|^2 \rangle.
\end{align*}
Here we define the universal gas constant $R$ as $R:=K_B N_a\approx 8.3145 \, J/mol$ and the molecular mass $m_s^m$ as $m_s^m=m_sN_a$. Note that the values of $u_{sk}^m$ and $T_{sk}^m$ are samely defined as in \eqref{uskTsk} by using $m_s^m$, $n_s^m$, $u_s^m$, $T_s^m$. For global macroscopic variables we use the following expressions:
\begin{align*}
	n^m=\sum_{s=1}^L n_s^m,\quad \rho^m=\sum_{s=1}^L \rho_s^m,\quad \rho_s^m = m_s^m n_s^m,\quad s=1,\cdots,L\cr
	u^m= \frac{1}{\rho^m} \sum_{s=1}^L \rho_s^m u_s^m, \quad 3n^mRT^m=3\sum_{s=1}^L n_s^mRT_s^m + \sum_{s=1}^L \rho_s^m|u_s^m-u^m|^2	
\end{align*}
Let us consider room temperature $T_s^m=300 K$. As in \cite{BGM}, here we use the collision frequencies corresponding to $T_s^m=300 K$ based on the following formula:
	\begin{align}\label{coll fre formula}
		\begin{split}
			\nu_0^{ss}&=\frac{4}{3} \frac{T}{\mu_s(T)}, \quad s=1,2\cr
			\nu_0^{12}&= \frac{2\sqrt{2}}{3}\frac{(m_1+m_2)^\frac{1}{4}}{(m_1m_2)^\frac{1}{2}}\frac{T}{(\mu_1(T)\mu_2(T))^\frac{1}{2}},
		\end{split}
	\end{align}
	where viscosity coefficients $\mu_s$ for noble gases are provided in \cite{KKMNRW}. In case of Helium and Argon, we have
	\begin{align*}
		&\nu_0^{11}= 19.96,\quad \nu_0^{12}= 1.153\cr 
		&\nu_0^{21}= 1.153,\quad \nu_0^{22}= 17.52.
	\end{align*}
Note that we set $\nu_{sk}=\nu_0^{sk}n_k$.
Now, we consider a shock problem by taking the Maxwellian as initial data which reproduces the following macroscopic variables:
\begin{align*}
	\left(\rho_0^m, u_0^m, T_0^m\right) =
	\begin{cases}
		\left(1.7628,\,0,\,300\right), \quad x<0.5\\
		\left(0.8814,\,0,\,300\right), \quad x>0.5
	\end{cases}
\end{align*}
The units of $\rho_0^m$, $u_0^m$, $T_0^m$ are $kg/m^3$, $m/s$, $K$, respectively. Now, we set 
\begin{align*}
	\begin{split}
		\left(\rho_{01}^m, \rho_{02}^m\right) &=
		\begin{cases}
			\left(0.1598,\,1.6030\right), \quad x<0.5\\
			\left(0.0799,\,0.8015\right), \quad x>0.5\\
		\end{cases}.
	\end{split}
\end{align*}
Note that the density of Helium and Argon gases for $T_{0}^m=300K$ for $p^m=1 \,bar$ are given by  
\begin{align*}
	\rho_1^m=0.1598 \, kg/m^3,\quad  	\rho_2^m=1.603 \, kg/m^3.
\end{align*}
For numerical simulation, we assume the free-flow boundary condition on the spatial domain $x\in [-6,6]$ with velocity domain $v\in [-160,160]$. We compute numerical solutions with $N_x=600$ and  $N_v=320$ up to $t_f=0.06$. Here we use CFL=$1.5$ and take different values of $\varepsilon=10^{-q}$, $q=3,4,5$. 

\begin{figure}[htbp]
	\centering
	\begin{subfigure}[b]{0.43\linewidth}
		\includegraphics[width=1\linewidth]{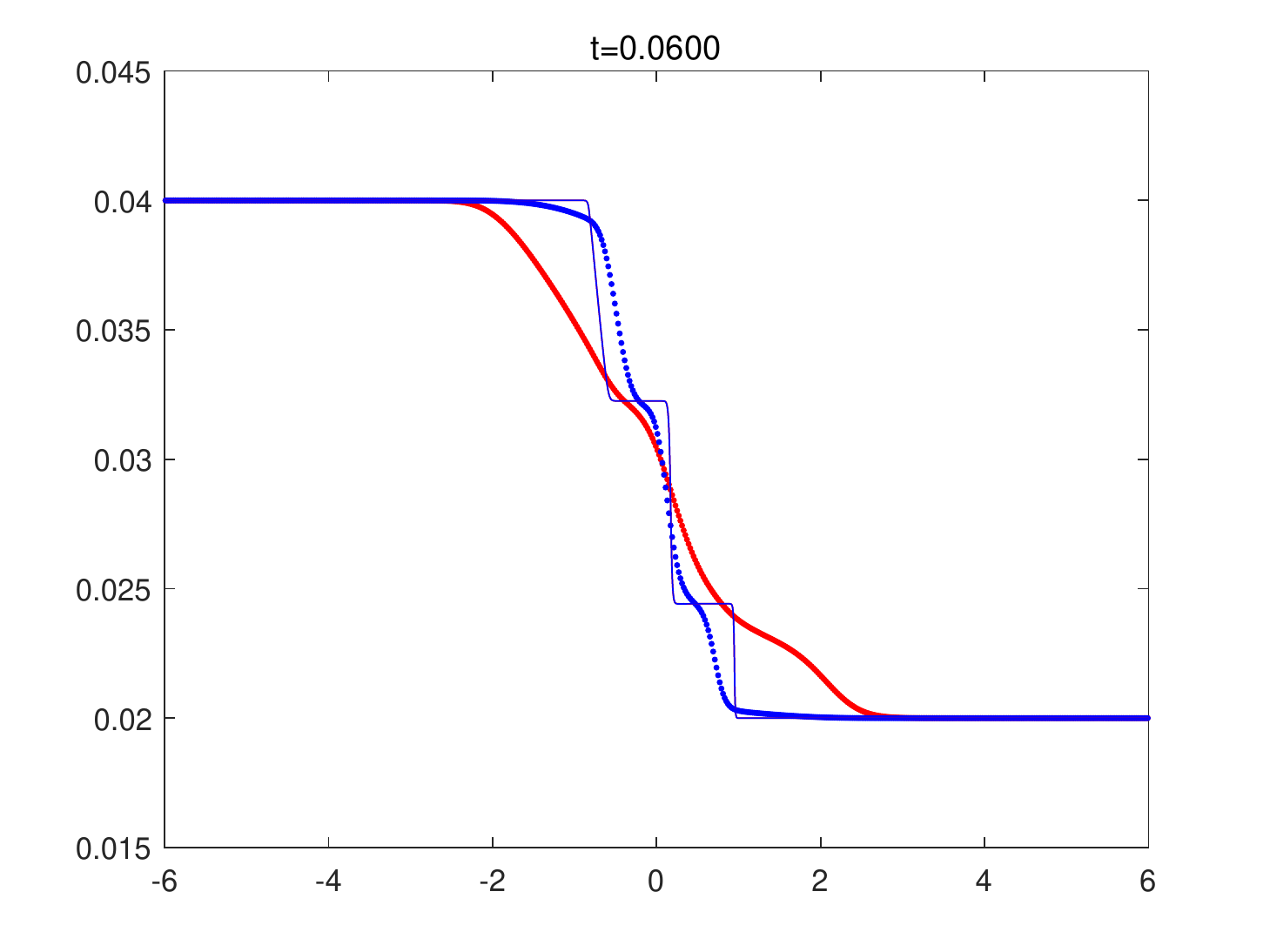}
		\subcaption{Number density $n_s$, $s=1,2$}
	\end{subfigure}
	\begin{subfigure}[b]{0.43\linewidth}
		\includegraphics[width=1\linewidth]{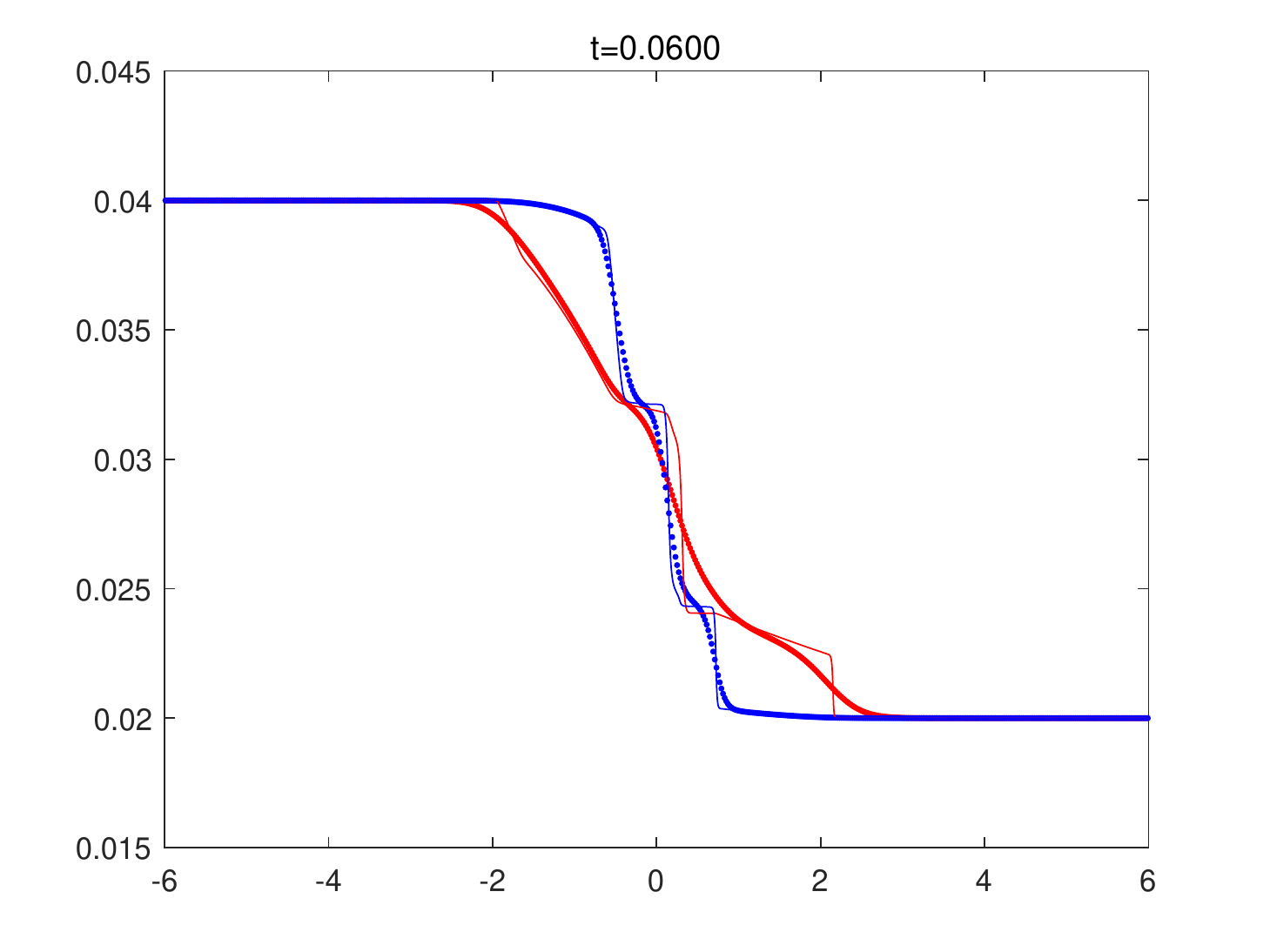}
		\subcaption{Number density $n_s$, $s=1,2$}
	\end{subfigure}
	\begin{subfigure}[b]{0.43\linewidth}
		\includegraphics[width=1\linewidth]{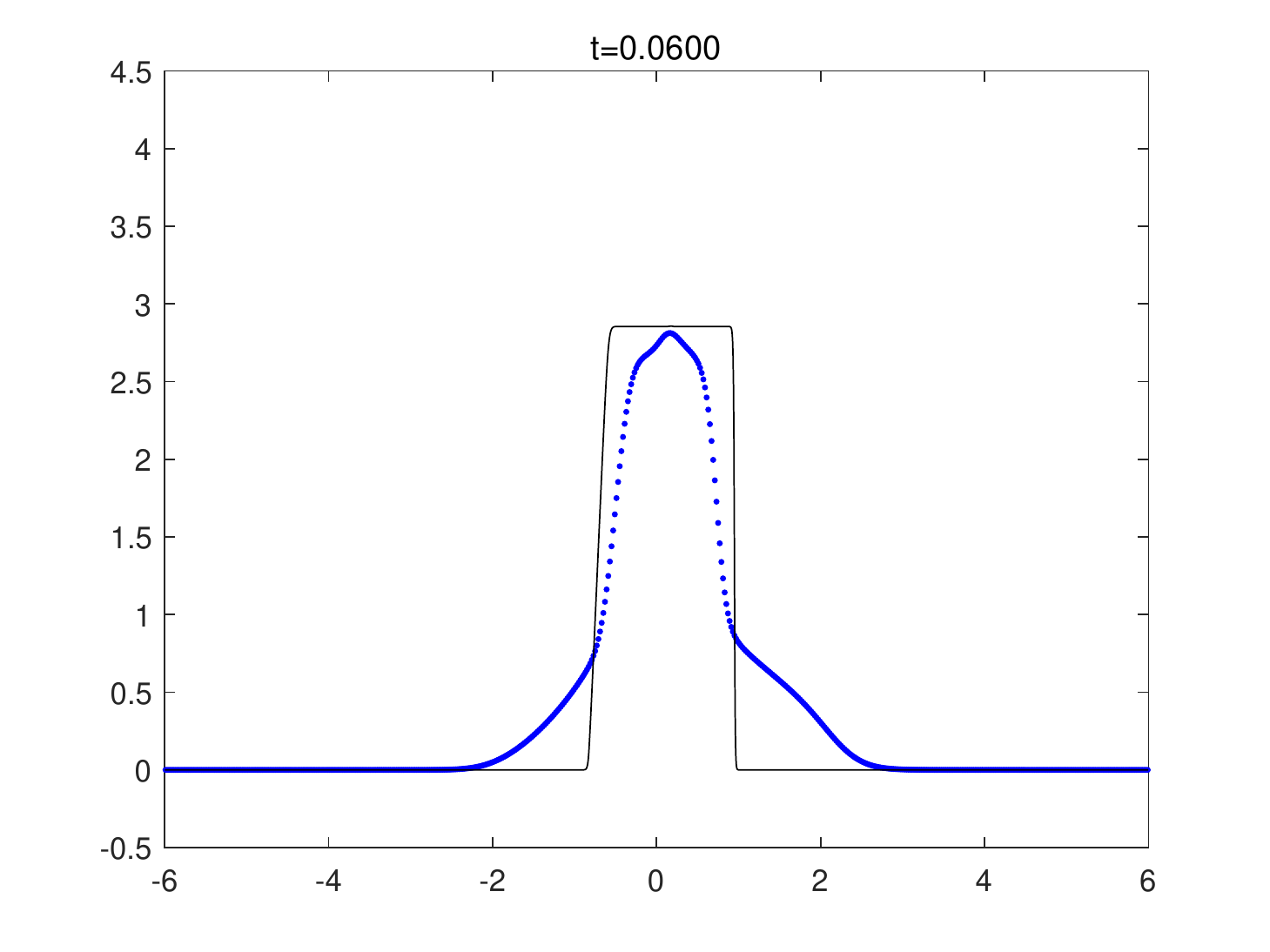}
		\subcaption{Velocity $u$}
	\end{subfigure}
	\begin{subfigure}[b]{0.43\linewidth}
		\includegraphics[width=1\linewidth]{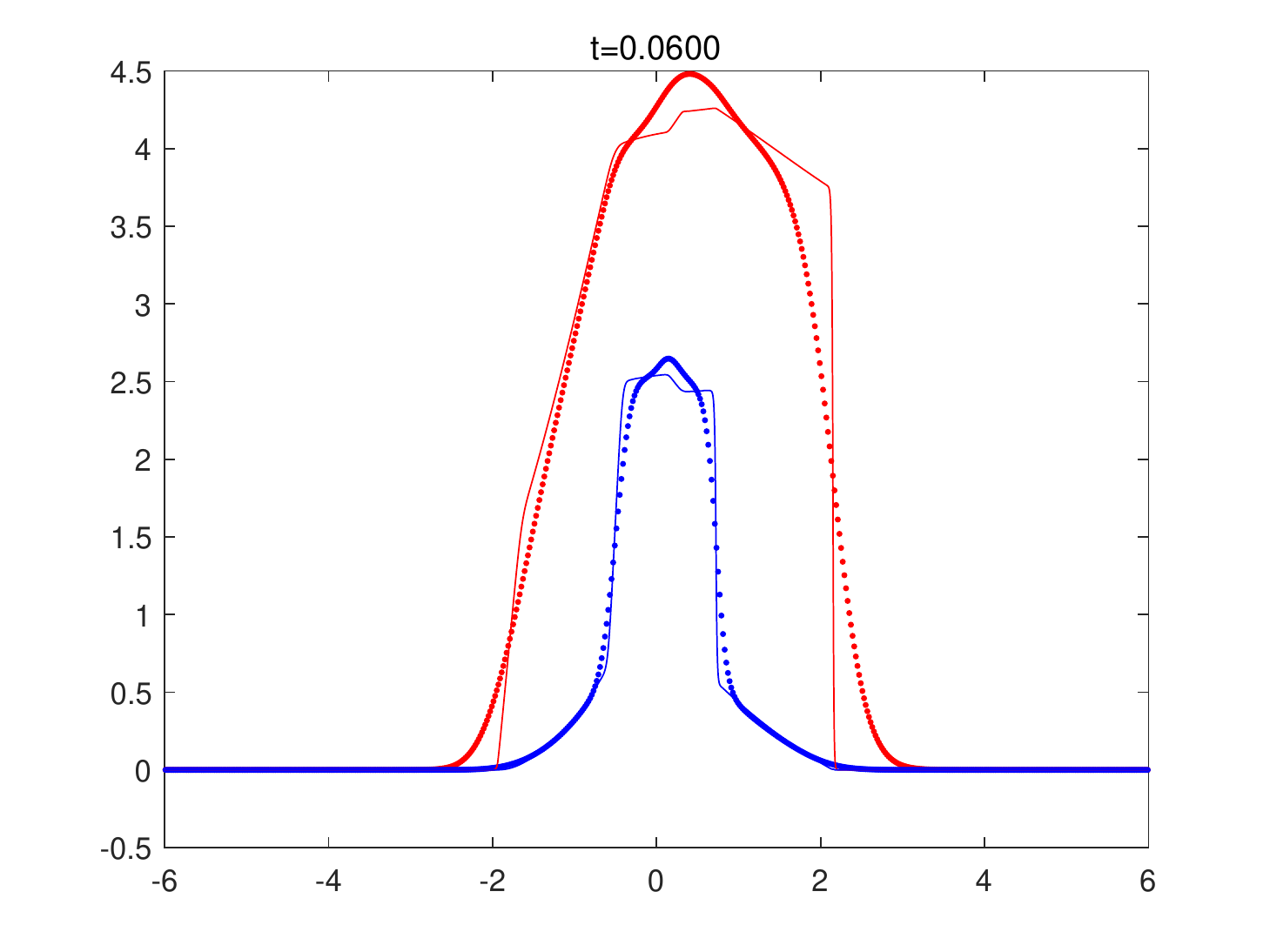}
		\subcaption{Velocity $u$}
	\end{subfigure}
	\begin{subfigure}[b]{0.43\linewidth}
		\includegraphics[width=1\linewidth]{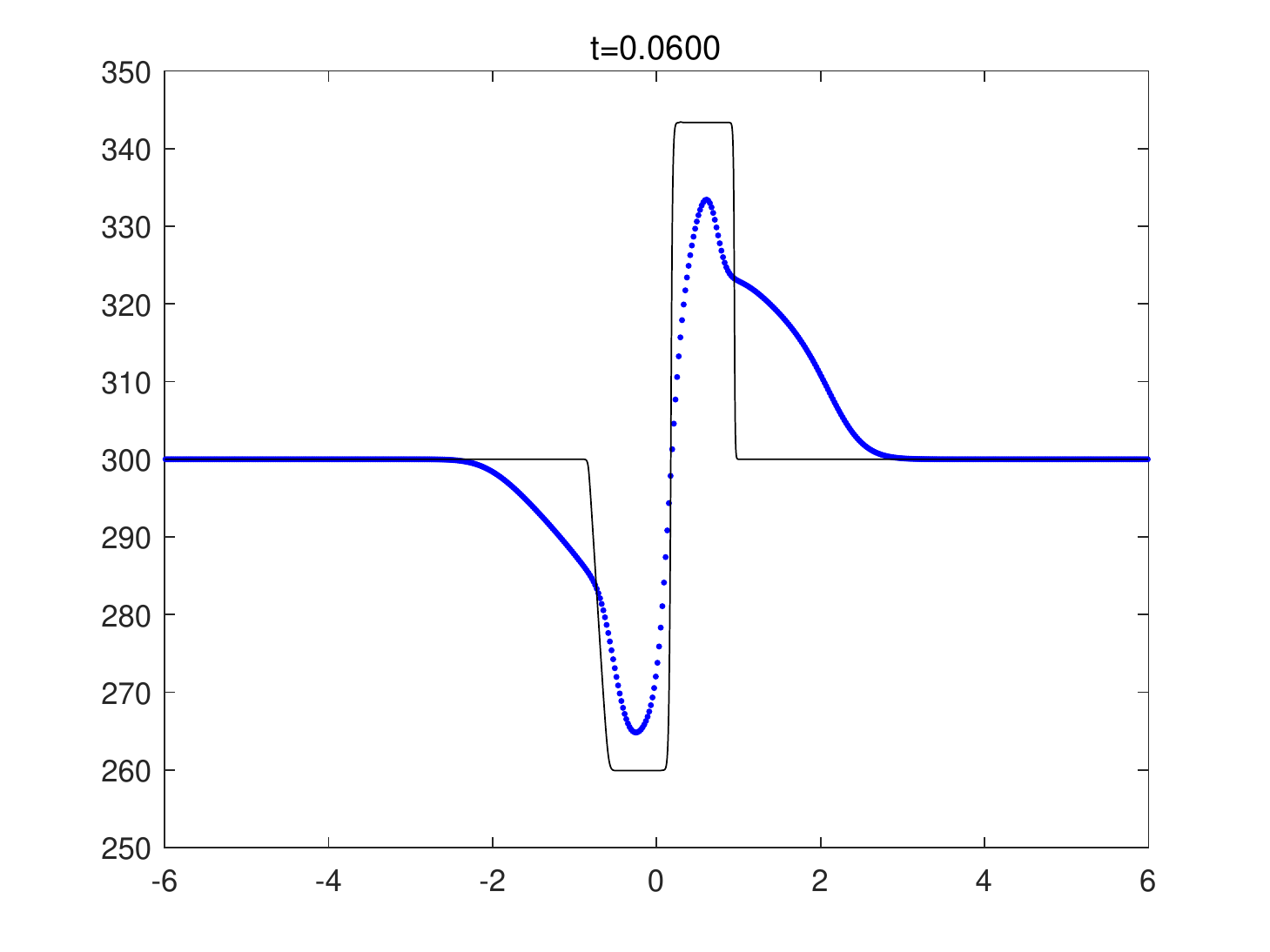}
		\subcaption{Temperature $T$}
	\end{subfigure}			
	\begin{subfigure}[b]{0.43\linewidth}
		\includegraphics[width=1\linewidth]{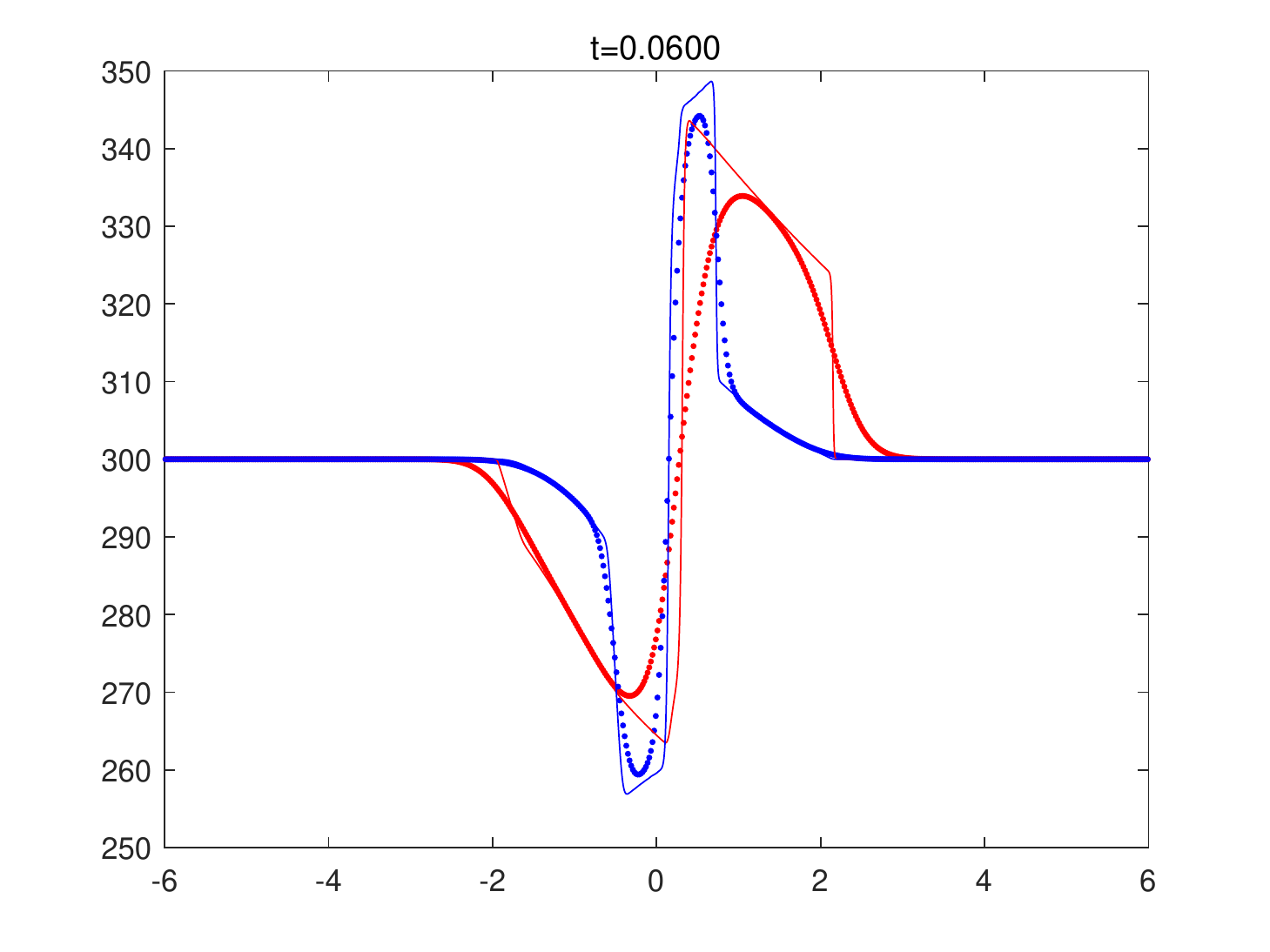}
		\subcaption{Temperature $T$}
	\end{subfigure}		
	\caption{Comparison of the scaled BBGSP model \eqref{bgk bbgsp multi} for $\varepsilon=\kappa=10^{-3}$ with: (left) global velocity and temperature Euler system \eqref{NSE} for $\varepsilon=0$ and (right) multi-velocity and multi-temperature Euler system \eqref{NSE multi} for $\varepsilon=0$, $\kappa=10^{-3}$. We use the initial data in \eqref{sec real parameter}.
	}\label{real parameter 3}
\end{figure}

In Figure \ref{real parameter 3}, for relatively large values of $\kappa=10^{-3}$, the panels on the left column show that the global velocity and temperature is closely related to the dynamics of Argon gas. This is because its density is ten times bigger than that of Helium gas. Thus, it is difficult to describe the dynamics of mixtures involving
Helium gas with global velocity and temperature description. On the other hand, the panels on the right column enables us to capture the behaviors of Helium gas and this is the case where multi-velocity and multi-temperature description can be a suitable model for a better description of this dynamics. In the following Figures \ref{real parameter 4}-\ref{real parameter 5}, both species velocities and temperatures are close to global velocity and temperature. Here the global velocity and temperature Euler system can be already a suitable choice for describing the dynamics of binary mixtures.

In the right panels of Figures \ref{real parameter 3}-\ref{real parameter 5}, we observe that the species velocity and temperature for light gas show very different behaviors as both $\varepsilon$ and $\kappa$ becomes smaller. For a better understanding of these observations with the scaled BBGSP model \eqref{bgk bbgsp multi}, let us consider binary gas mixtures with $m_2=rm_1$ where $r<1$ denotes their mass ratio. 
Then, in the limit $r\rightarrow 0$, we obtain
\begin{align*}
	M_{12} &\rightarrow M\left({\bf v};u_1,\frac{K_BT_1}{m_1}\right),\quad M_{21} \rightarrow M\left({\bf v};u_1,\frac{K_BT_2}{m_2} + \frac{1}{3}|u_1-u_2|^2\right),
\end{align*}
while $M_{ss} = M\left(v;u_s,\frac{K_BT_s}{m_s}\right)$ for $s=1,2$.
The form of $M_{12}$ and $M_{21}$ implies that as the intra-species collisions becomes dominant, light gas tends to follow the behavior of heavy gas:
\begin{align*}
	\frac{\partial{f_2}}{\partial{t}} + {\bf v} \cdot \nabla_{\textbf{x}}{f_2} &=  \frac{\nu_{21}}{\kappa} \left(n_{2}M\left({\bf v};u_1,\frac{K_BT_2}{m_2} + \frac{1}{3}|u_1-u_2|^2\right)-f_2\right) + \frac{\nu_{22}}{\varepsilon} \left(n_{2}M_{22}-f_2\right),
\end{align*} 
while heavy gas tends to behave like a single gas:
\begin{align*}
	\frac{\partial{f_1}}{\partial{t}} + {\bf v} \cdot \nabla_{\textbf{x}}{f_1} &=  \frac{\nu_{11}}{\varepsilon} \left(n_{1}M_{11}-f_1\right) + \frac{\nu_{12}}{\kappa} \left(n_{1}M\left({\bf v};u_1,\frac{K_BT_1}{m_1}\right)-f_1\right).
\end{align*}
%
Furthermore, the formula \eqref{coll fre formula} for noble gases gives
		\begin{align*}
			\nu_0^{12}
			&= \frac{2\sqrt{2}}{3}\left(\frac{m_2}{m_1}\right)^{\frac{1}{4}}\frac{\left(1+\frac{m_2}{m_1}\right)^\frac{1}{4}}{ (m_2)^\frac{1}{2}}\frac{T}{(\mu_1(T)\mu_2(T))^\frac{1}{2}}
			= \frac{2\sqrt{2}}{3}r^{\frac{1}{4}}\frac{(1+r)^\frac{1}{4}}{ (m_2)^\frac{1}{2}}\frac{T}{\left(\mu_1(T)\mu_2(T)\right)^\frac{1}{2}},
		\end{align*}
		in which, for a fixed value of $m_2>0$ the limit $r\rightarrow 0$ implies $\nu_0^{12}\rightarrow 0$. If $\nu_0^{12}<<\kappa$, the scaled BBGSP model \eqref{bgk bbgsp multi} formally reduces to two independent equations:
		\begin{align*}
			\frac{\partial{f_1}}{\partial{t}} + {\bf v} \cdot \nabla_{\textbf{x}}{f_1} &=  \frac{\nu_{11}}{\varepsilon} \left(n_{1}M\left({\bf v};u_1,\frac{K_BT_1}{m_1}\right)-f_1\right)\cr
			\frac{\partial{f_2}}{\partial{t}} + {\bf v} \cdot \nabla_{\textbf{x}}{f_s} &=  \frac{\nu_{22}}{\varepsilon} \left(n_{2}M\left({\bf v};u_2,\frac{K_BT_2}{m_2}\right)-f_2\right).
		\end{align*}
	where $\nu_{ss}=\nu_0^{ss}n_s$.

\begin{figure}[htbp]
	\centering
	\begin{subfigure}[b]{0.43\linewidth}
		\includegraphics[width=1\linewidth]{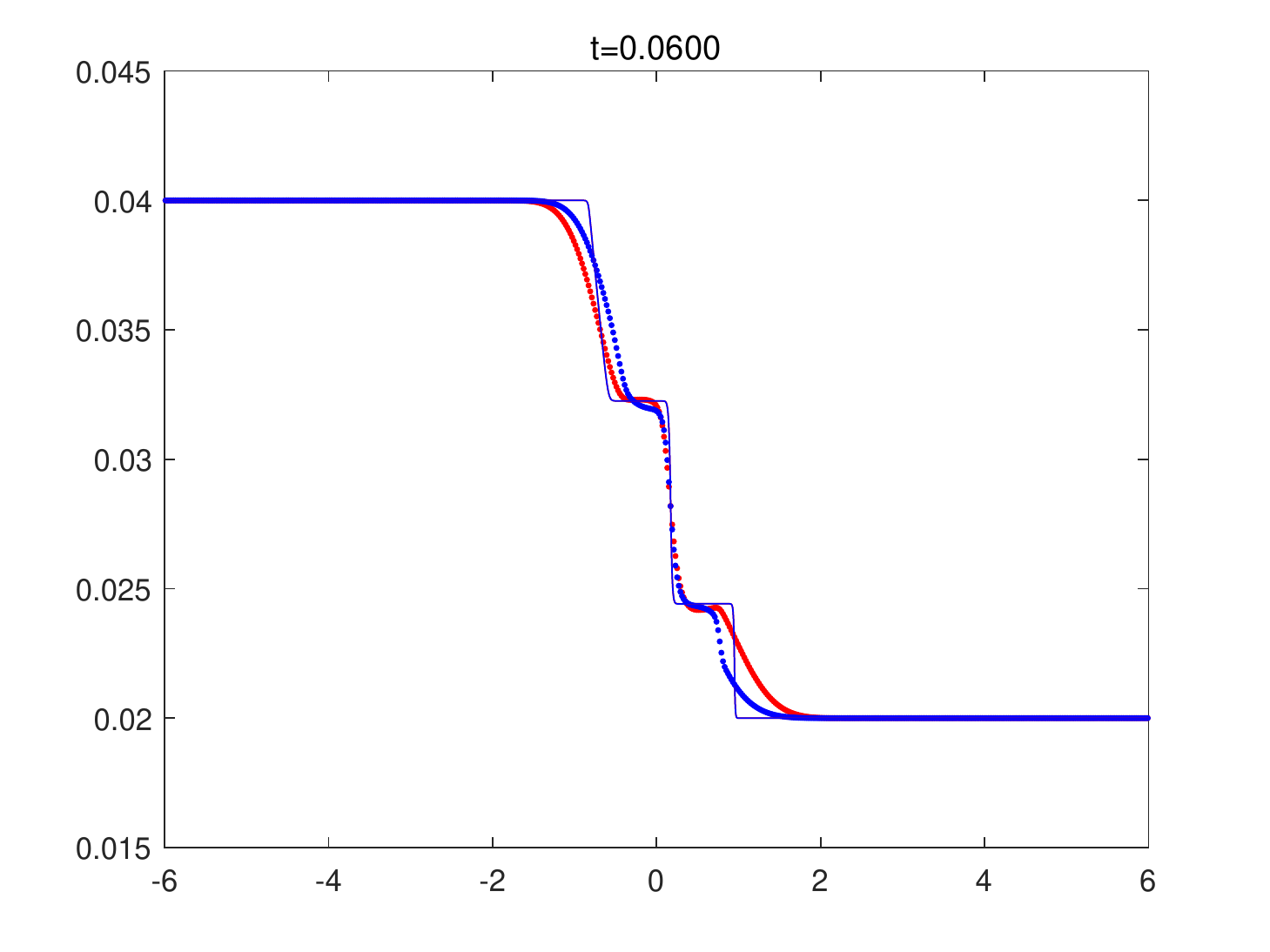}
		\subcaption{Number density $n_s$, $s=1,2$}
	\end{subfigure}
	\begin{subfigure}[b]{0.43\linewidth}
		\includegraphics[width=1\linewidth]{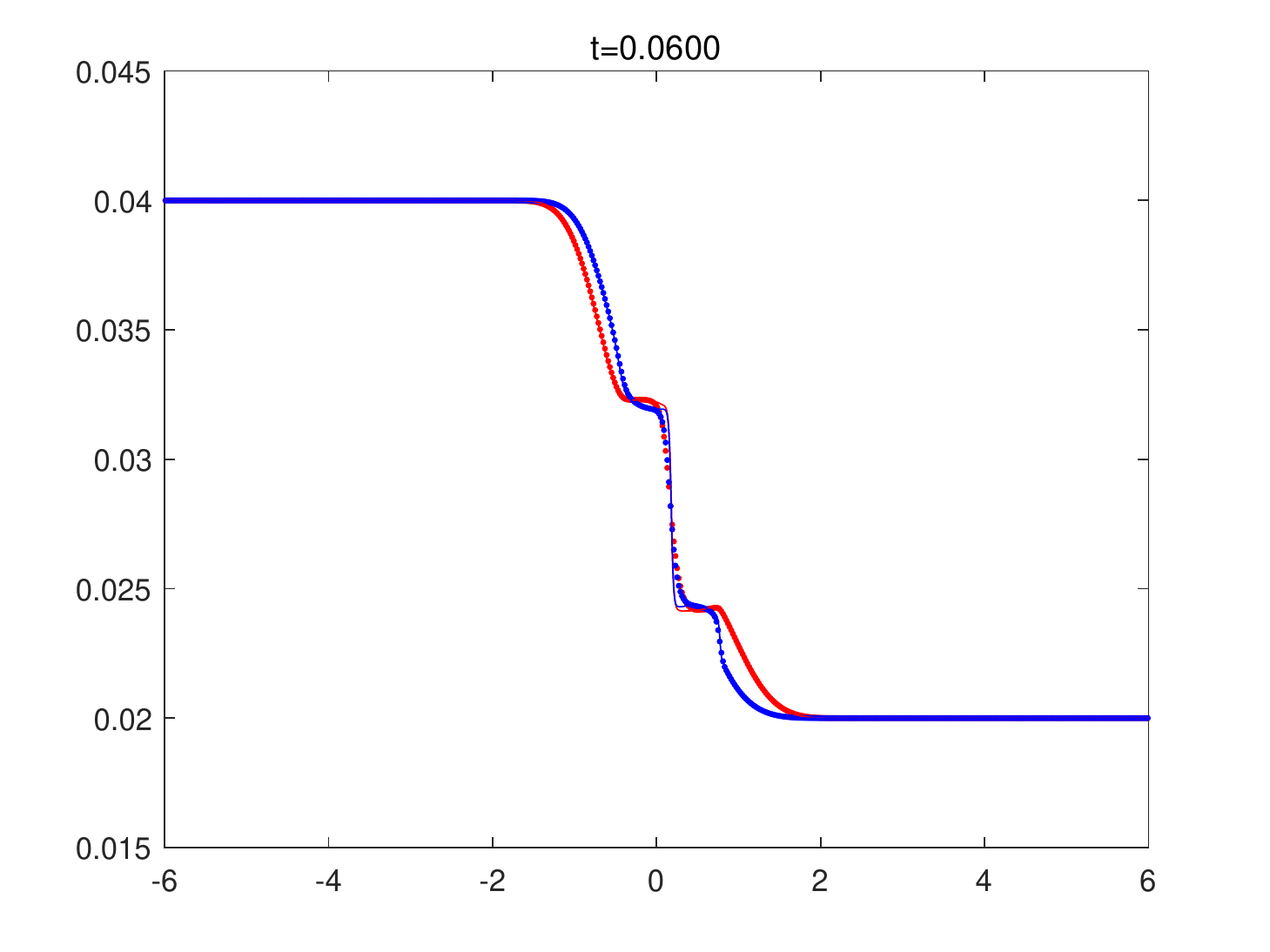}
		\subcaption{Number density $n_s$, $s=1,2$}
	\end{subfigure}
	\begin{subfigure}[b]{0.43\linewidth}
		\includegraphics[width=1\linewidth]{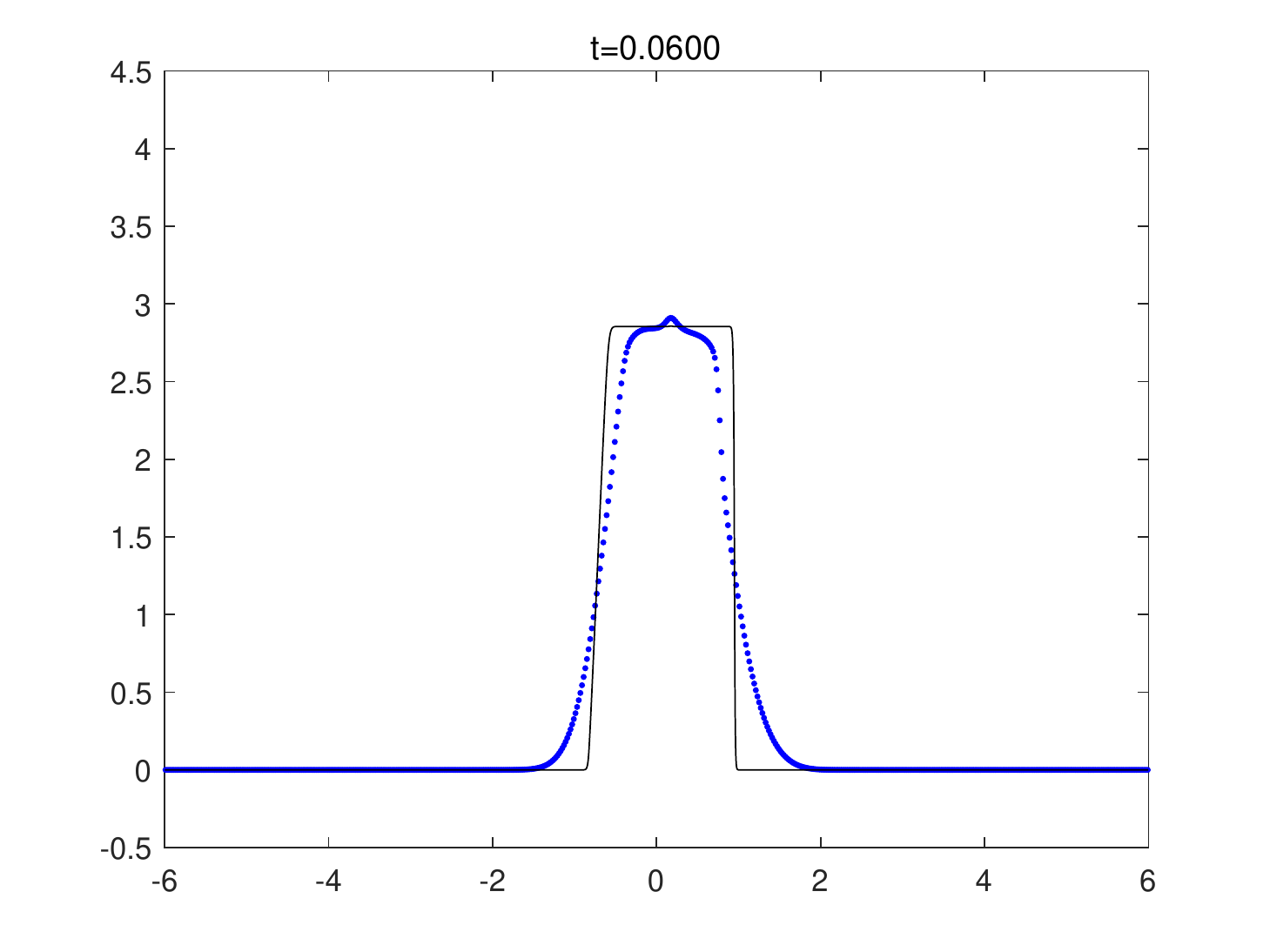}
		\subcaption{Velocity $u$}
	\end{subfigure}
	\begin{subfigure}[b]{0.43\linewidth}
		\includegraphics[width=1\linewidth]{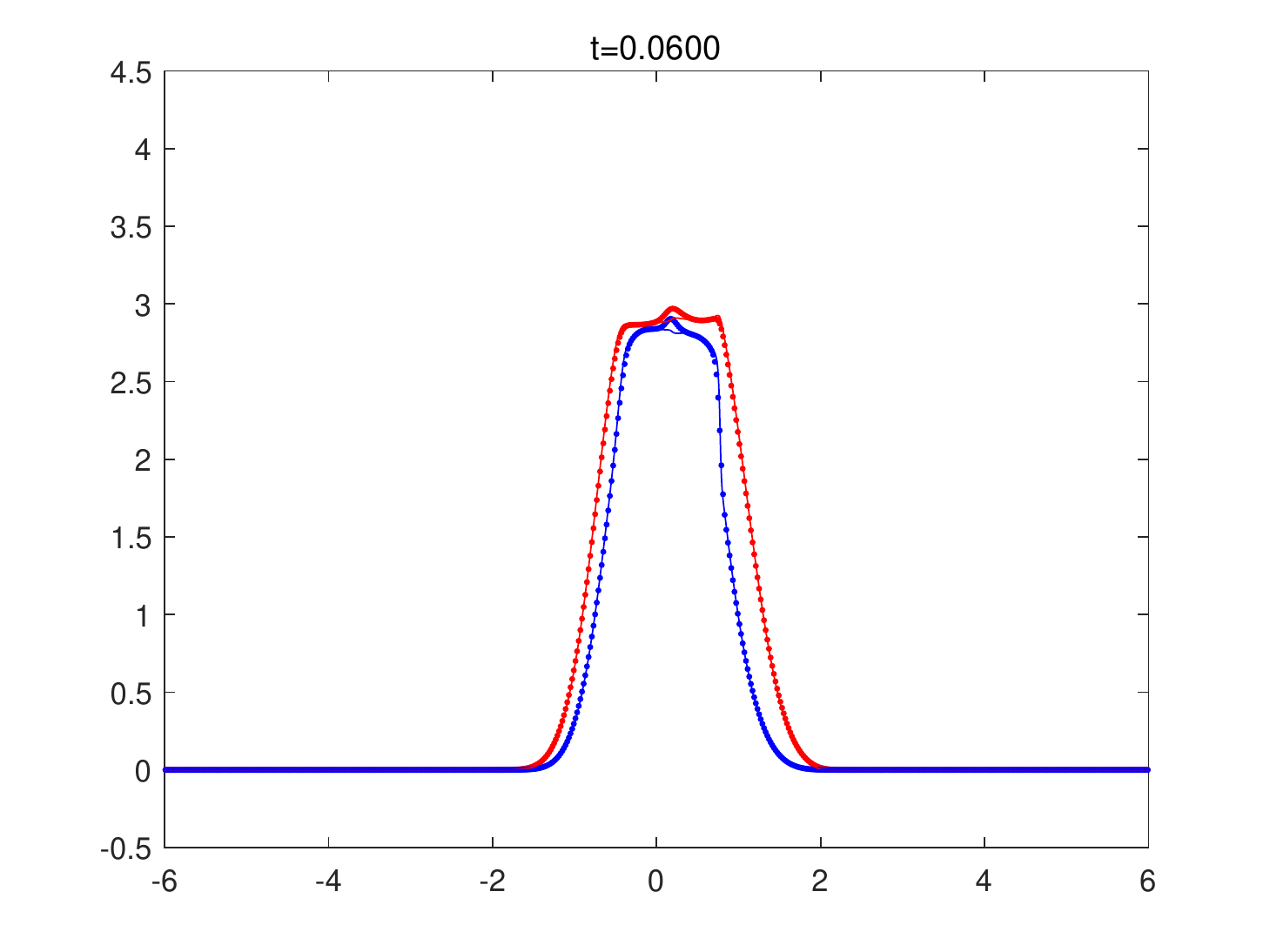}
		\subcaption{Velocity $u$}
	\end{subfigure}
	\begin{subfigure}[b]{0.43\linewidth}
		\includegraphics[width=1\linewidth]{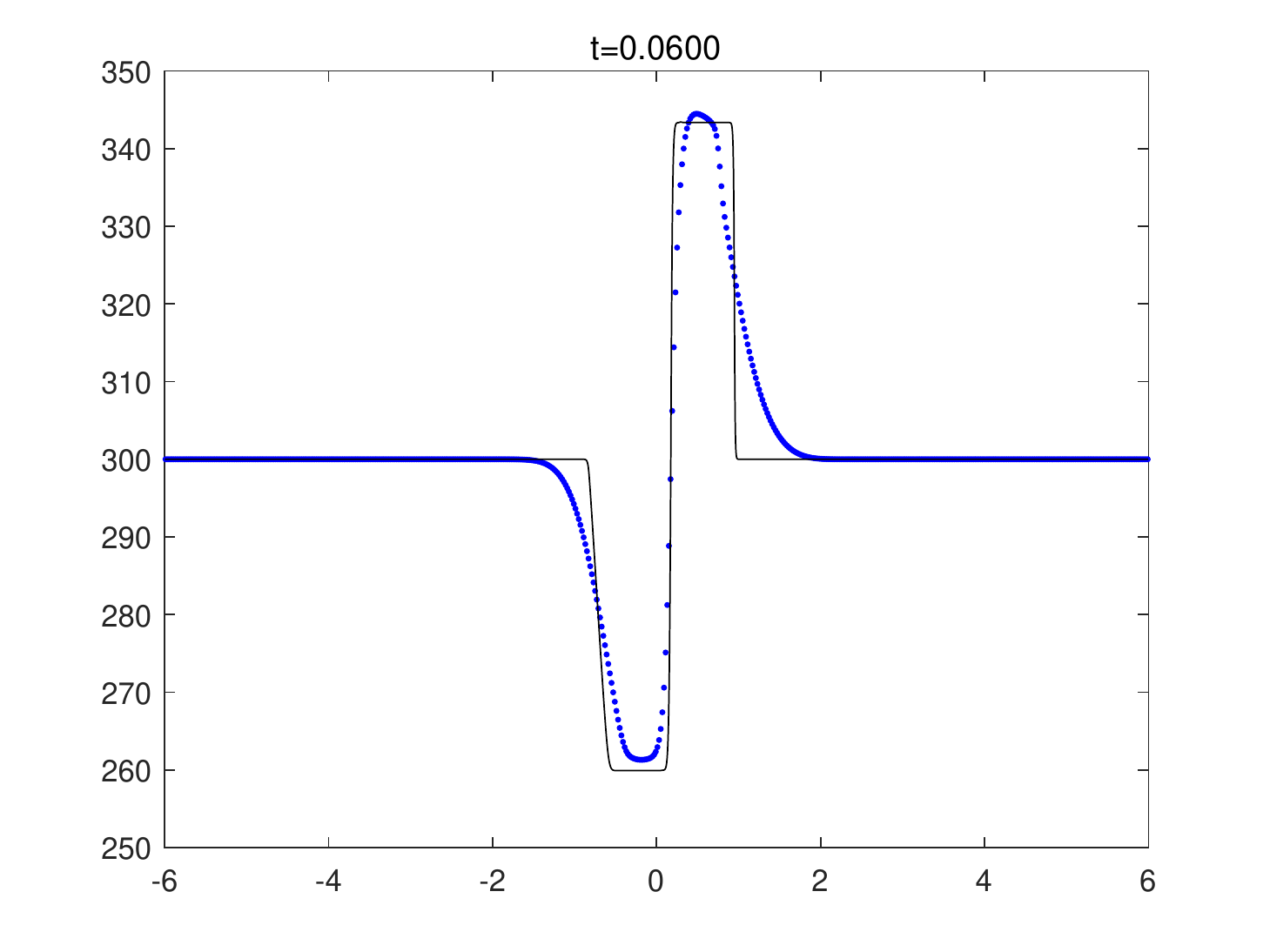}
		\subcaption{Temperature $T$}
	\end{subfigure}			
	\begin{subfigure}[b]{0.43\linewidth}
		\includegraphics[width=1\linewidth]{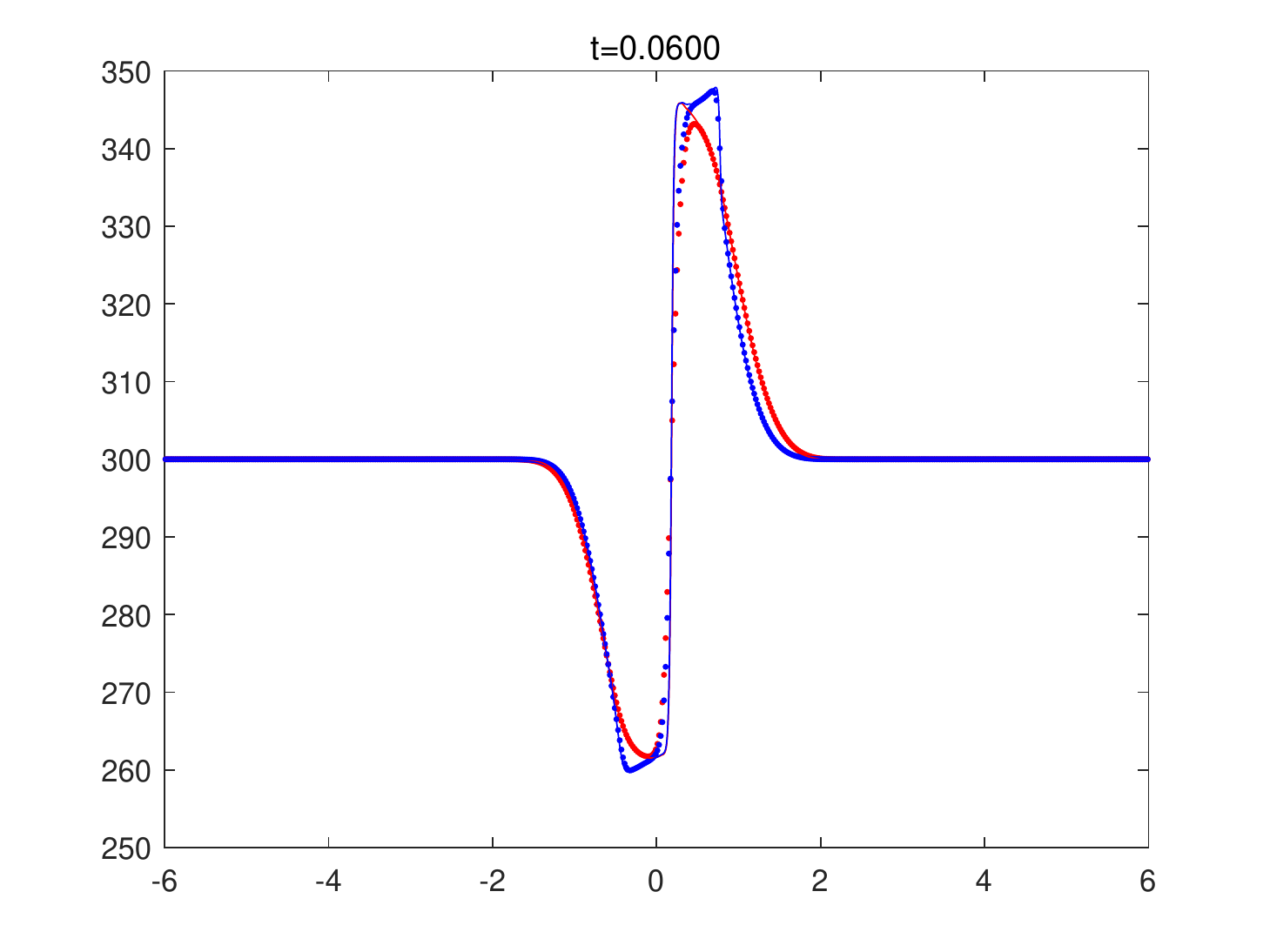}
		\subcaption{Temperature $T$}
	\end{subfigure}		
	\caption{Comparison of the scaled BBGSP model \eqref{bgk bbgsp multi} for $\varepsilon=\kappa=10^{-4}$ with: (left) global velocity and temperature Euler system \eqref{NSE} for $\varepsilon=0$ and (right) multi-velocity and multi-temperature Euler system \eqref{NSE multi} for $\varepsilon=0$, $\kappa=10^{-4}$. We use the initial data in \eqref{sec real parameter}.
	}\label{real parameter 4}
\end{figure}

\begin{figure}[htbp]
	\centering
	\begin{subfigure}[b]{0.43\linewidth}
		\includegraphics[width=1\linewidth]{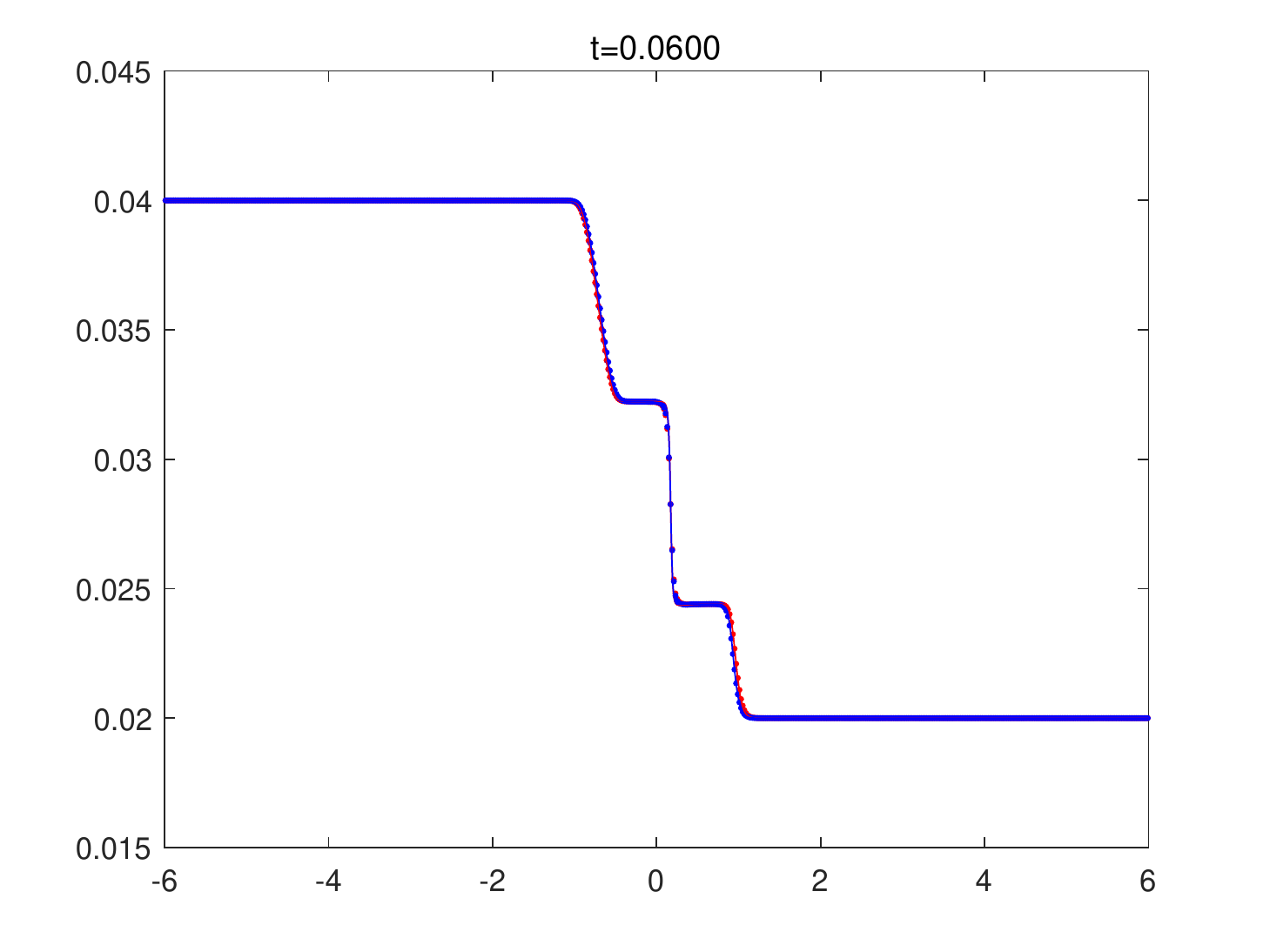}
		\subcaption{Number density $n_s$, $s=1,2$}
	\end{subfigure}
	\begin{subfigure}[b]{0.43\linewidth}
		\includegraphics[width=1\linewidth]{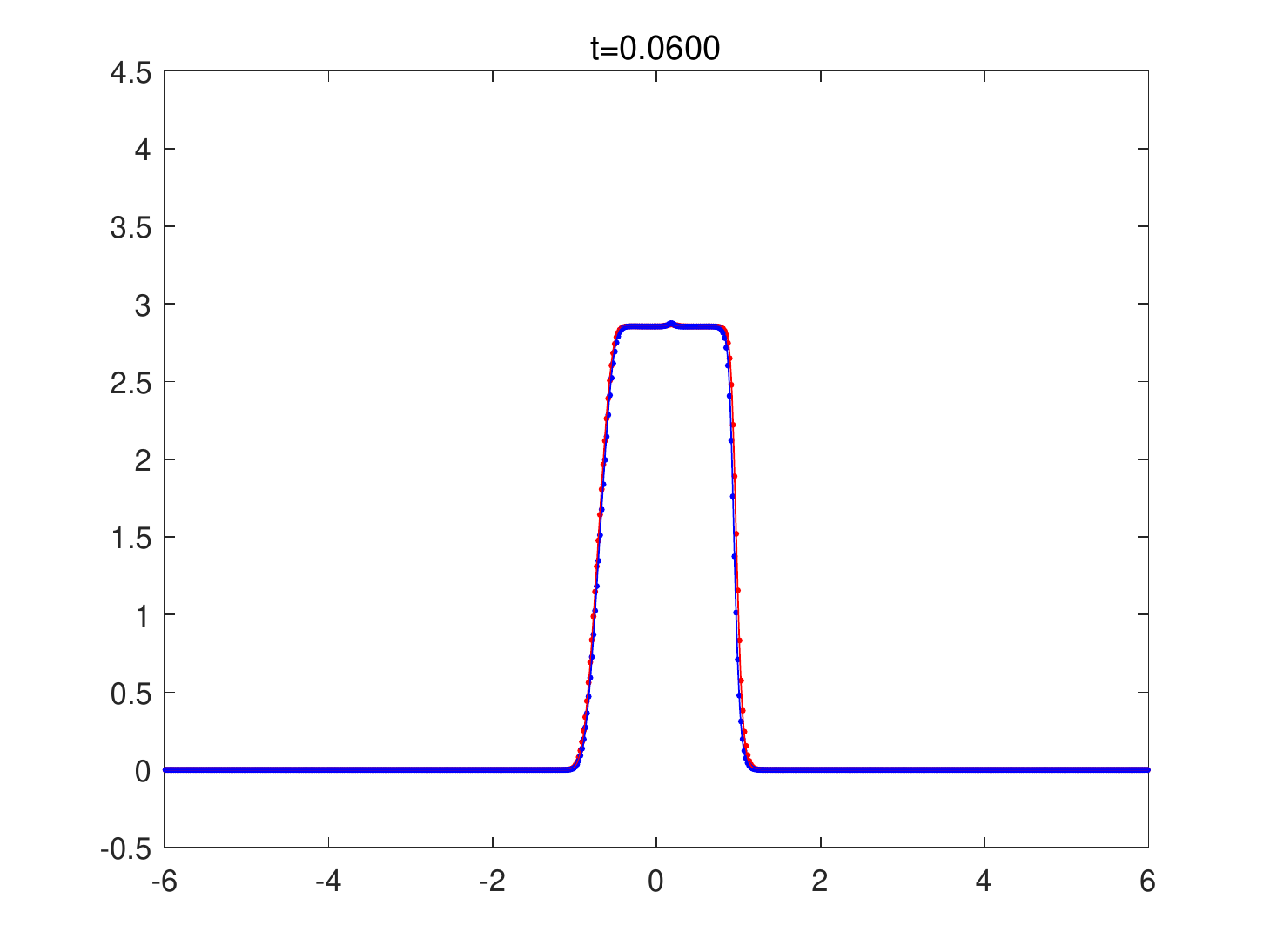}
		\subcaption{Velocity $u$}
	\end{subfigure}
	\begin{subfigure}[b]{0.43\linewidth}
		\includegraphics[width=1\linewidth]{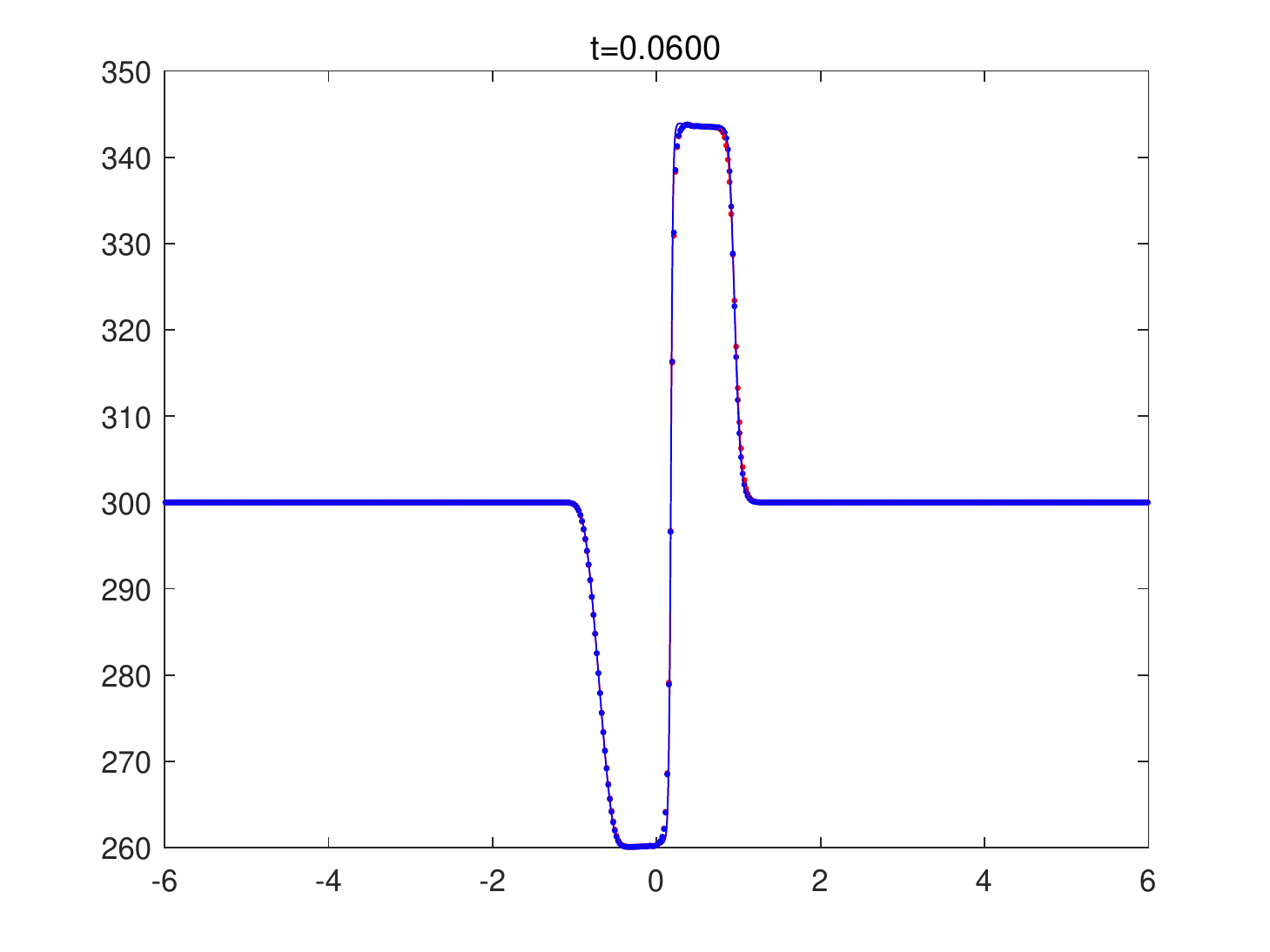}
		\subcaption{Temperature $T$}
	\end{subfigure}		
	\caption{Comparison of the scaled BBGSP model \eqref{bgk bbgsp multi} for $\varepsilon=\kappa=10^{-5}$ with multi-velocity and multi-temperature Euler system \eqref{NSE multi} for $\varepsilon=0$, $\kappa=10^{-5}$. We use the initial data in \eqref{sec real parameter}.
	}\label{real parameter 5}
\end{figure}

\subsection{Stationary shock}

In this test, we consider a classical problem of gas dynamics, namely the shock wave structure obtained by the Navier-Stokes description in a binary mixture of two noble gases: Neon (Ne) and Argon (Ar). This problem has been faced in \cite{BGM} by using the qualitative theory of dynamical systems applied to the time-independent version of the NS equations, which can be rewritten as a suitable system of first order ODEs. Here we alternatively obtain the stationary shock solution as asymptotic solution of the time-dependent Navier-Stokes equations \eqref{NSE} and also by solving the BBGSP model.

For this test, we consider molecular masses: 
\[
m_1= 20 \text{(Ne)},\quad  m_2 = 40 \text{(Ar)}.
\]
Based on \eqref{coll fre formula}, we set collision frequencies for $T=300$K 
by
\begin{align*}
&\nu_0^{11}= 12.46,\quad \nu_0^{12}= 15.22\cr 
&\nu_0^{21}= 15.22,\quad \nu_0^{22}= 17.52.
\end{align*}
We take initial data by the Maxwellian whose macroscopic fields reproduce
\begin{align*}
\begin{split}
\begin{pmatrix}
n_s(x,0)\\
u_s(x,0)\\
T_s(x,0)
\end{pmatrix}
= E_L^s+ \frac{E_R^s-E_L^s}{2}\big(\tanh(ax) +1\big),\quad x\in[-20,20]
\end{split}
\end{align*}
where $a$ is a parameter which adjusts the slope of the smooth jump associated to the initial data (we take $a=2$). The two states $E_L^s$ and $E_R^s$, $s=1,2$, are chosen according to the Rankine-Hugoniot conditions as follows \cite{BGM}: 
\begin{align*} 
\begin{split}
E_R^s&= \left(n_s^{\infty},u^{\infty},T^{\infty}\right),\\
E_L^s&= \left(\frac{4M\!a^2}{M\!a^2+3}n_s^\infty,\,\frac{M\!a^2+3}{4M\!a^2}u^\infty,\,\frac{(5M\!a^2-1)(M\!a^2+3)}{16M\!a^2}T^\infty\right)\cr
c^{\infty}&= \sqrt{\frac{5n^{\infty} T^{\infty}}{3\rho^{\infty}}} = \sqrt{\frac{5(n_1^{\infty}+n_2^{\infty}) T^{\infty}}{3(\rho_1^{\infty}+\rho_2^{\infty})}}\cr
M\!a&= \sqrt{\frac{u^{\infty}}{c^{\infty}}},
\end{split}
\end{align*}
where $M\!a$ is the Mach number.
We consider concentrations $\displaystyle \chi_1=\frac{n_1}{n}=0.1$ and $\displaystyle \chi_2=\frac{n_2}{n}=0.9$, and set 
\begin{align*}
	&n_1= \chi_1 n,\quad n_2= \chi_2 n,\quad M\!a= \sqrt{0.6},\quad T^\infty=300.
\end{align*}
In this problem, we impose the inflow and outflow boundary conditions.
We consider velocity domain $[-32,32]$ and compute numerical solutions with $N_v=80$, $N_x=200$ and CFL$=0.5$. For comparison, we set $\varepsilon=1$ for this problem.

\begin{figure}[htbp]
	\centering
	\begin{subfigure}[b]{1\linewidth}
		\includegraphics[width=1\linewidth]{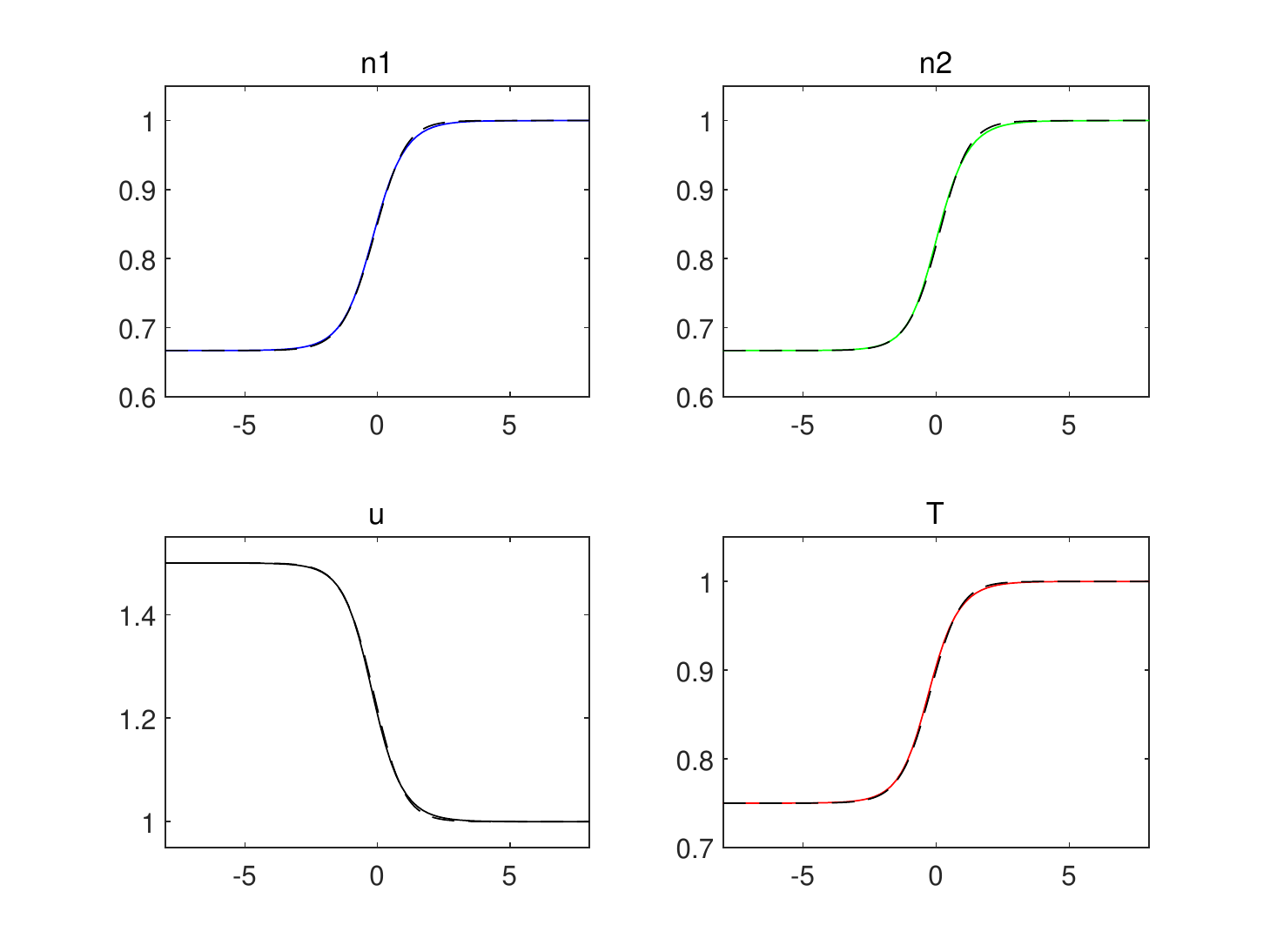}
	\end{subfigure}	
	\caption{BDF3-QCWENO35 for $\varepsilon=\kappa=10^{-0}$. Neon and Argon with $n_1= 0.1m_1,\quad n_2= 0.9m_2$. Black dashed lines are reference NS solutions and solid lines are BGK solutions.}\label{fig neon argon 0 case 1}
\end{figure}
In Figure \ref{fig neon argon 0 case 1}, we plot normalized macroscopic fields:
\[
n_s/n_s^\infty, \quad u/u^\infty, \quad  T/T^\infty.
\]
Our solution shows very good agreement between the long-time solution of the BBGSP model and the reference solution asymptotically obtained from the NS equations \eqref{NSE} discretized by  a MacCormack scheme $(N_x=800)$. Moreover, the results are in accordance with the steady shock wave solution that can be obtained by solving the stationary (ODEs) system of Navier-Stokes equations \cite{BGM}. For other relevant tests, we refer to \cite{BGM}.

%
%



\section*{Acknowledgement}
S. Y. Cho has been supported by ITN-ETN Horizon 2020 Project ModCompShock, Modeling and Computation on Shocks and Interfaces, Project Reference 642768. S. Y. Cho, S. Boscarino and G. Russo would like to thank the Italian Ministry of Instruction, University and Research (MIUR) to support this research with funds coming from PRIN Project 2017 (No. 2017KKJP4X entitled ``Innovative numerical methods for evolutionary partial differential equations and applications"). 
S. Boscarino has been supported by the University of Catania (``Piano della Ricerca 2016/2018, Linea di intervento 2"). S. Boscarino and G. Russo are members of the INdAM Research group GNCS. M. Groppi thanks the support by the
University of Parma, by the Italian National Group of Mathematical
Physics (GNFM-INdAM), and by the Italian National Research Project ``Multiscale phenomena in Continuum Mechanics:
singular limits, off-equilibrium and transitions" (PRIN 2017YBKNCE).


\appendix
\section{Leading error terms in Proposition \ref{prop error} and Remark \ref{remark 4.1}}

\subsection{Proof of Proposition \ref{prop error}}\label{Appendix A1}
\begin{proof}
	\noindent {\bf $\bullet$ Proof of (1)}: To show this, we first express $u^s$ and $u_{sk}$ in terms of $u_s$ and $u_k$ in $\mathcal{E}_u$:
	\begin{align*}
		\sum_{k=1}^L \nu_{sk}  (u^s-u_{sk}) &=  \sum_{k=1}^L \nu_{sk}  \bigg[ \left( u_s +  \frac{1}{m_sn_s\nu_s}\sum_{r=1}^{L}\xi^{sr}u_r\right)-\left((1-a_{sk})u_s + a_{sk}u_k\right) \bigg] \cr
		&=  \sum_{k=1}^L \nu_{sk}  \bigg[ \left( \frac{1}{m_sn_s\nu_s}\sum_{r=1}^{L}\xi^{sr}u_r\right)-a_{sk}(u_k-u_s) \bigg].
	\end{align*}
	By the definition of $\xi^{sr}$ in \eqref{xi gamma} and $a_{sk}$ in \eqref{abgamma original},
	we obtain
	\begin{align*}
		\sum_{k=1}^L \nu_{sk}  (u^s-u_{sk})&=\sum_{k=1}^L \nu_{sk}  \bigg[ \left( \frac{1}{m_sn_s\nu_s}\sum_{r=1}^{L}\xi^{sr}u_r\right)-a_{sk}(u_k-u_s) \bigg] \cr
		&=\sum_{r=1}^{L}\left( \frac{\nu_1^{sr}m_rn_r}{m_s + m_r} - \delta_{sr}\sum_{\ell= 1}^L \frac{\nu_1^{s\ell}m_\ell n_\ell}{m_s + m_\ell}
		\right)u_r-\sum_{k=1}^L   \bigg[ \frac{\lambda_{sk}m_kn_k}{m_s + m_k}(u_k-u_s) \bigg]=0.
	\end{align*}
	In the last line, we use the relation $\nu_1^{s\ell}=\lambda_{s\ell}$ in \eqref{collision fre rel}.\\	
	\noindent {\bf Proof (2)}: To prove this, we 
	begin by 
	splitting in $\mathcal{E}_T$ \eqref{leading order terms} into two parts:
	\begin{align*}
		\sum_{k=1}^L \nu_{sk}  (T^s-T_{sk}) 
		&=I + II + III,
	\end{align*}
	where
	\begin{align*}
		I&=  \sum_{k=1}^L \nu_{sk}  \bigg[ \frac{2}{3n_s K_B\nu_s}\sum_{r=1}^{L}\gamma^{sr}T_r  -b_{sk}(T_k-T_s)\bigg]\cr
		II&=\frac{2m_s}{3 K_B}\sum_{r=1}^{L}\nu_{sr}\bigg[\nu_1^{sr}\frac{m_rn_r}{\nu_{sr}(m_s + m_r)^2}\left(m_su_s + m_ru_r\right)\left(u_r-u_s\right)\bigg]- \sum_{k=1}^L \nu_{sk}\frac{\gamma_{sk}}{K_B}|u_s-u_k|^2\cr
		III&=- \frac{m_s}{3K_B}\sum_{k=1}^L \nu_{sk}  \bigg[ |u^s|^2-|u_s|^2 \bigg]
	\end{align*}
	Inserting $\gamma^{sr}$ in \eqref{xi gamma} into $I$, we have
	\begin{align*}
		I&=\frac{2}{3n_s K_B}\sum_{r=1}^{L}\gamma^{sr}T_r - \sum_{k=1}^L\nu_{sk}b_{sk}(T_k-T_s)\cr
		&=2\sum_{r=1}^{L}\left(\nu_1^{sr}\frac{m_sm_rn_r}{(m_s + m_r)^2} -\delta_{sr} \sum_{\ell = 1}^L \nu_1^{s\ell}\frac{m_s m_\ell n_\ell}{(m_s + m_\ell)^2}\right)T_r - \sum_{k=1}^L\nu_{sk}b_{sk}(T_k-T_s)\cr
		&=0.
	\end{align*}
	In the last line, we use $\displaystyle	b_{sk}=\frac{2a_{sk} m_s}{m_s + m_k}= \frac{2 \lambda_{sk}m_sm_kn_k}{\nu_{sk}(m_s+m_r)^2}$ and $\nu_1^{sk}=\lambda_{sk}.$
Next, we simplify $II$ as
	\begin{align*}
	II
	&=\frac{2m_s}{3 K_B}\sum_{r=1}^{L}\nu_{sr}\bigg[a_{sr}\left(u_s + \frac{m_r}{m_s+m_r}(u_r-u_s
	)\right)\cdot\left(u_r-u_s\right)\bigg]- \sum_{k=1}^L \nu_{sk}\frac{\gamma_{sk}}{K_B}|u_s-u_k|^2\cr
	&=\frac{2m_s}{3 K_B}\sum_{r=1}^{L}\nu_{sr}\bigg[ a_{sr}u_s \cdot \left(u_r-u_s\right) + a_{sr}\frac{m_r}{m_s + m_r}|u_r-u_s|^2 \bigg]- \sum_{k=1}^L \nu_{sk}\frac{\gamma_{sk}}{K_B}|u_s-u_k|^2
\end{align*}
This combined with $\gamma_{sk}$ in \eqref{abgamma original}
gives
	\begin{align*}
		II&=\frac{2m_s}{3 K_B}\sum_{r=1}^{L}\nu_{sk}a_{sk}\bigg[ u_s \cdot \left(u_k-u_s\right)  \bigg] + \frac{m_s}{3K_B} \sum_{k=1}^L \nu_{sk}(a_{sk})^2|u_s-u_k|^2.
	\end{align*}
	Now, the following relation
	\begin{align*}
		|u_{sk}|^2 = |a_{sk} (u_s-u_k) - u_s|^2= (a_{sk})^2 |u_s-u_k|^2 - 2a_{sk} (u_s-u_k)\cdot u_s  + |u_s|^2 
	\end{align*}
	implies
	\begin{align*}
		II+III&=-\frac{m_s}{3K_B}\sum_{k=1}^L \nu_{sk}  \bigg[ |u^s|^2-|u_{sk}|^2 \bigg].
	\end{align*}
	To simplify further this, recall the relation
	\begin{align}\label{us explicit}
		u^s
		&=u_s +\frac{1}{\nu_s}\sum_{r=1}^{L}\nu_{sr}a_{sr}(u_r-u_s),
	\end{align}
	and use this to get
	\begin{align*}
		II+III&=-\frac{m_s}{3K_B}\sum_{k=1}^L \nu_{sk}  \left[ 2u_s + \frac{1}{\nu_s}\sum_{r=1}^{L}\nu_{sr}a_{sr}(u_r-u_s) - a_{sk} (u_s-u_k)\right]\cr
		&\qquad\qquad\qquad\cdot \left[   \frac{1}{\nu_s}\sum_{r=1}^{L}\nu_{sr}a_{sr}(u_r-u_s) + a_{sk} (u_s-u_k)\right]\cr
		&=A + B
	\end{align*}
	where
	\begin{align*}
		A&=-\frac{m_s}{3K_B}\sum_{k=1}^L \nu_{sk}  \left[ 2u_s +  \frac{1}{\nu_s}\sum_{r=1}^{L}\nu_{sr}a_{sr}(u_r-u_s) \right] \left[   \frac{1}{\nu_s}\sum_{r=1}^{L}\nu_{sr}a_{sr}(u_r-u_s) \right]\cr
		&=-\frac{m_s}{3\nu_sK_B}\left[ 2u_s \nu_s +  \sum_{r=1}^{L}\nu_{sr}a_{sr}(u_r-u_s) \right] \cdot \left[   \sum_{r=1}^{L}\nu_{sr}a_{sr}(u_r-u_s) \right]\cr
		B&=-\left(\frac{2m_s u_s}{3K_B} \cdot \sum_{k=1}^L \nu_{sk}   a_{sk} (u_s-u_k) -\frac{m_s}{3K_B}\sum_{k=1}^L \nu_{sk}  (a_{sk})^2 |u_s-u_k|^2\right).
	\end{align*}
	To be more concise, we can rewrite $A$ and $B$ as
	\begin{align*}
		A&=-\frac{2 m_s u_s}{3K_B} \cdot X -\frac{m_s}{3\nu_sK_B} |X|^2\cr
		B&=\frac{2m_s u_s}{3K_B} \cdot X +\frac{m_s}{3K_B}Y	\end{align*}
	where
	\[
	X= \sum_{r=1}^{L}\nu_{sr}a_{sr}(u_r-u_s),\quad Y= \sum_{r=1}^{L}\nu_{sr}(a_{sr})^2|u_r-u_s)|^2.
	\]
	Then, we have
	\begin{align*}
		&II + III\cr
		&= \frac{m_s}{3K_B}\sum_{k=1}^L \nu_{sk}  (a_{sk})^2 |u_s-u_k|^2 - \frac{m_s}{3\nu_sK_B} \left( \sum_{r=1}^{L}\nu_{sr}a_{sr}(u_r-u_s) \right)\cdot \left( \sum_{r=1}^{L}\nu_{sr}a_{sr}(u_r-u_s) \right),
	\end{align*}
	which gives the desired result.
\end{proof}

\subsection{Calculation of $\bar{\mathcal{E}}_u$ and $\bar{\mathcal{E}}_T$ terms in \eqref{leading order terms GS}.}\label{appendix AAPBG}

\begin{proof}
	\noindent {\bf $\bullet$ Proof of (1)}: 
We first rewrite $\bar{u}$ as
\begin{align*}
	\bar{u}&=u_s +  \frac{1}{\sum_{r=1}^L \nu_r m_r n_r}\sum_{r\neq s}\nu_r m_r n_r(u_r-u_s).
\end{align*}
This, combined with the expression of $u^s$ in \eqref{us explicit}, gives
\begin{align*}
	\bar{\mathcal{E}}_u&=\nu_s(u^s-\bar{u})\cr
	&=\nu_s\left(\frac{1}{\nu_s}\sum_{r\neq s}\left(\nu_1^{sr}\frac{m_rn_r}{m_s + m_r}\right)(u_r-u_s)- \frac{1}{\sum_{r=1}^L \nu_r m_r n_r}\sum_{r\neq s}\nu_r m_r n_r(u_r-u_s)\right)\cr
	&=\sum_{r\neq s}\left(\nu_1^{sr}\frac{m_rn_r}{m_s + m_r}-\frac{\nu_s\nu_r m_r n_r}{\sum_{r=1}^L \nu_r m_r n_r}\right)(u_r-u_s).
\end{align*}
	
	\noindent {\bf Proof (2)}: 

%
From the definition of $T^s$ and $\bar{T}$, we have 
	\begin{align*}
		\bar{\mathcal{E}}_T&=
		\nu_s(T^s-\bar{T}) \cr
		&=    \nu_s\bigg[T_s - \frac{m_s}{3K_B} \left(|u^s|^2-|u_s|^2\right)  + \frac{2}{3n_s K_B\nu_s}\sum_{r=1}^{L}\gamma^{sr}T_r\cr
		&\qquad\qquad + \frac{2}{3n_s K_B\nu_s}\sum_{r=1}^{L}\nu_1^{sr}\frac{m_sm_rn_sn_r}{(m_s + m_r)^2}\left(m_su_s + m_ru_r\right)\left(u_r-u_s\right)\cr
		&\qquad\qquad-\left(T_s -\frac{\sum_{r=1}^L \nu_r n_r m_r \left(|\bar{u}|^2-|u_r|^2   \right)  }{3K_B\sum_{r=1}^L\nu_r n_r}  + \frac{\sum_{r\neq s} \nu_r n_r (T_r-T_s) }{\sum_{r=1}^L\nu_r n_r}\right)\bigg]
		 \cr
		&=J_1 + J_2 +J_3,
	\end{align*}
	where
	\begin{align*}
		J_1&= \nu_s\left[T_s+ \frac{2}{3n_s K_B\nu_s}\sum_{r=1}^{L}\gamma^{sr}T_r-\left(T_s -\frac{\sum_{r=1}^L \nu_r n_r m_r \left(|\bar{u}|^2-|u_r|^2   \right)  }{3K_B\sum_{r=1}^L\nu_r n_r}  + \frac{\sum_{r\neq s} \nu_r n_r (T_r-T_s) }{\sum_{r=1}^L\nu_r n_r}\right)\right]\cr
		J_2&= \nu_s\left[\frac{\sum_{r=1}^L \nu_r n_r m_r \left(|\bar{u}|^2-|u_r|^2   \right)  }{3K_B\sum_{r=1}^L\nu_r n_r} \right]\cr
		J_3&=  \sum_{k=1}^L \nu_{sk}  \bigg[- \frac{m_s}{3K_B} \left(|u^s|^2-|u_s|^2\right)+ \frac{2}{3 K_B\nu_s}\sum_{r=1}^{L}\nu_1^{sr}\frac{m_sm_rn_r}{(m_s + m_r)^2}\left(m_su_s + m_ru_r\right)\left(u_r-u_s\right)\bigg].
	\end{align*}
For $J_1$, we use
\begin{align*}
	\gamma^{sr}= 3K_B\nu_1^{sr}\frac{m_sm_rn_sn_r}{(m_s + m_r)^2} -\delta_{sr} 3K_B\sum_{\ell= 1}^L \nu_1^{s\ell}\frac{m_s m_\ell n_s n_\ell}{(m_s + m_\ell)^2},
\end{align*}
to obtain
\begin{align*}
	J_1
	&= \nu_s\bigg[  \frac{2}{\nu_s}\left(\nu_1^{ss}\frac{m_sm_sn_s}{(m_s + m_s)^2} - \sum_{\ell= 1}^L \nu_1^{s\ell}\frac{m_s m_\ell n_\ell}{(m_s + m_\ell)^2}\right)T_s+\frac{2}{\nu_s}\sum_{r\neq s}\nu_1^{sr}\frac{m_sm_rn_r}{(m_s + m_r)^2}T_r\cr
	&\quad-\left( \frac{\sum_{r\neq s} \nu_r n_r (T_r-T_s) }{\sum_{r=1}^L\nu_r n_r}\right)\bigg]\cr
	&=  \sum_{r\neq s}\left(2\nu_1^{sr}\frac{m_sm_rn_r}{(m_s + m_r)^2}- \frac{\nu_s\nu_r n_r}{\sum_{r=1}^L\nu_r n_r}\right)(T_r-T_s).
\end{align*}
Next, for $J_2$, we use
\begin{align*}
|\bar{u}|^2-|u_r|^2&=\left(\frac{\sum_{\ell=1}^L \nu_\ell m_\ell n_\ell (u_\ell- u_r)}{\sum_{\ell=1}^L \nu_\ell m_\ell n_\ell}\right)\cdot \left(\frac{\sum_{\ell=1}^L \nu_\ell m_\ell n_\ell (u_\ell+ u_r)}{\sum_{\ell=1}^L \nu_\ell m_\ell n_\ell} \right).
\end{align*}
Then, we have
\begin{align*}
	J_1+J_2&=  \sum_{r\neq s}\left(2\nu_1^{sr}\frac{m_sm_rn_r}{(m_s + m_r)^2}- \frac{\nu_s\nu_r n_r}{\sum_{r=1}^L\nu_r n_r}\right)(T_r-T_s)\cr
	&\quad+\nu_s\frac{\sum_{r=1}^L \nu_r n_r m_r \left[\left(\sum_{\ell=1}^L \nu_\ell m_\ell n_\ell (u_\ell- u_r)\right)\cdot \left(\sum_{\ell=1}^L \nu_\ell m_\ell n_\ell (u_\ell+ u_r) \right)  \right] }{3K_B\sum_{r=1}^L\nu_r n_r\left|\sum_{\ell=1}^L \nu_\ell m_\ell n_\ell)\right|^2}
\end{align*}
For $J_3$, we use
\begin{align*}
	|u^s|^2-|u_s|^2=\left( \frac{1}{\nu_s}\sum_{r\neq s}\left(\nu_1^{sr}\frac{m_rn_r}{m_s + m_r}\right)(u_r-u_s)\right)\cdot\left(2u_s + \frac{1}{\nu_s}\sum_{r\neq s}\left(\nu_1^{sr}\frac{m_rn_r}{m_s + m_r}\right)(u_r-u_s)\right) 
\end{align*}
to get
\begin{align*}
	J_3&=  \nu_{s}  \bigg[- \frac{m_s}{3K_B} \left(|u^s|^2-|u_s|^2\right)+ \frac{2}{3 K_B\nu_s}\sum_{r=1}^{L}\nu_1^{sr}\frac{m_sm_rn_r}{(m_s + m_r)^2}\left(m_su_s + m_ru_r\right)\left(u_r-u_s\right)\bigg]\cr
	&=  - \frac{m_s}{3K_B} \left( \sum_{r\neq s}\left(\nu_1^{sr}\frac{m_rn_r}{m_s + m_r}\right)(u_r-u_s)\right)\cdot\left(2u_s + \frac{1}{\nu_s}\sum_{r\neq s}\left(\nu_1^{sr}\frac{m_rn_r}{m_s + m_r}\right)(u_r-u_s)\right) \cr
	&\quad+   \frac{2m_s}{3 K_B}\sum_{r\neq s}\nu_1^{sr}\frac{m_rn_r}{m_s + m_r}\left(u_s + \frac{m_r}{m_s + m_r}(u_r-u_s)\right)\left(u_r-u_s\right)\cr
	&=  - \frac{m_s}{3K_B} \left( \sum_{r\neq s}\left(\nu_1^{sr}\frac{m_rn_r}{m_s + m_r}\right)(u_r-u_s)\right)\cdot\left( \frac{1}{\nu_s}\sum_{r\neq s}\left(\nu_1^{sr}\frac{m_rn_r}{m_s + m_r}\right)(u_r-u_s)\right) \cr
	&\quad+   \frac{2m_s}{3 K_B}\sum_{r\neq s}\nu_1^{sr}\frac{m_r^2n_r}{(m_s + m_r)^2}\left|u_r-u_s\right|^2.
\end{align*} 
To sum up, we rewrite $\bar{\mathcal{E}}_u$ in terms of  $u_s$ and $T_s$ with $\bar{\mathcal{E}}_u=J_1+J_2+J_3$.
\end{proof}
\section{Representation of NS equations for $n_s$, $u$ and $T$}
\subsection{Derivation of \eqref{NSE rewrite}}	
	We begin with the first equation in \eqref{NSE}:
	\begin{align*}
	\begin{split}
	&\frac{\partial n_s}{\partial t} =- \nabla \cdot(n_s u) - \varepsilon \nabla \cdot(n_s u_s^{(1)}), \quad s=1,\cdots,L.
	\end{split}
	\end{align*}
	The sum of these $L$ equations leads to
	\begin{align}\label{single dn drho}
	\begin{split}
	&\frac{\partial n}{\partial t} =- \nabla \cdot(n u) -  \nabla \cdot\left(\sum_{s=1}^L\varepsilon n_s u_s^{(1)}\right),\cr
	&\frac{\partial \rho}{\partial t} =- \nabla \cdot(\rho u) -  \nabla \cdot\left(\sum_{i=1}^L\varepsilon \rho_s u_s^{(1)}\right).
	\end{split}
	\end{align}
	Next, we rewrite the second equation in \eqref{NSE} as
	\begin{align*}
	\begin{split}
	&u\frac{\partial \rho}{\partial t} + \rho\frac{\partial u}{\partial t} + u\nabla \cdot(\rho u) + \rho u\nabla \cdot u +  \nabla(n K_B T) + \varepsilon \nabla \cdot(P^{(1)})=0.
	\end{split}
	\end{align*}
	This together with \eqref{single dn drho} gives 
	\begin{align}\label{single du}
	\begin{split}
	\frac{\partial u}{\partial t}&=\frac{u}{\rho} \left(  \nabla \cdot\left(\sum_{i=1}^L\varepsilon \rho_s u_s^{(1)}\right)\right) - u\nabla \cdot u -  \frac{\nabla(n K_B T)}{\rho} - \frac{\varepsilon \nabla \cdot(P^{(1)})}{\rho}.
	\end{split}
	\end{align}
	For the third equation in \eqref{NSE}, we transform it into the following form:
	\begin{align}\label{NSE 3rd rewrite}
	\begin{split}
	\frac{\rho u}{2}  \cdot \frac{\partial u}{\partial t} + \frac{u}{2} \cdot \frac{\partial}{\partial t}\left(\rho u\right) + \frac{3K_BT}{2}\frac{\partial n}{\partial t} + \frac{3nK_B}{2}\frac{\partial T}{\partial t}&\cr 
	+ \nabla\left(\frac{1}{2}\rho |u|^2\right) \cdot u  + \left(\frac{1}{2}\rho |u|^2\right) \nabla \cdot u &\cr
	+ \nabla\left( \frac{5}{2}nK_BT\right) \cdot u + \left( \frac{5}{2}nK_BT\right) \nabla \cdot u&\cr
	+ \varepsilon \nabla \cdot(P^{(1)}\cdot u) + \varepsilon \nabla \cdot q^{(1)}&=0.
	\end{split}
	\end{align}
	To simplify this, we use
	\begin{align}\label{del rel}
		\nabla\left(\textbf{a}\cdot \textbf{b}\right)= \textbf{a} \times \left(\nabla\times \textbf{b}\right) + \textbf{b} \times \left(\nabla\times \textbf{a}\right) + (\textbf{a} \cdot \nabla)\textbf{b} + (\textbf{b} \cdot \nabla)\textbf{a},
	\end{align} 
	to have
	\begin{align*}
	\begin{split}
		&\nabla\left(\frac{1}{2}\rho |u|^2\right) =\nabla\left(\frac{\rho u}{2} \cdot u\right) =\left(\frac{\rho u}{2} \times \left(\nabla\times u\right) + u \times \left(\nabla\times \frac{\rho u}{2}\right) +  \left(\frac{\rho u}{2} \cdot \nabla \right)  u + \left(u \cdot \nabla\right)\frac{\rho u}{2}\right).
	\end{split}
\end{align*}
Also, we use the following decomposition:
\begin{align*}
	\begin{split}
		&\nabla\left( \frac{5}{2}nK_BT\right) \cdot u=\frac{1}{\rho}\nabla\left( nK_BT\right) \cdot \frac{\rho u}{2} + \nabla\left( \frac{3}{2}nK_BT\right) \cdot u + \nabla\left( \frac{1}{2}nK_BT\right) \cdot u.
	\end{split}
\end{align*}
Using these relations, we write \eqref{NSE 3rd rewrite} in the following form:
	\begin{align*}
	\begin{split}
	&\frac{\rho u}{2}  \cdot \left(\frac{\partial u}{\partial t} + u \cdot \nabla u + \frac{1}{\rho}\nabla\left( nK_BT\right)\right) + \frac{u}{2} \cdot \left( \frac{\partial}{\partial t}\left(\rho u\right) + \nabla \cdot(\rho u \otimes u)+   \nabla\left(nK_BT\right)\right)\cr 
	&+ \frac{3K_BT}{2}\left(\frac{\partial n}{\partial t} + \nabla n \cdot u  + n \nabla \cdot u\right) + \frac{3nK_B}{2}\frac{\partial T}{\partial t}\cr 
	&+ \left(\frac{\rho u}{2} \times \left(\nabla\times u\right) + u \times \left(\nabla\times \frac{\rho u}{2}\right) \right) \cdot u\cr
	&+ \left(\left(u \cdot \nabla\right)\frac{\rho u}{2}\right) \cdot u + \left(\frac{1}{2}\rho |u|^2\right) \nabla \cdot u - \nabla \cdot(\rho u \otimes u) \cdot \frac{u}{2}\cr
	&+ \left(\frac{3}{2}nK_B \nabla T\right) \cdot u + nK_BT \nabla \cdot u + \varepsilon \nabla \cdot(P^{(1)}\cdot u) + \varepsilon \nabla \cdot q^{(1)}=0.
	\end{split}
	\end{align*}
	Recalling \eqref{single du}, the second equation in \eqref{NSE}, \eqref{single dn drho} and 
	\begin{align*}
	\left(\frac{\rho u}{2} \times \left(\nabla\times u\right) + u \times \left(\nabla\times \frac{\rho u}{2}\right) \right) \cdot u&=0\cr
	\left(\left(u \cdot \nabla\right)\frac{\rho u}{2}\right) \cdot u + \left(\frac{1}{2}\rho |u|^2\right) \nabla \cdot u &= \nabla \cdot(\rho u \otimes u) \cdot \frac{u}{2}
	\end{align*}
	we finally derive
	\begin{align*}
	\begin{split}
	&\frac{u}{2}  \cdot \left(u \left(  \nabla \cdot\left(\sum_{i=1}^L\varepsilon \rho_s u_s^{(1)}\right)\right) \right)- u \cdot \left( \varepsilon \nabla \cdot(P^{(1)})\right)\cr 
	&+ \frac{3K_BT}{2}\left(-  \nabla \cdot\left(\sum_{s=1}^L\varepsilon n_s u_s^{(1)}\right) \right)+ \frac{3nK_B}{2}\frac{\partial T}{\partial t}\cr 
	&+ \left(\frac{3}{2}nK_B \nabla T\right) \cdot u + nK_BT \nabla \cdot u + \varepsilon \nabla \cdot(P^{(1)}\cdot u) + \varepsilon \nabla \cdot q^{(1)}=0.
	\end{split}
	\end{align*}
This gives the expression in \eqref{NSE rewrite}.

\subsection{Derivation of \eqref{NSE multi rewrite}}\label{App 4}
	We begin with the first equation in \eqref{NSE multi}:
	\begin{align}\label{multi dn}
	\begin{split}
	&\frac{\partial n_s}{\partial t} =- \nabla \cdot(n_s u_s), \quad s=1,\cdots,L.
	\end{split}
	\end{align}
	which implies
	\begin{align*}
	\begin{split}
	&\frac{\partial \rho_s}{\partial t} =- \nabla \cdot(\rho_s u_s).
	\end{split}
	\end{align*}
	Next, we use this to simplify the second equation in \eqref{NSE multi}
as
	\begin{align*}
	\begin{split}
	-\rho_s\frac{\partial u_s}{\partial t}&=u_s \left(\frac{\partial \rho_s}{\partial t}  + \nabla \cdot(\rho_s u_s)\right) + \rho_s u_s\nabla \cdot u_s +  \nabla(n_s K_B T_s) + \varepsilon \nabla \cdot(P_s^{(1)}) - \sum_{k\neq s }^{L}\mathcal{R}_{sk}\cr
	&=\rho_s u_s\nabla \cdot u_s +  \nabla(n_s K_B T_s) + \varepsilon \nabla \cdot(P_s^{(1)}) - \sum_{k\neq s }^{L}\mathcal{R}_{sk}.
	\end{split}
	\end{align*}
	This reduces to 
	\begin{align}\label{multi du}
	\begin{split}
	\frac{\partial u_s}{\partial t}&=- u_s\nabla \cdot u_s -  \frac{\nabla(n_s K_B T_s)}{\rho_s} - \frac{\varepsilon \nabla \cdot(P_s^{(1)})}{\rho_s} + \frac{1}{\rho_s}\sum_{k\neq s }^{L}\mathcal{R}_{sk}.
	\end{split}
	\end{align}
	For the third equation in \eqref{NSE multi}, 
	we rewrite it as
	\begin{align}\label{NSE multi 3rd rewrite}
	\begin{split}
	\frac{\rho_s u_s}{2}  \cdot \frac{\partial u_s}{\partial t} + \frac{u_s}{2} \cdot \frac{\partial}{\partial t}\left(\rho_s u_s\right) + \frac{3K_BT_s}{2}\frac{\partial n_s}{\partial t} + \frac{3n_sK_B}{2}\frac{\partial T_s}{\partial t}&\cr 
	+ \nabla\left(\frac{1}{2}\rho_s |u_s|^2\right) \cdot u_s  + \left(\frac{1}{2}\rho_s |u_s|^2\right) \nabla \cdot u_s &\cr
	+ \nabla\left( \frac{5}{2}n_sK_BT_s\right) \cdot u_s + \left( \frac{5}{2}n_sK_BT_s\right) \nabla \cdot u_s&\cr
	+ \varepsilon \nabla \cdot(P_s^{(1)}\cdot u_s) + \varepsilon \nabla \cdot q_s^{(1)}&=\sum_{k\neq s }^{L}\mathcal{S}_{sk}.
	\end{split}
	\end{align}
	To simplify this, we use \eqref{del rel}
	to obtain
	\begin{align*}
		\begin{split}
&\nabla\left(\frac{1}{2}\rho_s |u_s|^2\right) \cdot u_s\cr
&=\left(\frac{\rho_s u_s}{2} \times \left(\nabla\times u_s\right) + u_s \times \left(\nabla\times \frac{\rho_s u_s}{2}\right) +  \left(\frac{\rho_s u_s}{2} \cdot \nabla \right) u_s + \left(u_s \cdot \nabla\right)\frac{\rho_s u_s}{2}\right)\cdot u_s.
		\end{split}
	\end{align*}
Using this and the following decomposition:
\begin{align*}
	 \nabla\left( \frac{5}{2}n_sK_BT_s\right) \cdot u_s= \frac{1}{\rho_s}\nabla\left( n_sK_BT_s\right) \cdot \frac{\rho_s u_s}{2} + \nabla\left( \frac{3}{2}n_sK_BT_s\right) \cdot u_s + \nabla\left( \frac{1}{2}n_sK_BT_s\right) \cdot u_s&.
\end{align*}
we can rewrite \eqref{NSE multi 3rd rewrite} as
	\begin{align*}
	\begin{split}
	&\frac{\rho_s u_s}{2}  \cdot \left(\frac{\partial u_s}{\partial t} +  u_s \cdot \nabla u_s + \frac{1}{\rho_s}\nabla\left( n_sK_BT_s\right)\right)\cr
	&+ \frac{u_s}{2} \cdot \left( \frac{\partial}{\partial t}\left(\rho_s u_s\right) + \nabla \cdot(\rho_s u_s \otimes u_s)+   \nabla\left(n_sK_BT_s\right)\right)\cr 
	&+ \frac{3K_BT_s}{2}\left(\frac{\partial n_s}{\partial t} + \nabla n_s \cdot u_s  + n_s \nabla \cdot u_s\right) + \frac{3n_sK_B}{2}\frac{\partial T_s}{\partial t}\cr 
	&+ \left(\frac{\rho_s u_s}{2} \times \left(\nabla\times u_s\right) + u_s \times \left(\nabla\times \frac{\rho_s u_s}{2}\right) \right) \cdot u_s\cr
	&+ \left(\left(u_s \cdot \nabla\right)\frac{\rho_s u_s}{2}\right) \cdot u_s + \left(\frac{1}{2}\rho_s |u_s|^2\right) \nabla \cdot u_s - \nabla \cdot(\rho_s u_s \otimes u_s) \cdot \frac{u_s}{2}\cr
	&+ \left(\frac{3}{2}n_sK_B \nabla T_s\right) \cdot u_s + n_sK_BT_s \nabla \cdot u_s + \varepsilon \nabla \cdot(P_s^{(1)}\cdot u_s) + \varepsilon \nabla \cdot q_s^{(1)}=\sum_{k\neq s }^{L}\mathcal{S}_{sk}.
	\end{split}
	\end{align*}
	Finally, we use \eqref{multi du}, the second equation in \eqref{NSE multi}, \eqref{multi dn} and the following relations: 
	\begin{align*}
	\left(\frac{\rho_s u_s}{2} \times \left(\nabla\times u_s\right) + u_s \times \left(\nabla\times \frac{\rho_s u_s}{2}\right) \right) \cdot u_s&=0\cr	\left(\left(u_s \cdot \nabla\right)\frac{\rho_s u_s}{2}\right) \cdot u_s + \left(\frac{1}{2}\rho_s |u_s|^2\right) \nabla \cdot u_s &= \nabla \cdot(\rho_s u_s \otimes u_s) \cdot \frac{u_s}{2}
	\end{align*}
	to derive
	\begin{align*}
	\begin{split}
	&u_s \cdot \left( -\varepsilon \nabla \cdot(P_s^{(1)}) + \sum_{k\neq s }^{L}\mathcal{R}_{sk}\right)+ \frac{3n_sK_B}{2}\frac{\partial T_s}{\partial t}\cr 
	&+ \left(\frac{3}{2}n_s K_B \nabla T_s\right) \cdot u_s + n_s K_B T_s \nabla \cdot u_s + \varepsilon \nabla \cdot(P_s^{(1)}\cdot u_s) + \varepsilon \nabla \cdot q_s^{(1)}=\sum_{k\neq s }^{L}\mathcal{S}_{sk}.
	\end{split}
	\end{align*}
This gives the expression in \eqref{NSE multi rewrite}.


\bibliographystyle{amsplain}

\end{document}